\documentclass[twocolumn,showpacs,showkeys,preprintnumbers,superscriptaddress,amsmath,floatfix,amssymb,secnumarabic,nofootinbib]{revtex4-1}
\usepackage[colorlinks=true]{hyperref}
\usepackage{graphicx}
\usepackage{dblfloatfix}    
\usepackage{placeins}
\usepackage{braket}
\usepackage{amsmath}
\usepackage{epsfig}
\usepackage{float}
\usepackage{nicefrac}
\usepackage{xcolor}
\usepackage{color}
\usepackage{bm}
\usepackage{subfigure}
\usepackage{chngcntr}
\usepackage{ulem}

\def\({\left(}
\def\){\right)}
\def\[{\left[}
\def\]{\right]}

\newcommand{\ve}[1]{\boldsymbol{#1}}

\newcommand{\diag}[1]{ {\rm diag} \, \left( #1 \right) }

\newcommand{\beq} {\begin{eqnarray}}
\newcommand{\eeq} {\end{eqnarray}}

\newcommand{\comment}[1]{}

\begin{document}
\sloppy

\title{Mitigating  spikes in fermion Monte  Carlo  methods  by  reshuffling  measurements }

\author{Maksim~Ulybyshev}
\email{Maksim.Ulybyshev@physik.uni-wuerzburg.de}
\affiliation{Institut f\"ur Theoretische Physik und Astrophysik, Universit\"at W\"urzburg, 97074 W\"urzburg, Germany}

\author{Fakher~Assaad}
\email{Fakher.Assaad@physik.uni-wuerzburg.de}
\affiliation{Institut f\"ur Theoretische Physik und Astrophysik, Universit\"at W\"urzburg, 97074 W\"urzburg, Germany}
\affiliation{W\"urzburg-Dresden Cluster of Excellence ct.qmat, Am Hubland, 97074 W\"urzburg, Germany}

\begin{abstract}
    
We propose a method to mitigate  heavy-tailed distributions  in  fermion Quantum Monte Carlo simulations  originating  from zeros of  the  fermion determinant. In this case the second moment of the observables might be not well defined, and we show that  by  merely changing  the synchronization between local updates and computation of observables, one can reduce the prefactor of the  heavy-tailed distribution, thus substantially suppressing statistical fluctuations of observables. We also show that  the average,   or  the  first moment, is well  defined  and hence  is  independent on  our   measuring scheme. The method is especially suitable for local observables similar to e.g. double occupancy, where the resulting speedup can reach two orders of magnitude. For observables, containing spatial correlators, the speedup is more moderate, but still ranges between five and ten.     Our  results are independent on the  nature  of the  auxiliary 
field,  discrete or  continuous,  and  pave  the  way  to improve  measurement  strategies for  Hybrid Monte  Carlo   simulations.

\end{abstract}

\maketitle

\section{\label{sec:Intro}Introduction}
\begin{figure}[]
   \centering
   \subfigure[]{\label{fig:histories_global}\includegraphics[width=0.48\textwidth , angle=0]{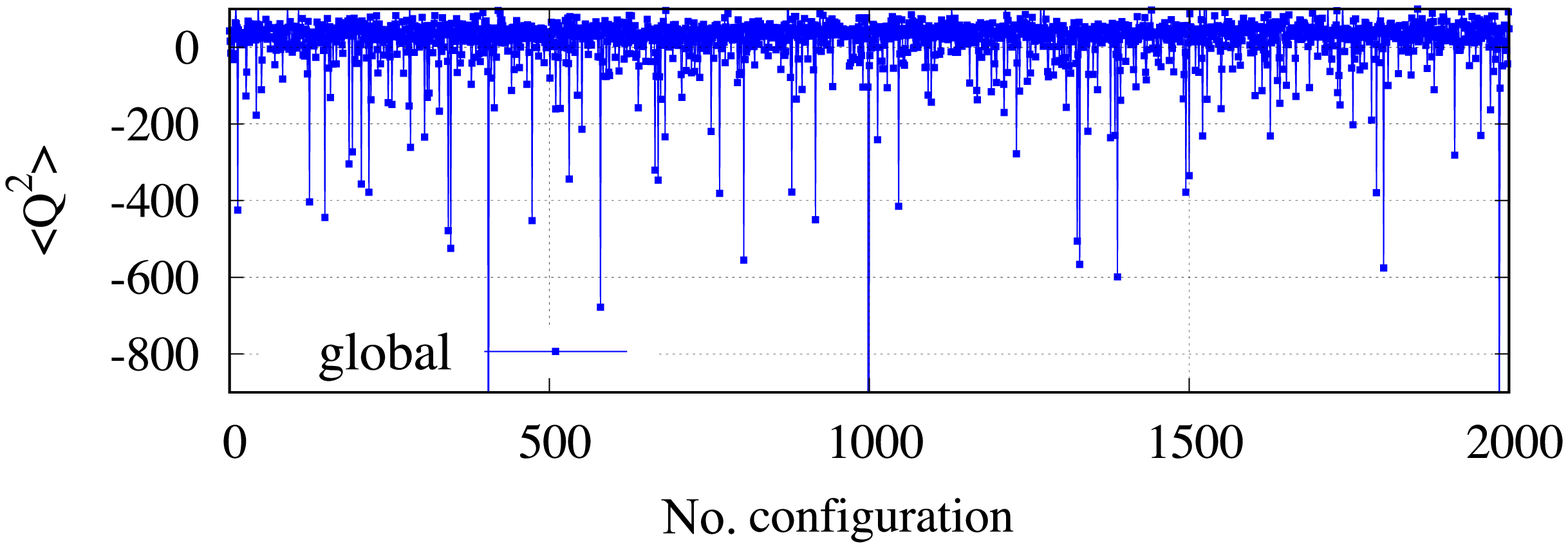}}
   \subfigure[]{\label{fig:histories_Tlocal}\includegraphics[width=0.48\textwidth , angle=0]{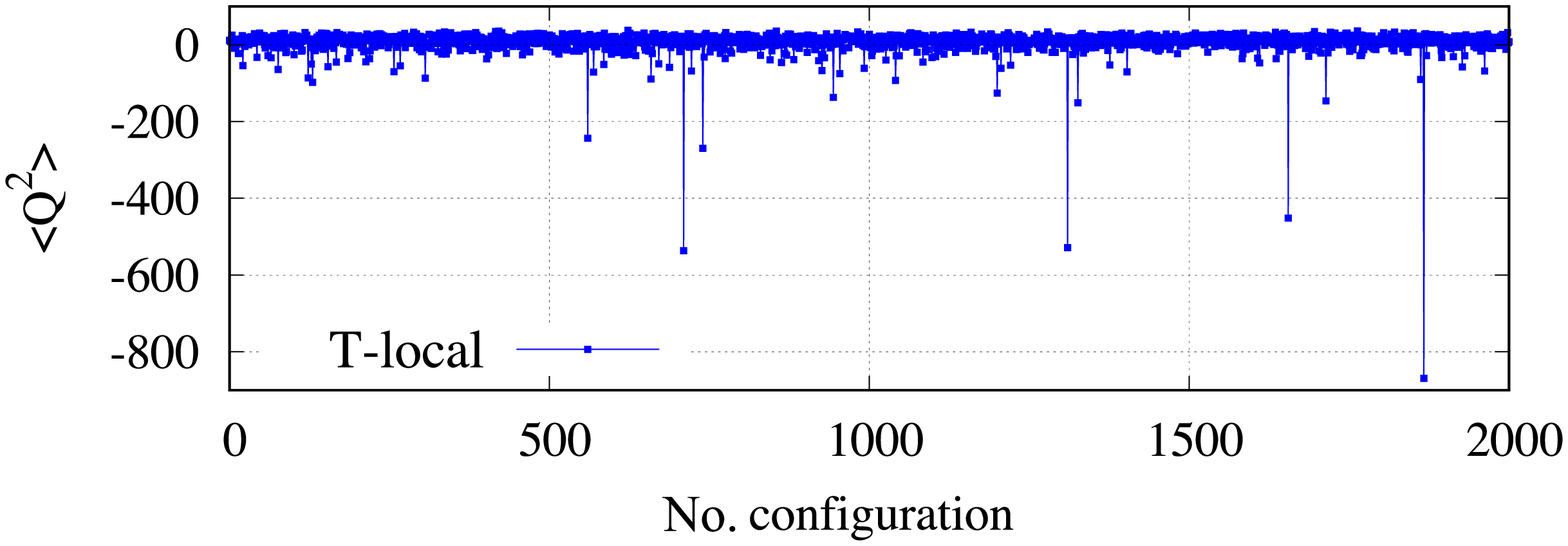}}
    \subfigure[]{\label{fig:histories_STlocal}\includegraphics[width=0.48\textwidth , angle=0]{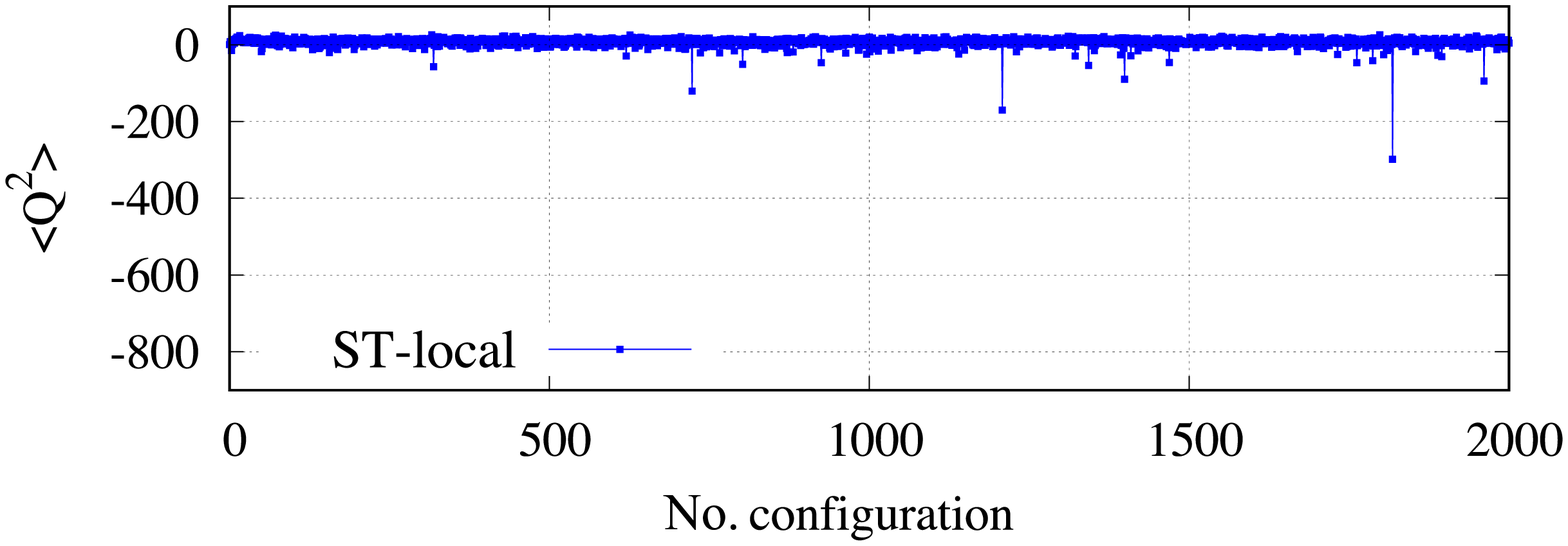}}
       \caption{Sublattice  charge  fluctuations of Eq.~\ref{eq:squared_charge}   for  the   global, Eq.~\ref{eq:Global} (upper plot), 
         T-local,  Eq.~\ref{eq:Tlocal} (middle  plot)  and  ST-local Eq.~\ref{eq:STlocal} (lower plot)  schemes. 
         Calculations were done on a  $6\times6\times256$ hexagonal lattice at $U=5.0$ and $\beta=20.0$ using the SU(2) symmetrical discrete 
         Hubbard  Stratonovich (HS) field 
         within a  standard local  update BSS-QMC  algorithm \cite{ALF_v2}. 
           }
   \label{fig:histories}
\end{figure}

Many numerical and  analytical  schemes  for  identical  fermions   involve the fermion determinant.    This  quantity   possesses  zeros  that  are  
at the  origin of   many issues  in   Quantum Monte Carlo (QMC) methods \cite{Blankenbecler:1981jt,Sugiyama86,Duane87,White89,Sorella89,Assaad08_rev}.  This  stems from the fact  that  although the  weight of  the 
configuration is  proportional to the  fermion determinant,   the Green  function -- required  to compute   observables --   corresponds to the inverse of the fermion matrix. Hence,  if the stochastic walk is   close  to a  zero of the fermion determinant, then 
observables   will show  spikes (see Fig. \ref{fig:histories}),   thus potentially  leading to a fat tailed  distribution \cite{Hao16}.    Here we  show that the tail of  this distribution  can  be  substantially  reduced in  magnitude, at  zero cost, by   merely   reshuffling the  sequence of measurements.

Our  intuition  is  based on the following  argument.    The  acceptance  of  local updates in  fermion Monte  Carlo is based on  the local Green function.   If  this quantity  is big,  due to proximity of a zero mode,  then  the  acceptance will be close to unity,   and  the Monte Carlo  walk  will  move away  from the zero  mode.  Hence if  we  measure local  observables  just  after  the  update of  the local  site,     we   conjecture  that  this synchronization will mitigate  spikes   as  opposed   to   measurement   schemes  where  updates and  the calculation of  observables are not  correlated.

We now   state  our  conjecture   more   precisely. Fermion Monte  Carlo methods rely on  a path integral formulation of the partition function: 
\begin{equation}
	   Z    =  \int   D\left\{ s_{\ve{i}, \tau}    \right\}    e^{ -  S \left(  \left\{ s_{\ve{i}, \tau}    \right\}  \right)  }. \label{eq:basic_parititon_func}
\end{equation}
Here,    $(\ve{i}, \tau)$ is    a  space-time  index and    the  integration over  the fields $s_{\ve{i}, \tau} $     is carried  out  with Monte Carlo methods.   We will not dwell on the negative sign problem and  thereby assume that the action  $S$  is a  real function of the fields. Let $C^{n}_{\ve{i},\tau}$  be  the configuration obtained  after updating the  field   at    space-time   site $ {\ve{i},\tau} $  of the  $n^{th}$  sweep  and   $O_{\ve{i},\tau}(C)$  the   value   of a  local  variable  for the   configuration $C$.  Note that a sweep  corresponds  to  updating  all fields in the space  time lattice. 
We will consider three different ways of synchronizing  local updates of the fields $s_{\ve{i}, \tau}$ and measurement of observables.
 \begin{itemize}
\item Global scheme:
   \begin{equation}
\label{eq:Global}
	\langle  O \rangle   =   \frac{1}{N}  \sum_{n=1}^{N}    \sum_{\ve{i},\tau }O_{\ve{i},\tau}(C^{n}_{V,L_\tau}).
\end{equation} 
    One   carries  out  the  measurement   after  having  updated  all the   sites $ \ve{i} : 1  \cdots  V $  and   time   slices $\tau:1 \cdots L_{\tau} $.  
    This measurement scheme is  used in  Langevin  \cite{Batrouni85,ALF_v2,Goetz21} or Hybrid Monte  Carlo (HMC)  calculations  \cite{Duane87,Assaad_complex,PhysRevD.101.014508}. 
\item Time(T)-local scheme (standard for present-day BSS-QMC algorithms \cite{ALF2017}):
       \begin{equation}
\label{eq:Tlocal}
	\langle  O \rangle   =   \frac{1}{N}  \sum_{n=1}^{N}    \sum_{\ve{i},\tau }O_{\ve{i},\tau}(C^{n}_{V,\tau}).
\end{equation} 
    One   carries  out  the  measurement of the observable  $O_{\ve{i},\tau}$ at a  given time slice $\tau$ after  having  updated  all fields on this time slice $ \ve{i} : 1  \cdots  V $. 
\item Space-Time(ST)-local scheme:
\begin{equation}
\label{eq:STlocal}
	\langle  O \rangle   =   \frac{1}{N}  \sum_{n=1}^{N}    \sum_{\ve{i},\tau }O_{\ve{i},\tau}(C^{n}_{\ve{i},\tau}).
\end{equation}
    In this case the measurement of local observables $O_{\ve{i},\tau}$ is carried out after each local  update on the same site. 
\end{itemize}

Our claim is that in the  presence of  zeros of the fermion determinant,  the estimates ~\ref{eq:Global},~\ref{eq:Tlocal} and ~\ref{eq:STlocal}  perform substantially different in terms of the statistical error bars, despite yielding identical values for the averages of observables.  Note that in each  scheme,  we  carry  out  the  same amount of  measurements such  that no  additional  computational  cost  is  involved. We  merely  reshuffle  the  ordering of the measurements.
The worst strategy is the global one, \ref{eq:Global}, and the best one is the ST-local reshuffling  \ref{eq:STlocal}, which is to the best of our knowledge not yet employed in existing QMC codes. One can readily observe the reduction in the intensity of spikes in the comparison of the three schemes in Fig. \ref{fig:histories}. 

To  place  our observation on a  firm  basis,  we will  prove   in Sec.~\ref{sec:generalities},   that if  the  m$^{th}$ moment  of  the   
probability  distribution of the considered  observables exists,   then  the  reshuffling  scheme  leaves  this  quantity  invariant.  However, as  shown in Appendix  \ref{AppendixB}  zeros  of  the  fermion determinant  generically  lead  to the fat  tailed  distributions   with  power law   that depends   
upon  the  dimensionality of the manifold of the field configurations where the fermion determinant vanishes.   Hence in this  case, the m$^{th}$ moment does not exist for sufficiently large m and the reshuffling will 
affect the  probability  distribution, the property we demonstrate in the toy model in Sec.~\ref{sec:toy_model}. It demonstrates the reduction of spikes depending upon the synchronization between measurements and Monte Carlo updates.  If  the  variance  of  the observable  does not exist  then  the central limit  theorem  cannot be   used  any more   for   the  error  analysis  and  alternative  schemes  for heavy tailed  distributions, presented  in Appendix \ref{AppendixA},  have  to  be used.  We apply these schemes to the error estimation in real-world QMC calculations described in Sec.~\ref{sec:formalism} showing clear connection between spikes, fat-tail distributions and zeros of the determinant. In section \ref{sec:main_results}  we demonstrate the reduction of spikes depending on the measurement scheme in QMC calculations when the fat-tail distribution appear due to the zeros of determinant. Finally, we also give some estimates for possible speedup.  The conclusion is  devoted to  discussions and further perspectives.

\section{\label{sec:generalities}  General  framework }
In this section,   we  show  that  our   reshuffling approach does not  alter  the m$^{\text{th}}$  moment of the  observable  provided  
that this quantity  exists.    We  start   with a  local  update  at   space  time  point   $x=( \ve{i},\tau) $.   This local  update of  a  configuration $C'$  is  described by  
a  stochastic  matrix   $\tilde{T}^{(x)} $  on   the  configuration space of  configurations  $\left\{ C   \right\} $,  that  satisfies   
$ \tilde{T}^{(x)}_{C,C'} \geq 0$,    $ \sum_{C} \tilde{T}^{(x)}_{C,C'}  = 1 $.   We  furthermore   require  stationarity to hold.    
 That is,   for  the  equilibrium probability  distribution 
\begin{equation}
     P(C)   =  \frac{1}{Z}    e^{- S(C) }    
\end{equation}
we  require,
\begin{equation}
	   \sum_{C'}     \tilde{T}^{(x)}_{C,C'} P(C')  = P(C).  
\end{equation}
Clearly,   a  single  spin  flip  cannot  be ergodic     so  that we  have  to  combine  the   local  updates  to  form a  sweep  that  visits   every  space time 
point.   In the BSS  algorithm,  this is  done   sequentially.    Let  us    choose  one  space-time  index  $x_0 $   as  the  origin  and  define  the  sweep   as
\begin{equation}
	      T^{(x_0)} =       \left( \prod_{x=1}^{x_0 }  \tilde{T}^{(x)}     \right)    \left(   \prod_{x=x_0 +  1 }^{V L_{\tau}}  \tilde{T}^{(x)}      \right).
\end{equation}
One  will show  that $ T^{(x_0)} $   is a  stochastic  matrix  that   satisfies    the  stationarity  condition.    We  will  furthermore  assume  that it is  ergodic,
such  that:
\begin{equation}
    \forall \,  C, C'    \, \,   \exists     \, n   \, \,  |    \, \,  \left(  T^{(x_0)}  \right)^n_{C,C'}   > 0. 
\end{equation}
 The   $n^{th}$   element of  the Markov  chain,     $C^{n}_{x_0}  $,    will  occur  with  probability   $  (T^{(x_0)})^n_{C^{n}_{x_0},C^{0}}  $  
 where $ C^{0} $  is  an  arbitrary    initial     configuration.   Owing  to the   stationarity  and  ergodicity  properties of  $T^{(x_0)} $     the  fraction of  
 time  that  $ C^{n}_{x_0}  $  occurs  in the   Markov  chain is given  by $ P( C^{n}_{x_0} )  $.   More  specifically, 
\begin{equation} 
\label{eq:MC}
	  \lim_{N \rightarrow \infty}   \frac{1}{N}   \sum_{n=1}^{N}    \delta_{C,C^{n}_{x_0} }     =  P(C).
\end{equation}
Importantly    the  right  hand  side of  Eq.~\ref{eq:MC}    is  independent  on the  choice   of  the  origin $x_0$.   We  refer  the  reader to  Ref.~\cite{Sokal97}   for  further   
reading on  the Monte  Carlo   method.   
Let us  now  define  a  local  observable   $O_{y} (C) $  on the configuration space.     This  could  for  example  corresponds  to   the 
double occupancy  of  the  Hubbard model  at  space-time  index $y$.   For the chain of   configurations   $C_{x_0}^{n}$,  the  probability 
  distribution   of  this  observable  reads: 
\begin{equation} 
	   P_{y,x_0} (O)    =  \lim_{N \rightarrow \infty } \frac{1}{N} \sum_{n=1}^{N}  \delta  \left(  O   -    O_{y}(C_{x_0}^{n})      \right). 
\end{equation} 
The  question we  are  asking is if   $ P_{y,x_0} (O) $  depends  on  $x_0$.    Let us  consider   the  m$^{\text{th}}$   moment: 
\begin{equation}
	 \int  d  O   P_{y,x_0} (O)   O^m       =     \sum_{C} \left(  \lim_{N \rightarrow \infty }   \frac{1}{N} \sum_{n=1}^{N}   \delta_{C,C_{x_0}^{n}}   \right) 
	 O_{y}^{m}(C).
\end{equation}
Given  Eq.~\ref{eq:MC},   we   conclude  that   $  \int  d  O   P_{y,x_0} (O)   O^m $     is  independent on the choice  of  the origin $x_0$   and  that  
our    reshuffling  procedure should  not  alter   the m$^{\text{th}}$  moment.    The implicit  assumption  in the  above  is  the  very  existence of  the 
 m$^{\text{th}}$  moment.      Hence  for  fat  tailed  distributions  where   higher  moments  are  not  defined    $  P_{y,x_0} (O)  $  may   very  well 
 have  an  $x_0$  dependency  albeit with  an   $x_0$  independent  first moment.   In this  case,  our  reshuffling  process    has  potentially  a  non-trivial  
 effect.     The  next  section  presents  an  explicit  toy  model  that   demonstrates  this.

 \begin{figure}[]
   \centering
   \subfigure[]{\label{fig:toy_model_action_K0.0}\includegraphics[width=0.35\textwidth , angle=0]{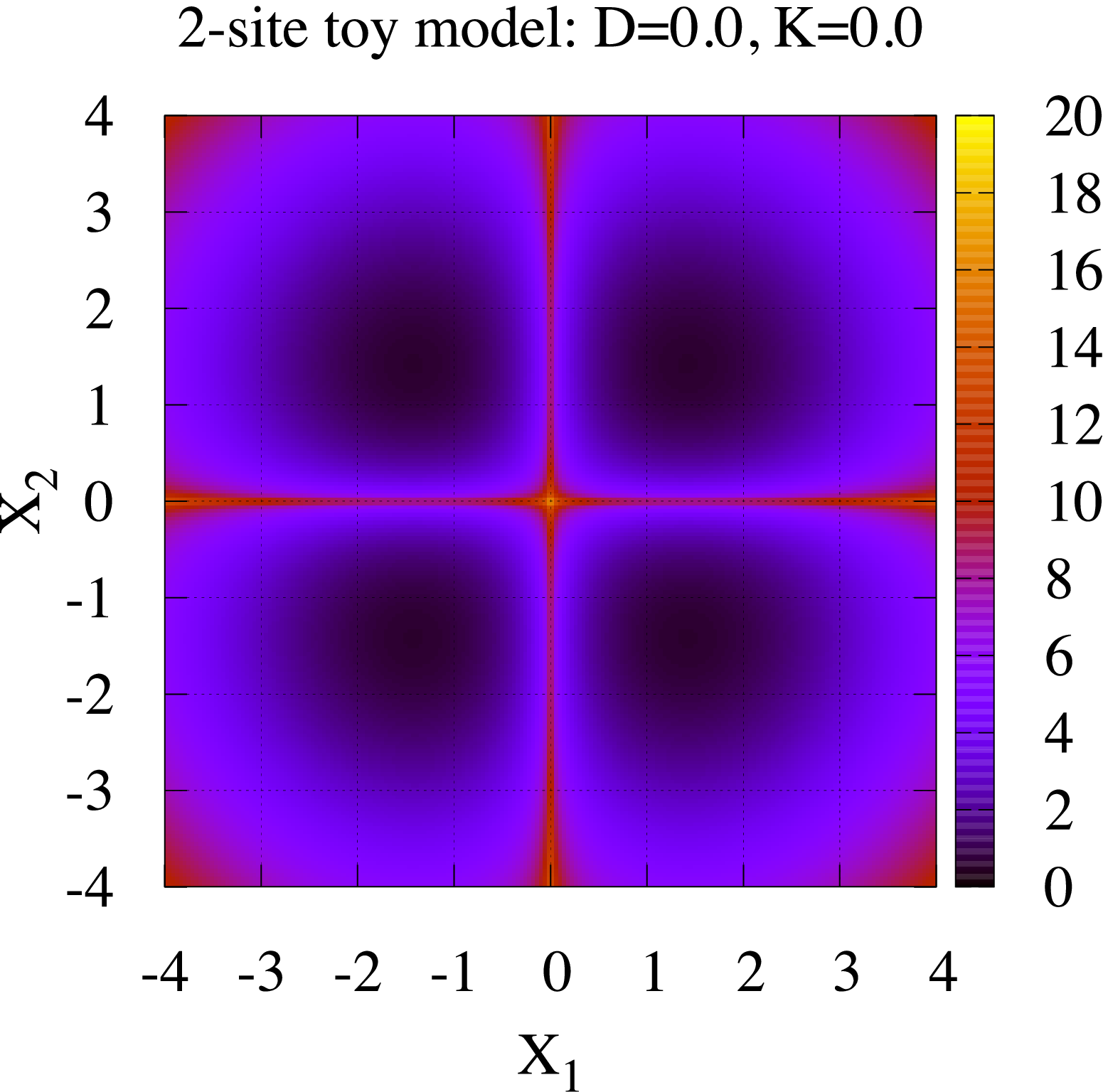}}
   \subfigure[]{\label{fig:toy_model_action_K0.5}\includegraphics[width=0.35\textwidth , angle=0]{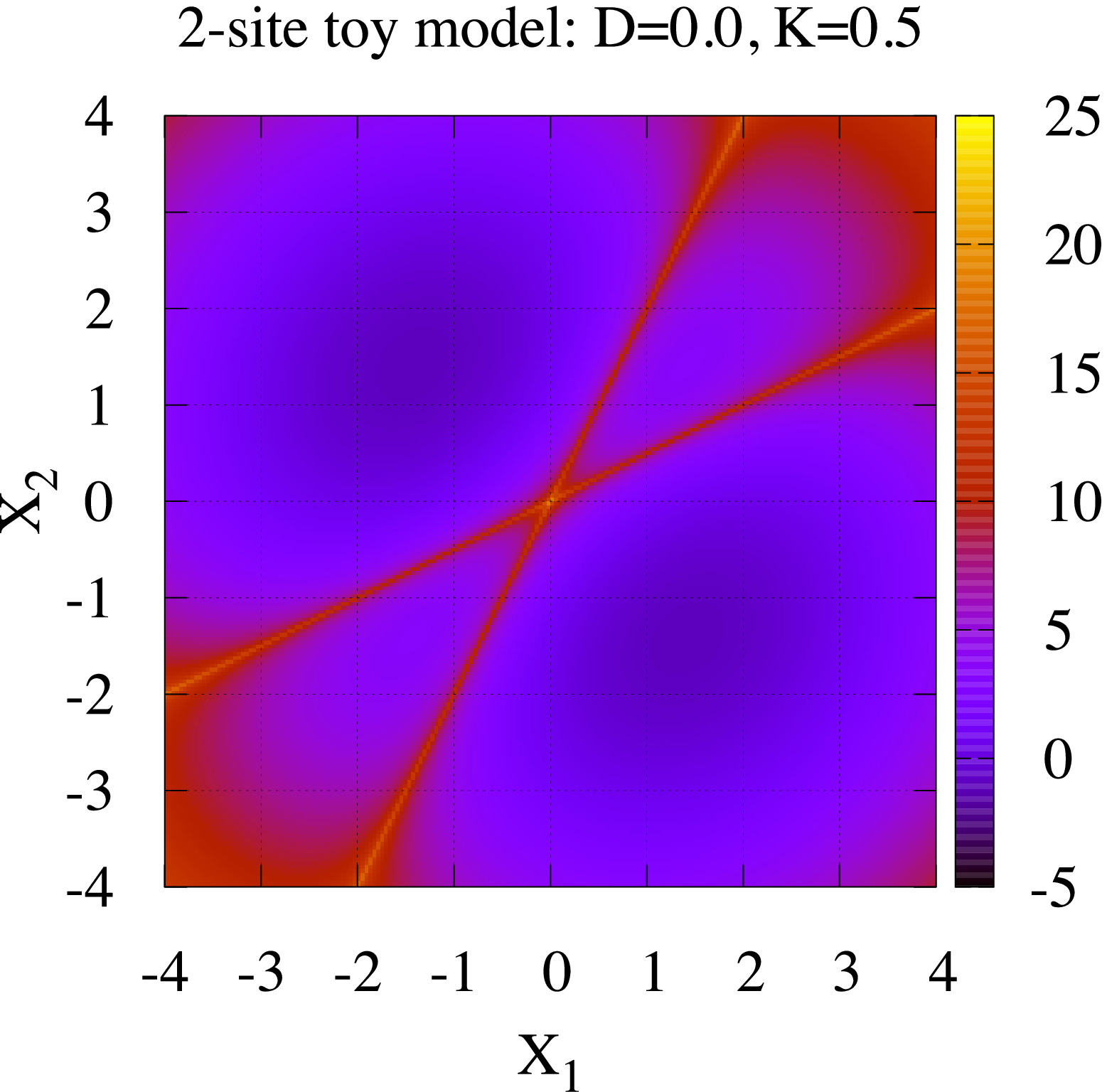}}
         \caption{Action of the two-dimensional toy model \ref{eq:toy_action}. (a) the action when $K=0.0$ and the fields $x_1$ and $x_2$ are uncorrelated; (b) the action when $K=0.5$. $D=0.0$ in both cases.}
    \label{fig:toy_model_action}
\end{figure}

  \begin{figure}[]
   \centering
   \subfigure[]{\label{fig:toy_res_K0.0}\includegraphics[width=0.30\textwidth , angle=-90]{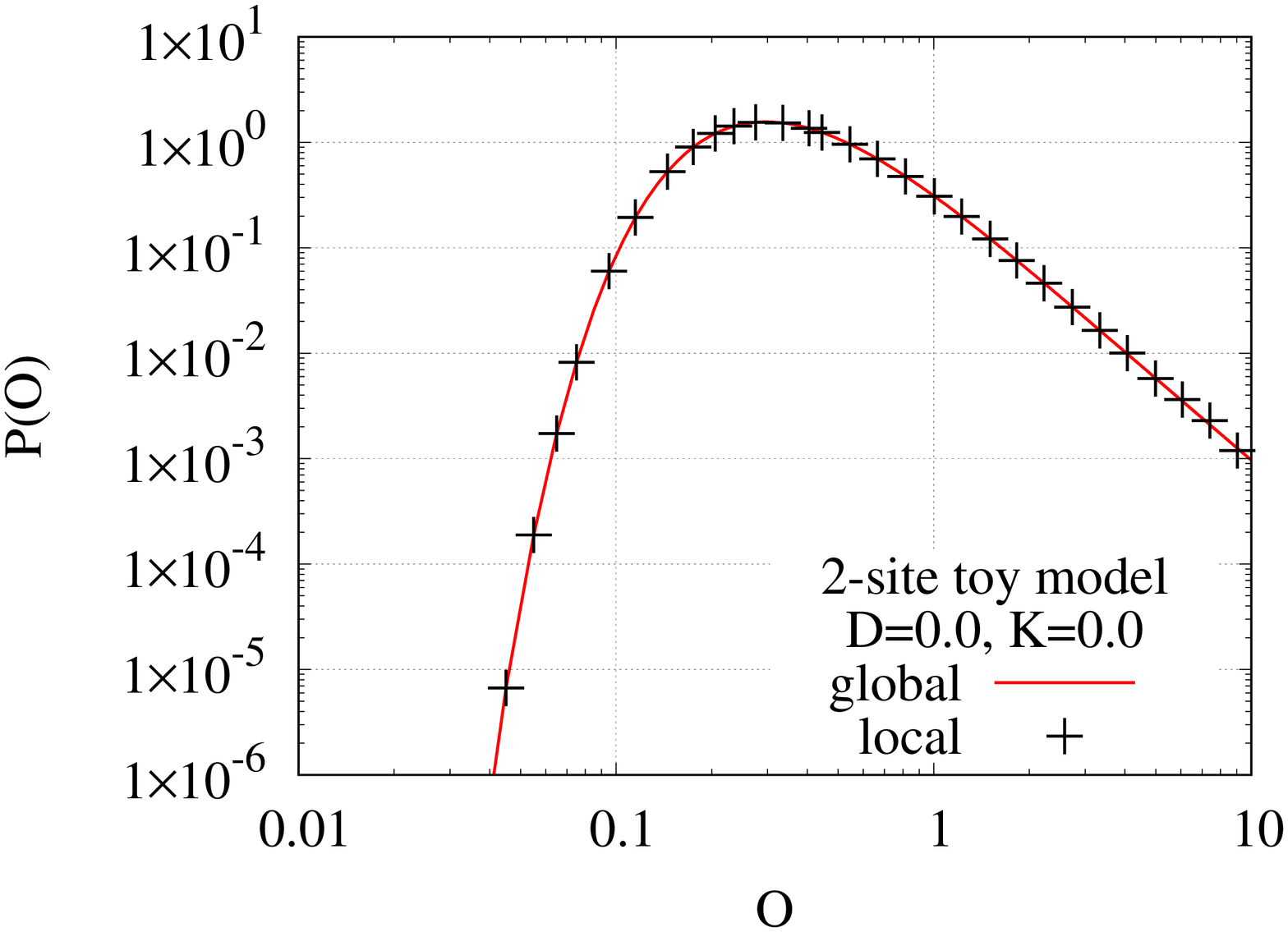}}
   \subfigure[]{\label{fig:toy_res_K0.9}\includegraphics[width=0.30\textwidth, angle=270]{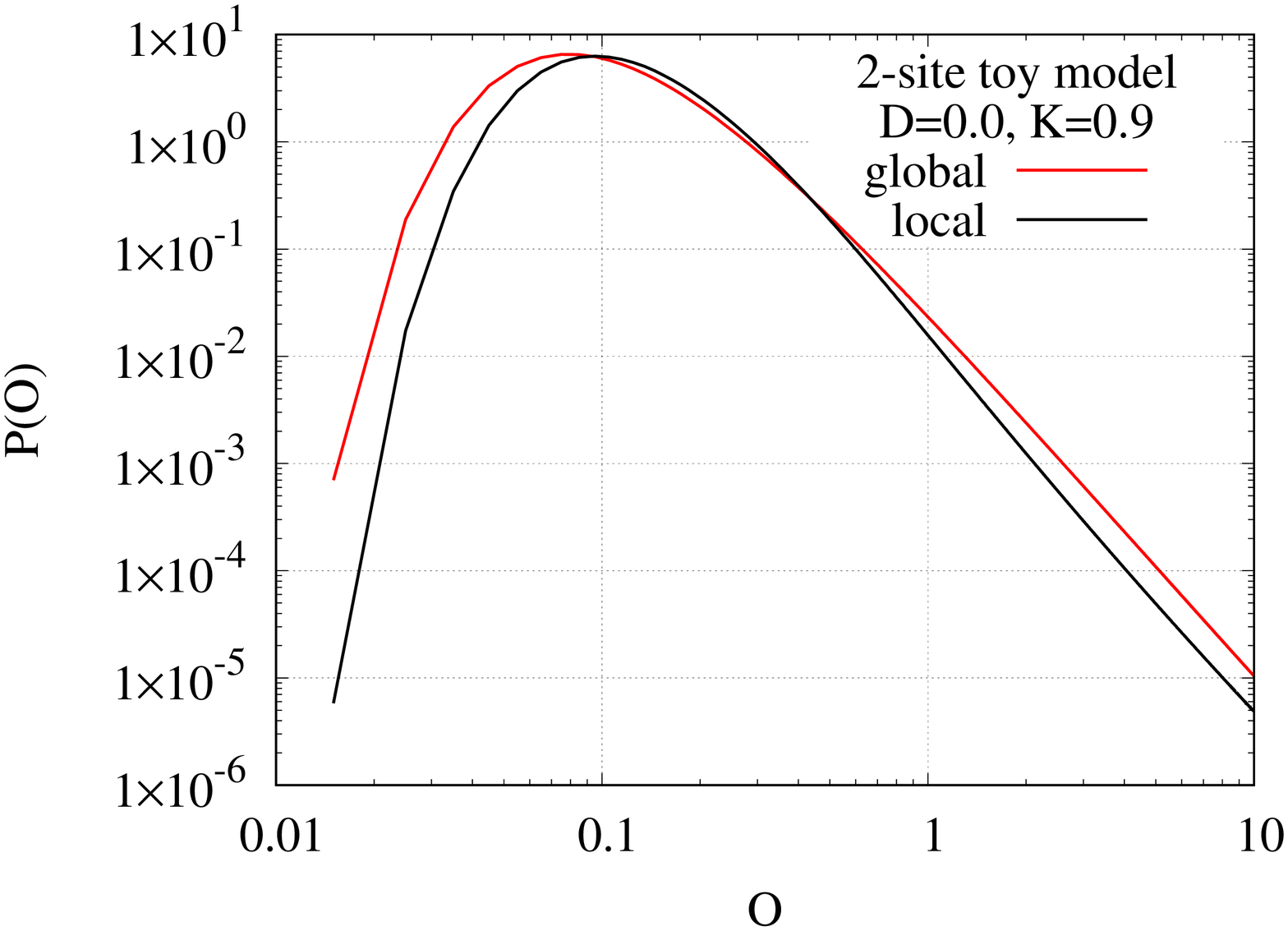}}
   \subfigure[]{\label{fig:toy_res_all}\includegraphics[width=0.30\textwidth , angle=-90]{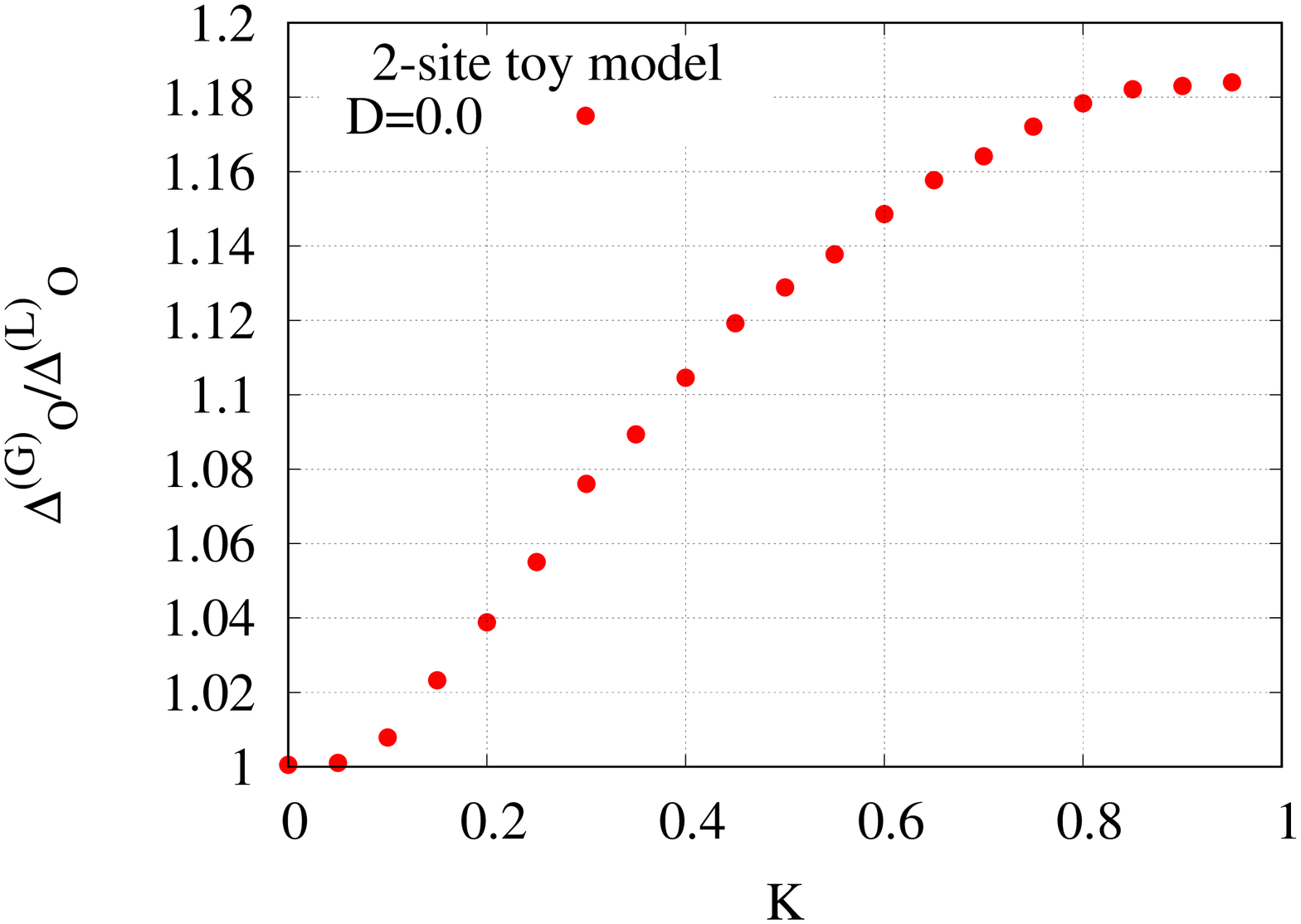}}
   
         \caption{Results for the toy model. Figure (a) shows the identical probability distributions for the observable when global or local measurements are used in the case of uncorrelated fields $x_1$ and $x_2$ ($K=0$). Figure (b) is the comparison of the probability distributions for local and global measurements at $K=0.9$. Figure (c) shows the ratio of the widths of the distributions for global and local schemes of measurements (see Eq. \ref{eq:main_95_conf}). Larger value of this ratio corresponds to bigger advantage from local measurements. $D=0.0$ in all cases. }
   \label{fig:toy_model_results}
\end{figure}

\section{\label{sec:toy_model}  Toy model }

Our toy model is inspired by the recent study of the structure of the path integral within the Lefschetz thimbles approach \cite{PhysRevD.101.014508}. We start from the probability distribution for two random variables $x_1$ and $x_2$:
\begin{equation}
	P(x_1, x_2)= C_0 e^{-\frac{1}{2}(x_1^2+x_2^2)} \left(x_1-f(x_2)\right)^2\left(x_2-f(x_1)\right)^2,
	\label{eq:toy_probability}
\end{equation}
where $C_0$ is the normalization coefficient. It corresponds to the exponent in the partition function  \ref{eq:basic_parititon_func}, thus the model action can be written as
\begin{equation}
	S(x_1, x_2)= \frac{1}{2}(x_1^2+x_2^2)- 2 \ln \left| \left(x_1-f(x_2)\right) \left(x_2-f(x_1)\right) \right|.
	\label{eq:toy_action}
\end{equation}
The action is symmetrical with respect  to the exchange $x_1 \Leftrightarrow x_2$, which reflects the translational symmetry of the real lattice models. $P(x_1, x_2)$ includes the Gaussian part and the remaining terms model fermion determinants.  This structure is inspired by the Hubbard-Stratonovich decomposition with continuous auxiliary fields routinely employed in determinantal QMC \cite{Assaad08_rev,Conformal_PhysRevB.99.205434, Buividovich:2018yar} (see section \ref{sec:formalism} below for the detailed explanation). Our toy "determinants" are equal to zero along the lines $x_1=f(x_2)$ and $x_2=f(x_1)$, and we need the second power of each multiplier since there are two equivalent fermion determinants for electrons and holes on a bipartite lattice at half-filling. 

The observables should model the fermionic green functions, thus we consider two expressions 
\begin{eqnarray}
	O_1(x_1, x_2)= \frac{1}{\left(x_1-f(x_2)\right)^2}, \\
	O_2(x_1, x_2)= \frac{1}{\left(x_2-f(x_1)\right)^2},
	\label{eq:toy_observables}
\end{eqnarray}
connected with each other through the "translational symmetry" $O_1(x_1, x_2)=O_2(x_2, x_1)$. Average over lattice is equivalent to the computation of the sum $O=\frac{1}{2}(O_1+O_2)$.   Importantly, and as in determinantal QMC observables  diverge at the zeros of the toy determinant.   

In order to be able to derive all probability distributions analytically, we consider a linear form $f(x)=D+K x$. Despite simplicity, this is enough to demonstrate all essential phenomena leading to the reduction of the tail in fat-tailed distributions.  
The action \ref{eq:toy_action} is plotted in Fig.~\ref{fig:toy_model_action} for two cases: $K=0, D=0$, when the two variables are independent, and for $K=0.5, D=0$. One can clearly observe the lines, where the  action diverges logarithmically.  These are the lines corresponding to the zeros of fermion determinant.

The global scheme of measurements is modelled  by the following procedure: first, we generate both  variables $x_1, x_2$ according to the probability density \ref{eq:toy_probability}; second, we measure the full observable $O$. After transition from the probability distribution \ref{eq:toy_probability} for $x_1$ and $x_2$ to the probability density $P_O(z)$ for the full observable $O$ we get the following expression:
\begin{equation}
	P^{(G)}_O(z)= \frac{C_0}{2} \sum_{\nu_1, \nu_2} \int_{-z}^{z} d \alpha \frac{e^{-\frac{1}{2}(x_1(z,\alpha, \nu_1, \nu_2)^2+x_2(z,\alpha, \nu_1, \nu_2)^2)}} {(z-\alpha)^{5/2}(z+\alpha)^{5/2} (1-K^2) }, 
	\label{eq:toy_O_global}
\end{equation}
where 
\begin{eqnarray}
      x_1 = \frac{1}{1-K^2}\left( \frac{ \nu_1}{\sqrt{z-\alpha}}+\frac{K \nu_2}{\sqrt{z+\alpha} } + D(K+1) \right) \nonumber  \\
      x_2=\frac{1}{1-K^2}\left( \frac{ K \nu_1}{\sqrt{z-\alpha}}+\frac{\nu_2}{\sqrt{z+\alpha} } + D(K+1) \right)
	\label{eq:toy_x_alpha_z}
\end{eqnarray}
and $\nu_{1,2}=\pm 1$. $\alpha$ parametrizes the curve of constant observable $O$ in $(x_1;x_2)$ plane. For simplicity, we considered only the case $K\in(0;1)$. 

In the local scheme of measurements, we compute observable $O_1$ after the update of $x_1$ field according to the conditional probability $P(x_1 | x_2)$, defined as
\begin{eqnarray}
	P(x_1 | x_2)=  e^{-\frac{1}{2}x_1^2} \times \nonumber  \\   \left(x_1-f(x_2)\right)^2 \left(x_2-f(x_1)\right)^2 \tilde C_0(x_2),
	\label{eq:conditional_probability}
\end{eqnarray}
with the additional normalization constant $\tilde C_0(x_2)$ 
\begin{eqnarray}
	{\tilde C}^{-1}_0(x_2) = \int_{-\infty}^{\infty} d \tilde x_1 e^{-\frac{1}{2}{\tilde x}^2_1}    \left(\tilde x_1-f(x_2)\right)^2  \times \nonumber \\  \left(x_2-f( \tilde x_1)\right)^2.
	\label{eq:normalization1}
\end{eqnarray}
Subsequently, measurement of $O_2$ is made after similar update of the $x_2$ field according to the conditional probability $P(x_2 | x_1)$. 

In order to derive the probability distribution for the observable, we  consider the full Markov   sequence  in the Monte Carlo process. 
The   sequence of configurations   consists of some fixed starting point $x^{(0)}_1$ and values of the fields after subsequent updates: $x^{(0)}_2, \,x^{(1)}_1, \,x^{(1)}_2, \,... $. Probability distribution for the full chain of $2 N+1$ local updates can be written as:
\begin{eqnarray}
 P_{2N+1} (x^{(0)}_1, x^{(0)}_2, x^{(1)}_1, x^{(1)}_2, ... x^{(N)}_1, \,x^{(N)}_2)  = \nonumber \\  P(x^{(0)}_2 | x^{(0)}_1) \prod^N_{i=1} \left( P(x^{(i)}_1 | x^{(i-1)}_2) P(x^{(i)}_2 | x^{(i)}_1) \right),
	\label{eq:full_probability}
\end{eqnarray}
where each local update is made with the conditional probability \ref{eq:conditional_probability}. According to this notation, $O_1$ is measured using $(x^{(N)}_1,  x^{(N-1)}_2)$ pair of fields and $O_2$ is measured using $(x^{(N)}_1,  x^{(N)}_2)$ fields. 

Due to the fact that the probability \ref{eq:toy_probability} is the stationary distribution for this Markov chain, the distribution of the pair  $(x^{(N)}_1,  x^{(N-1)}_2)$ converges to Eq. \ref{eq:toy_probability} in the limit $N\rightarrow\infty$:
\begin{eqnarray}
 \int dx^{(0)}_2 dx^{(1)}_1 ... dx^{(N-2)}_1 dx^{(N-2)}_2 dx^{(N-1)}_1  \times \nonumber \\ P_{2N-1} (x^{(0)}_1, x^{(0)}_2 ... x^{(N-1)}_1, \,x^{(N-1)}_2) \times \nonumber \\ P(x^{(N)}_1 | x^{(N-1)}_2)|_{N\rightarrow\infty}=P(x^{(N)}_1, x^{(N-1)}_2).
	\label{eq:double_probability}
\end{eqnarray}
Thus, after a sufficiently long warm-up Monte Carlo time, the local measurement scheme can be described as the computation of the observable $O$, which is a function of three random variables: 
\begin{eqnarray}
	O(x_1, x_2, \tilde x_2)= \frac{1}{2} \left( \frac{1}{\left(x_1-f(x_2)\right)^2}+ \frac{1}{\left(\tilde x_2-f(x_1)\right)^2}  \right). \nonumber \\
	\label{eq:toy_observable_local}
\end{eqnarray}
These variables are distributed according to the probability density
\begin{eqnarray}
	P^{(L)}(x_1, x_2, \tilde x_2)= P(x_1, x_2) P(\tilde x_2| x_1).
	\label{eq:toy_probability_localX}
\end{eqnarray}
The final probability distribution for the observable $O$ from \ref{eq:toy_observable_local} reads as
\begin{eqnarray}
	P^{(L)}_O(z)= \sum_{\nu=\pm 1}\int_0^\infty d \beta \int_{f(\beta)+1/\sqrt{2z}}^\infty d\alpha  C_0  \tilde C_0(\alpha) \times \nonumber  
	\\ e^{-\frac{1}{2}(\alpha^2+\beta^2+{\tilde x}^2_2 (\alpha, \beta, \nu) )}   \left(\alpha-f(\beta)\right)^2 \left(\beta-f(\alpha)\right)^2 \times \nonumber  
	\\ \left(\tilde x_2 (\alpha, \beta, \nu) -f(\alpha)\right)^2  \frac{\alpha-f(\tilde x_2(\alpha, \beta, \nu)) }{  \left( 2z-\frac{1}{(\alpha-f(\beta))^2} \right)^{3/2}},
	\nonumber \\
	\label{eq:toy_probability_localO}
\end{eqnarray}
where
\begin{equation}
\tilde x_2 (\alpha, \beta, \nu)=f(\alpha)+\nu \left( 2z-\frac{1}{(\alpha-f(\beta))^2} \right)^{-1/2}
    \label{fig:toy_x_tilde}
\end{equation}
and the pair $(\alpha, \beta)$ parametrizes the surface of constant observable $O$ in 3D space $(x_1; x_2; \tilde x_2)$. 

Both distributions have power law asymptotic in the limit $O\rightarrow \infty$:
\begin{equation}
    P^{(L/G)}_O(z)|_{z\rightarrow\infty}\approx A^{(L/G)}(D, K)/z^{\xi(D, K)} \label{eq:toy_asymptotic},
\end{equation}
where the power $\xi$ stays the same but the prefactors $A$ are different. $P^{(L)}_O(z)$ and $P^{(G)}_O(z)$ are plotted in Fig. \ref{fig:toy_model_results}. First we check that the local and global schemes are identical in case when $K=0$ (see Fig. \ref{fig:toy_res_K0.0}). Since $x_1$ and $x_2$ are completely independent in this case, and $O_i$ is dependent only on corresponding $x_i$, the order of updates and measurements does not matter. Indeed, the points for local and global schemes overlap precisely in Fig. \ref{fig:toy_res_K0.0}. The situation changes at $K \neq 0$. The example plot for $K=0.9$ (see Fig. \ref{fig:toy_res_K0.9}) shows that $P^{(L)}_O(z)$ and $P^{(G)}_O(z)$ are now different: both distributions still have the same power law tail which is evident from the double logarithm plot, but  the prefactor $A$ is slightly smaller for the local scheme. 

In order to quantify this difference, we compute the width of the distribution at the  95\% level (see Appendix \ref{AppendixA}):
\begin{equation}
\Delta^{(L/G)}_O=z^{(L/G)}_2-z^{(L/G)}_1, \label{eq:main_95_conf}
\end{equation}
where
\begin{eqnarray}
\int_{z^{(L/G)}_2}^\infty dz P^{(L/G)}_O(z) = 0.025, \label{eq:main_z2_def}  \\ 
\int_{0}^{z^{(L/G)}_1} dz P^{(L/G)}_O(z) = 0.025 \label{eq:main_z1_def}
\end{eqnarray}
The ratio $\Delta^{(G)}_O/\Delta^{(L)}_O$  is displayed in Fig. \ref{fig:toy_res_all} depending on $K$ at $D=0$. We can reach approximately 20\% reduction of the tail in the local scheme of measurements.

In the  toy model,   the improvement is minuscule.  We interpret this in  terms of the   small difference between local and global updates in this two-fields system: even in the local scheme we still update half of the fields. However, as we demonstrate further below, the effectiveness of the tail suppression grows with the dimensionality of the configuration space. Hence the effect becomes quite pronounced in real QMC calculations, where the number of the fields can easily exceed $10^4$.

\section{\label{sec:formalism}   Formalism  for the  repulsive   Hubbard model. }

Example QMC calculations will be carried out for the repulsive Hubbard model on bipartite hexagonal and square lattices:
\begin{equation}
	\hat{H}    =  -t  \sum_{ \langle \ve{i},\ve{j} \rangle, \sigma  }    \hat{c}^{\dagger}_{\ve{i},\sigma} \hat{c}^{\phantom\dagger}_{\ve{j},\sigma}   
	+  U \sum_{\ve{i}} \left(  \hat{n}_{\ve{i},\uparrow} -1/2  \right)   \left(  \hat{n}_{\ve{i},\downarrow} -1/2  \right).
	\label{eq:Hamiltonian_spin}
\end{equation}
Here, $\hat{c}^{\dagger}_{\ve{i},\sigma}$ are creation operators for electrons with spin $\sigma=\uparrow, \downarrow$ at site $\ve{i}$ and $\hat{n}_{\ve{i},\sigma} =\hat{c}^{\dagger}_{\ve{i},\sigma} \hat{c}^{\phantom\dagger}_{\ve{i},\sigma}  $. 
It is  beyond the  scope  of this article  to  provide a  detailed   review  of the  finite  temperature   auxiliary field QMC  algorithm  and  we  refer the reader to
Refs.~\cite{Assaad08_rev,ALF_v2}  for overviews  and  implementations of the  algorithm.  Here  we will concentrate on the   so called   SU(2)  invariant  Hubbard-Stratonovich (HS)  transformation \cite{Hirsch83}. In order to introduce it, we   carry  out  a  partial  particle-hole  transformation, 
\begin{equation}
\label{ParticleHole}
  \left\{ { {\hat{c}^{}_{{\ve{i}}, \uparrow}, \hat{c}^{\dagger}_{{\ve{i}}, \uparrow} \to \hat{a}_{\ve{i}}, \hat{a}^{\dagger}_{\ve{i}}, } \atop
{\hat{c}^{}_{{\ve{i}}, \downarrow}, \hat{c}^{\dagger}_{{\ve{i}}, \downarrow} \to \pm \hat{b}^{\dagger}_{\ve{i}}, \pm \hat{b}_{\ve{i}}} } \right. ,
 \end{equation}
where the sign in the second line alternates depending on the sublattice,  $\hat{a}^{\dagger}_{\ve{i}}$ and $\hat{b}^{\dagger}_{\ve{i}}$ are creation operators for electrons and holes correspondingly.
The Hamiltonian  \ref{eq:Hamiltonian_spin} acquires the following form:
\begin{eqnarray}
  \label{eq:Hamiltonian}
  \hat H = -t \sum_{\langle {\ve{i}},{\ve{j}}\rangle} (  \hat a^\dag_{{\ve{i}}} \hat a_{{\ve{j}}} + \hat b^\dag_{{\ve{i}}} \hat b_{{\ve{j}}} ) &+& \frac{U}{2} \sum_{{\ve{i}}} \hat q_{\ve{i}}^2,
\end{eqnarray}
where $\hat q_{\ve{i}} = \hat a^\dag_{\ve{i}} \hat a_{\ve{i}} - \hat b^\dag_{\ve{i}} \hat b_{\ve{i}}$ is the operator of electrical charge  measured with  respect to the  half-filled  case on site $\ve{i}$. 

In order to make an analytical prediction on the power law in the heavy-tailed distributions, we need to introduce the continuous auxiliary fields on the basis of the usual Gaussian HS transformation:
\begin{eqnarray}
\label{eq:continuous_HS_imag}
  e^{-\frac{\Delta \tau}{2} U \hat q^2} \cong \int D \phi  e^{- \frac{1}{2 U \Delta \tau}  \phi^2} e^{i \phi \hat q}.
\end{eqnarray}
  The partition function can be obtained as the integral
\begin{eqnarray}
  \mathcal{Z} = \int \mathcal{D} \phi_{{\ve{i}},\tau} e^ {-S_B}  \det M_{el.} \det M_{h.}, \nonumber \\
   S_B(\phi_{{\ve{i}},\tau}) = \sum_{{\ve{i}},\tau}  \frac  {\phi_{{\ve{i}},\tau}^2} {2 \Delta \tau U},
  \label{eq:Z_continuous}
\end{eqnarray}
where fermionic operators for continuous auxiliary fields are written as
\begin{eqnarray}
 M_{el.} = I +\prod^{L_\tau}_{\tau=1} \left({ e^{-\Delta \tau h} \diag{ e^{i \phi_{{\ve{i}},\tau}} } }\right), \nonumber \\
 M_{h.} = I +\prod^{L_\tau}_{\tau=1} \left({ e^{-\Delta \tau h} \diag{ e^{-i \phi_{{\ve{i}},\tau}} } }\right). 
  \label{eq:M_continuous}
\end{eqnarray}
Both fermionic operators are $V \times V$ matrices where $V$ is the number of lattice sites in space, $h$ is the matrix of single-particle Hamiltonian which defines the tight-binding part in the expression \ref{eq:Hamiltonian}.  The diagonal  $V \times V$ matrix $\diag{ e^{-i \phi_{\ve{i},\tau}}}$ includes all exponents with auxiliary fields belonging to a given Euclidean time slice $\tau$. From Eq. \ref{eq:Z_continuous} we conclude that the full action includes both quadratic form and the logarithms of the fermion determinants, pretty similar to the toy model \ref{eq:toy_action}:
\begin{eqnarray}
S=S_{B} - \ln (\det M_{el.} \det M_{h.}).
 \label{eq:action_continuous}
\end{eqnarray}

In practical QMC calculations, the exponent \ref{eq:continuous_HS_imag} is replaced by 
the Gauss-Hermite quadrature up to fourth order \cite{goth2020higher}:
\begin{equation}
	     e^{-\Delta  \tau   \frac{U}{2} {\hat{q}_{\ve{i}}^2}} =   
	      \sum_{s= \pm 1, \pm 2} \gamma(s)  e^{i \lambda \eta(s) \hat{q}_{\ve{i}}}  +  {\cal O} \left\{ \left( \Delta  \tau  U \right)^4  \right\}
	     \label{eq:discrete_HS}
\end{equation}
with  $\lambda =   \sqrt{ \Delta \tau U / 2 }$,  $\gamma(\pm 1)= 1 + \sqrt(6)/3$, $\gamma(\pm 2)= 1 - \sqrt(6)/3$  and $\eta(\pm 1)=\pm \sqrt{2 (3-\sqrt{6})}$, $\eta(\pm 2)=\pm \sqrt{2 (3+\sqrt{6})}$  \cite{ALF_v2}. 

In this case the partition function can be computed as a sum over all possible values of discrete fields $s_{{\ve{i}},\tau}$:
\begin{eqnarray}
  \mathcal{Z} = \sum_{s_{{\ve{i}},\tau}} \prod_{{\ve{i}},\tau}  \gamma(s_{{\ve{i}},\tau})  \det D_{el.}(s_{{\ve{i}},\tau}) \det D_{h.}(s_{{\ve{i}},\tau}),
  \label{eq:Z_discrete}
\end{eqnarray}
where $D_{el.}$ and $D_{h.}$ are the fermionic operators for electrons and holes respectively:
\begin{eqnarray}
 D_{el.}(s_{{\ve{i}},\tau}) = I +\prod^{L_\tau}_{\tau=1} \left({ e^{-\Delta \tau h} \diag{ e^{i \lambda \eta(s_{{\ve{i}},\tau}) s_{{\ve{i}},\tau}} } }\right), \nonumber \\
 D_{h.}(s_{x,\tau}) = I +\prod^{N_t}_{\tau=1} \left({ e^{-\Delta \tau h} \diag{ e^{-i \lambda  \eta(s_{{\ve{i}},\tau})s_{{\ve{i}},\tau}} } }\right). 
  \label{eq:M_discrete}
\end{eqnarray}

For the details of sampling algorithms and calculation of observables, we refer a reader to the description of ALF package \cite{ALF_v2}. Despite quite different expression for the partition function, we get quite similar results for discrete and continuous fields if the sampling algorithms are equivalent (see numerical proof and discussion in the end of the section \ref{sec:main_results}). 

In our examples, we consider four different observables:
\begin{itemize}
    \item Sublattice  charge fluctuations:
    \begin{equation}
        \hat Q^2= \left( \sum_{{\ve{i}} \in sublat. 1} \hat{ q}_{{\ve{i}}}  \right)^2
        \label{eq:squared_charge}
    \end{equation}
      \item Sublattice spin  fluctuations:
    \begin{equation}
        \hat S^2= \left( \sum_{{\ve{i}} \in sublat. 1} { (\hat n_{{\ve{i}}, \uparrow}}-{\hat  n_{{\ve{i}}, \downarrow}) }  \right)^2
        \label{eq:squared_spin}
    \end{equation}
    \item Double occupancy:
    \begin{equation}
        \hat d= \frac{1}{V} \sum_{{\ve{i}}} { \hat n_{{\ve{i}}, \uparrow}} {\hat  n_{{\ve{i}}, \downarrow} } 
        \label{eq:squared_d_occ}
    \end{equation}
     \item squared staggered  moment:
    \begin{equation}
        \Delta \hat S^2= \left( \sum_{{\ve{i}} \in sublat. 1} ( \hat n_{{\ve{i}}, \uparrow}-{\hat  n_{{\ve{i}}, \downarrow} )  - \sum_{{\ve{i}} \in sublat. 2} ( \hat n_{{\ve{i}}, \uparrow}}-\hat  n_{{\ve{i}}, \downarrow} )  \right)^2
        \label{eq:squared_delta_spin}
    \end{equation}
\end{itemize}

All these observables include the second powers of equal-time fermion green function $G(\tau)$. Due to the relation 
\begin{equation}
    \det {G(\tau)}^{-1}= \det  D_{el.},
    \label{eq:g_det}
\end{equation}
which  also holds for continuous auxiliary fields, $\det G(\tau)$ (thus at least some elements of the green function $G(\tau)$ itself)  diverges at zeros of the determinant. These divergences are the reason why  the  distributions of four-fermion observables become heavy-tailed, as observed in the toy model \ref{eq:toy_asymptotic}. 
As  shown in Appendix~\ref{AppendixB}, the power $\xi$ can be connected to the dimensionality of the manifolds, on which the determinants vanish.  The SU(2) invariant HS decomposition 
for the Hubbard model corresponds to the worst-case scenario \cite{PhysRevD.101.014508}.  Here, the zeros of the determinant form "domain walls" in 
 configuration space such that  the dimensionality the  zero-manifolds is  given $V L_\tau-1$, where $V L_\tau$ corresponds to the number 
 of auxiliary fields.
The connection to the power law in heavy-tailed distributions was considered in details in \cite{Ulybyshev:2017} and  for the particular case of zeros of the determinant forming domain walls, $\xi=5/2$. A detailed proof is  given in Appendix~\ref{AppendixB}.

\begin{figure}
    \centering
   \subfigure[]{\includegraphics[width=0.23\textwidth,angle=0]{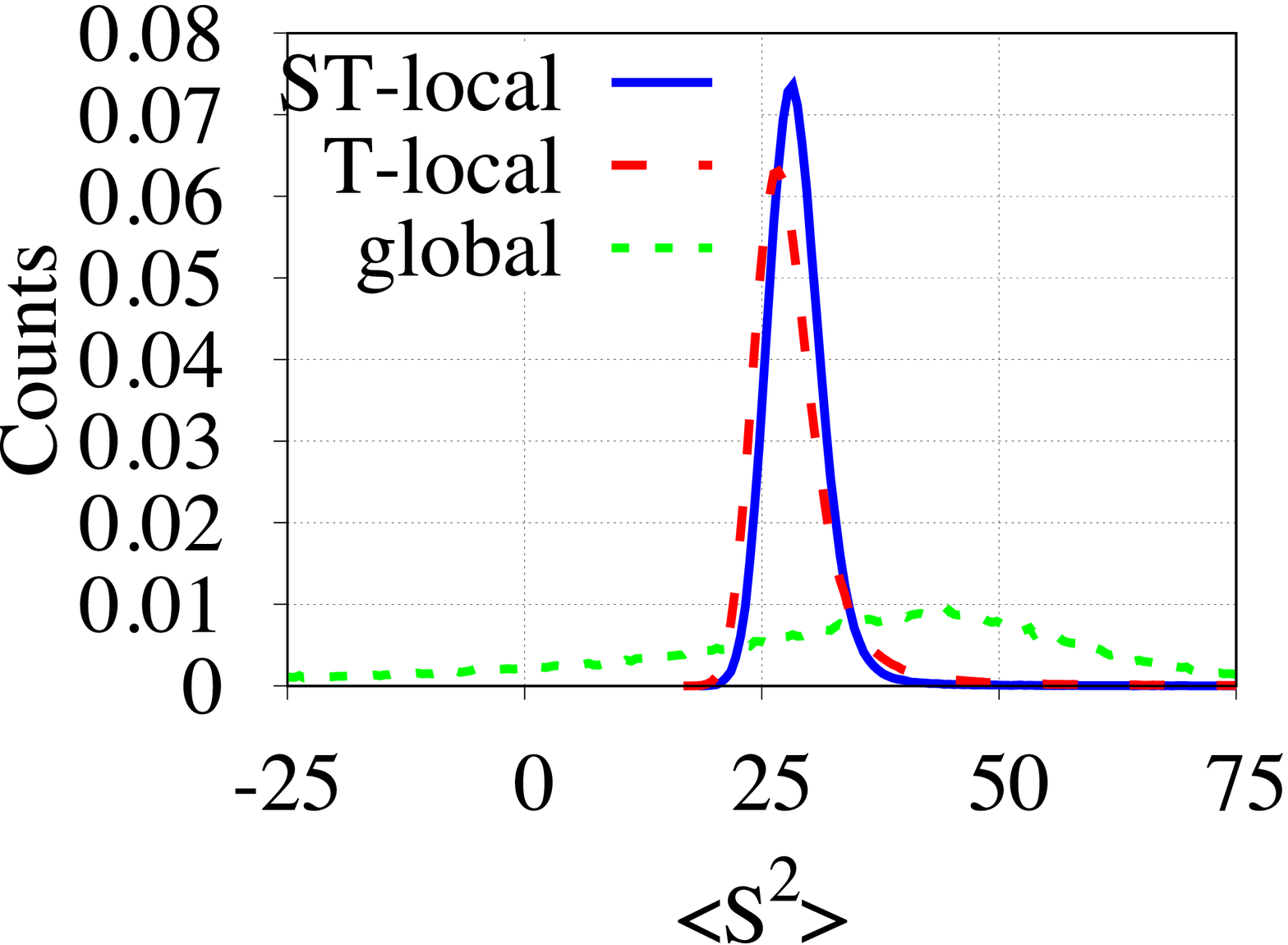}\label{fig:histogram_spin_6x6x256}} \subfigure[]{\includegraphics[width=0.23\textwidth,angle=0]{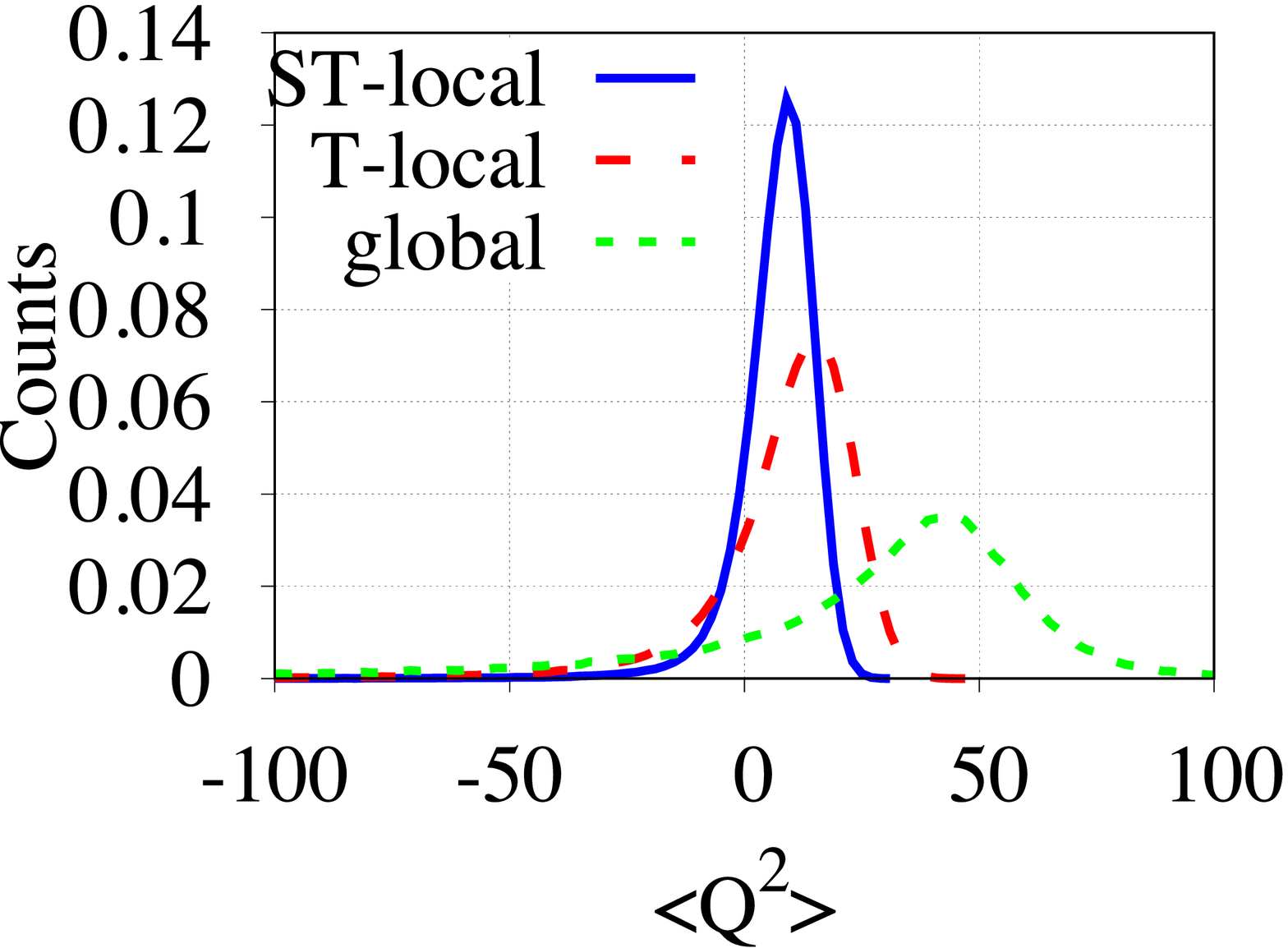}\label{fig:histogram_charge_6x6x256}} 
    \subfigure[]{\includegraphics[width=0.23\textwidth,angle=0]{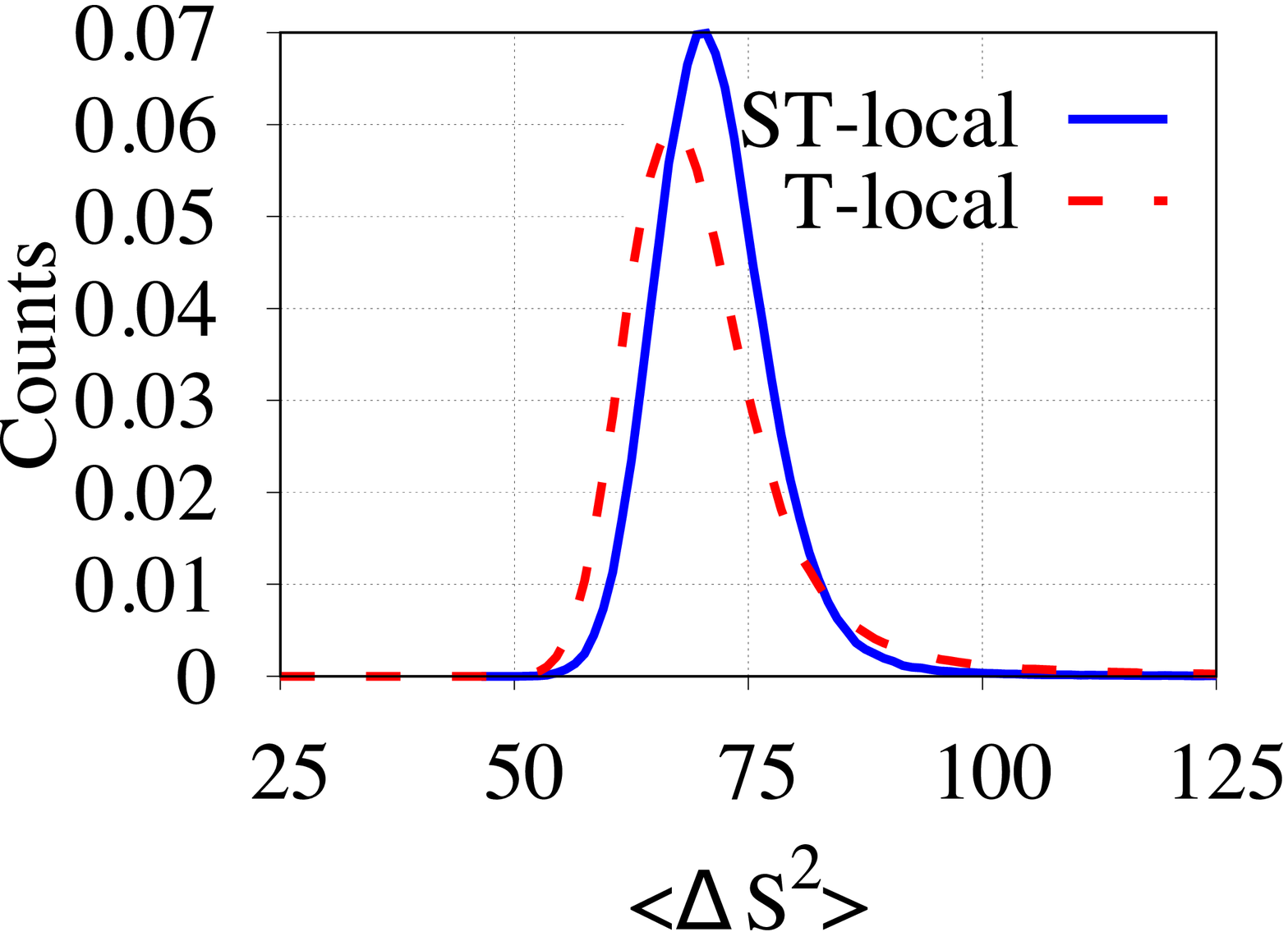}\label{fig:histogram_delta_spin_6x6x256}} 
    \subfigure[]{\includegraphics[width=0.23\textwidth,angle=0]{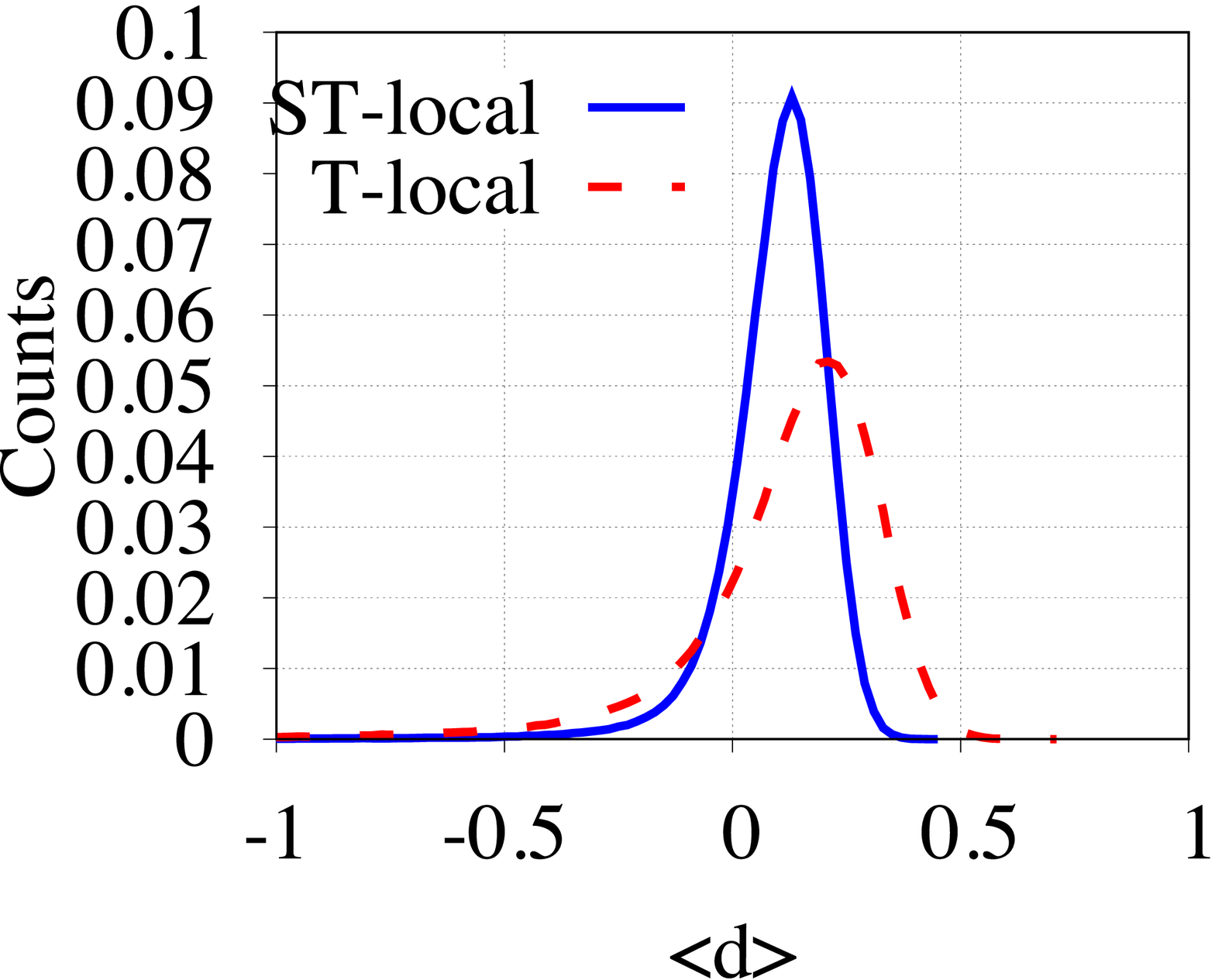}\label{fig:histogram_double_occ_6x6x256}} 
       \caption{Histograms for the distribution of observables after one full sweep. Each plot corresponds to one observable: (a) - sublattice spin fluctuations; (b) - sublattice charge fluctuations; (c) - squared  staggered moment; (d) - double occupancy. In each case several ways of synchronization between updates and calculation of observable are shown. Calculations were done on the  $6\times6\times256$ hexagonal lattice at $U=5.0$ and $\beta=20.0$ using the SU(2) symmetrical discrete HS field with standard local Blankenbecler, Scalapino, Sugar (BSS)-QMC updates. }
   \label{fig:full_histograms_6x6x256}
\end{figure}

  \begin{figure}[]
   \centering
 \includegraphics[width=0.3\textwidth , angle=-90]{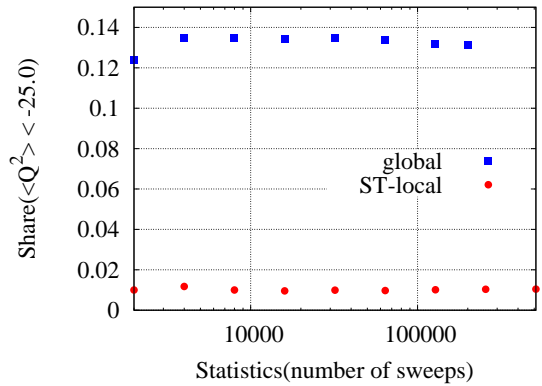}
   
      \caption{Evolution of the distributions for the sublattice charge fluctuations with increased number of sweeps. The plot shows the share of the configurations, where the observable is less than -25.0 (large negative spike) depending on the whole number of configurations. The share remains stable, thus signalling that the distributions shown in the figures \ref{fig:full_histograms_6x6x256}  and \ref{fig:full_histograms_comparison} are not dependent on the size of statistics. Calculations were done on the $6\times6\times256$ hexagonal lattice at  $U=5.0$ and  $\beta=20.0$.}
   \label{fig:autocorrelation_check}
\end{figure}

\begin{figure}[]
    \centering

 \subfigure[]{\label{fig:histogram_charge_12x12x256}\includegraphics[width=0.23\textwidth , angle=0]{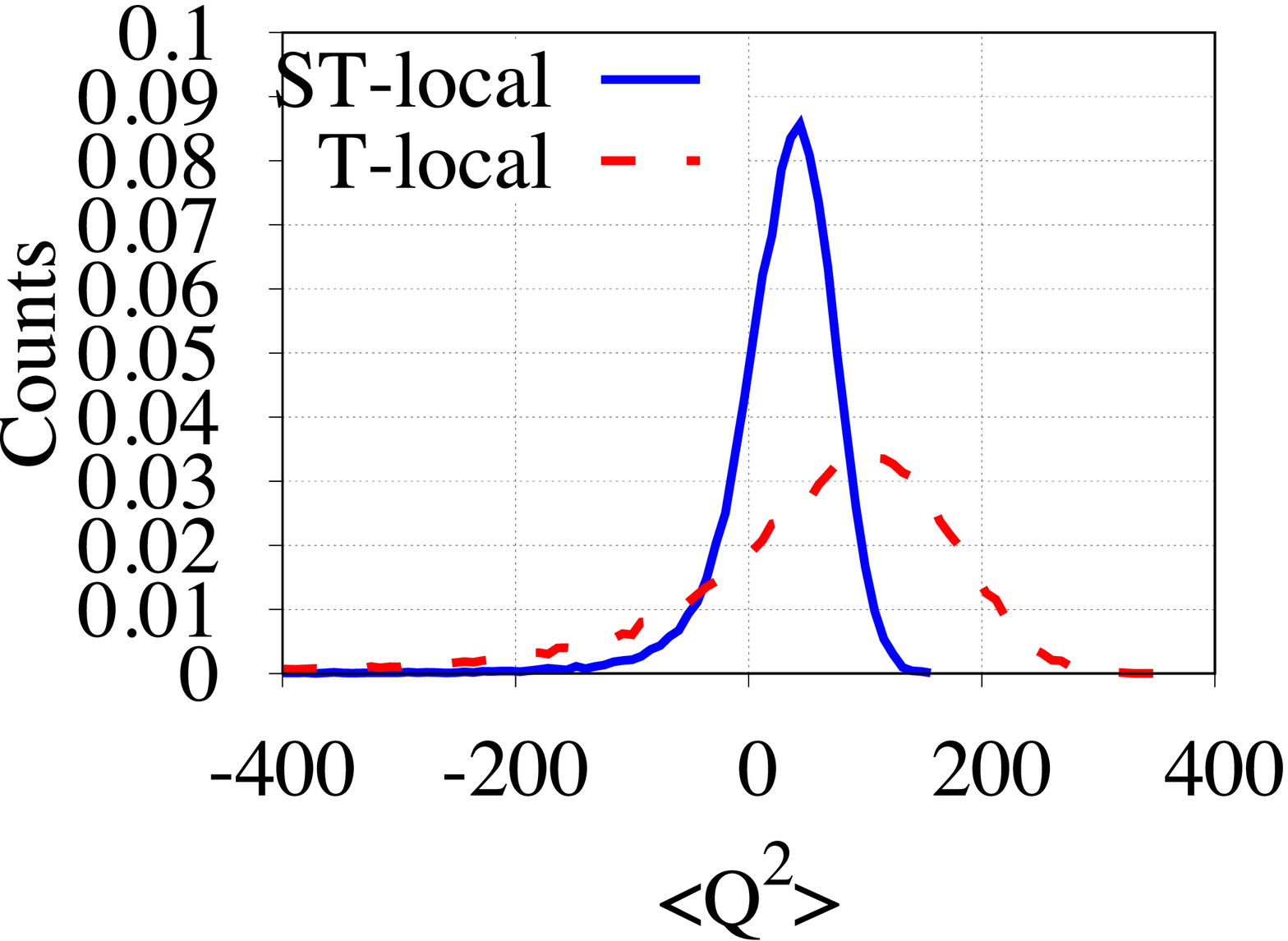}}
  \subfigure[]{\label{fig:histogram_spin_12x12x256}\includegraphics[width=0.23\textwidth , angle=0]{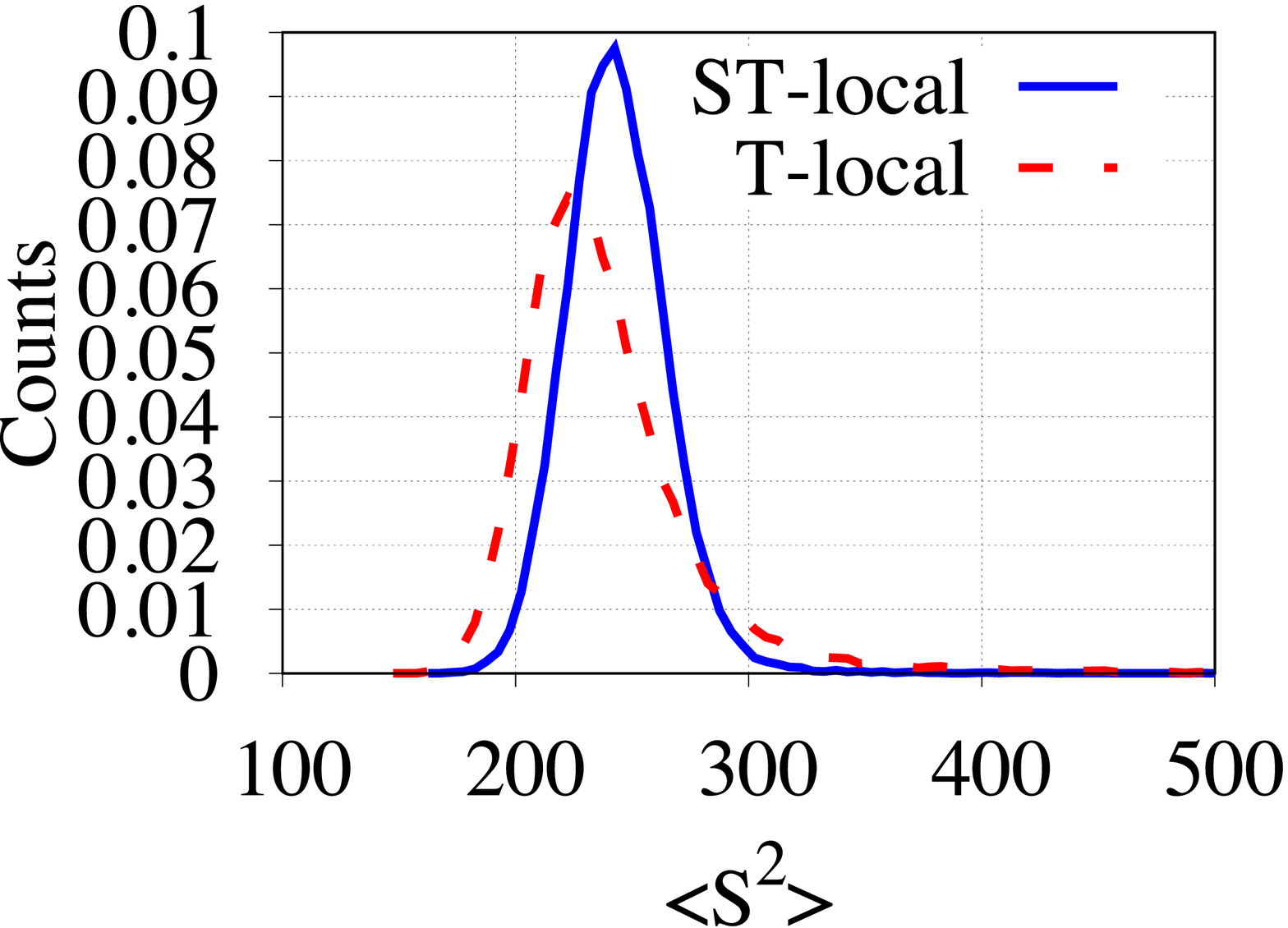}}
  \subfigure[]{\label{fig:histogram_charge_6x6x512}\includegraphics[width=0.23\textwidth , angle=0]{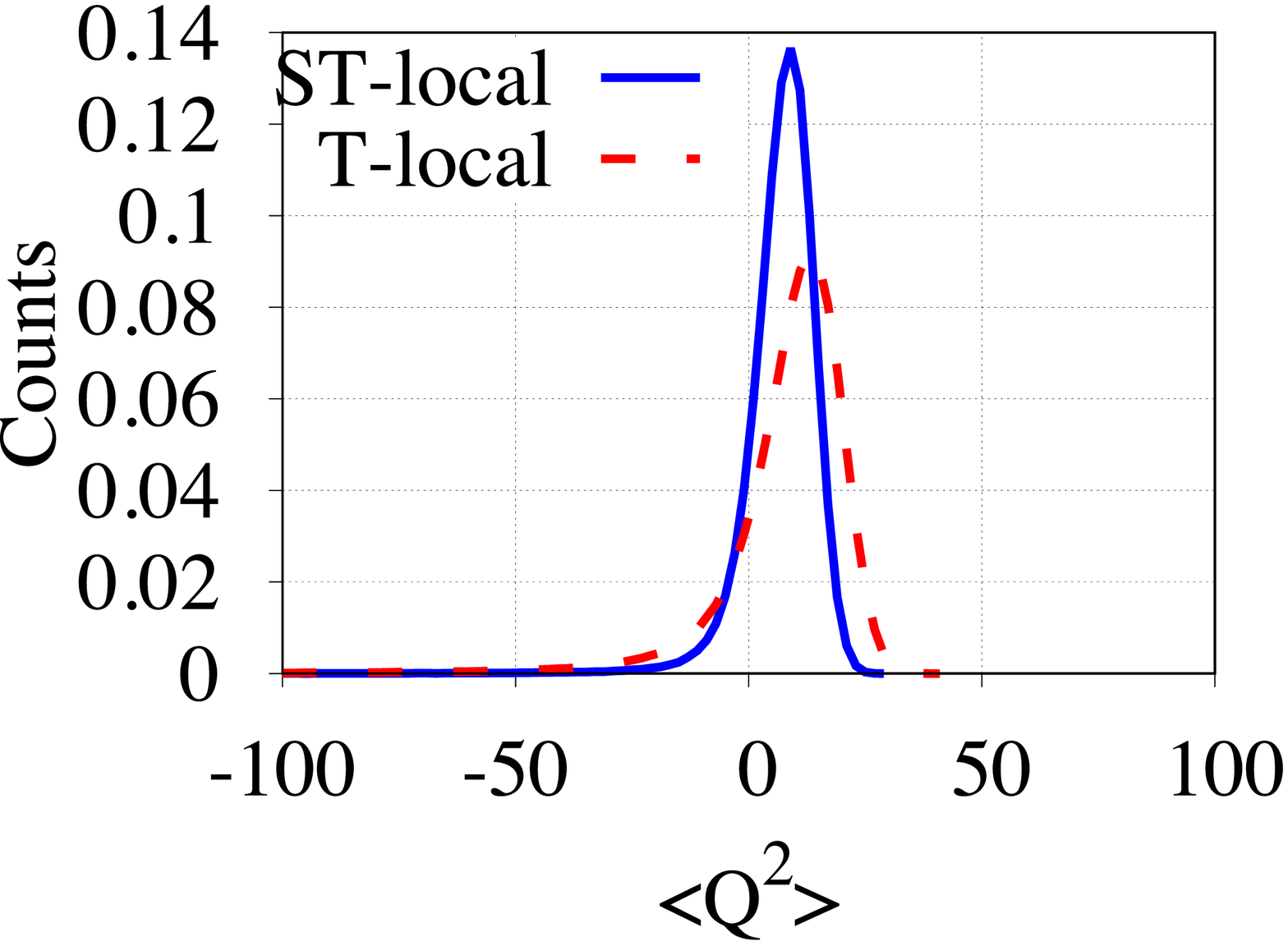}}
  \subfigure[]{\label{fig:histogram_spin_6x6x512}\includegraphics[width=0.23\textwidth , angle=0]{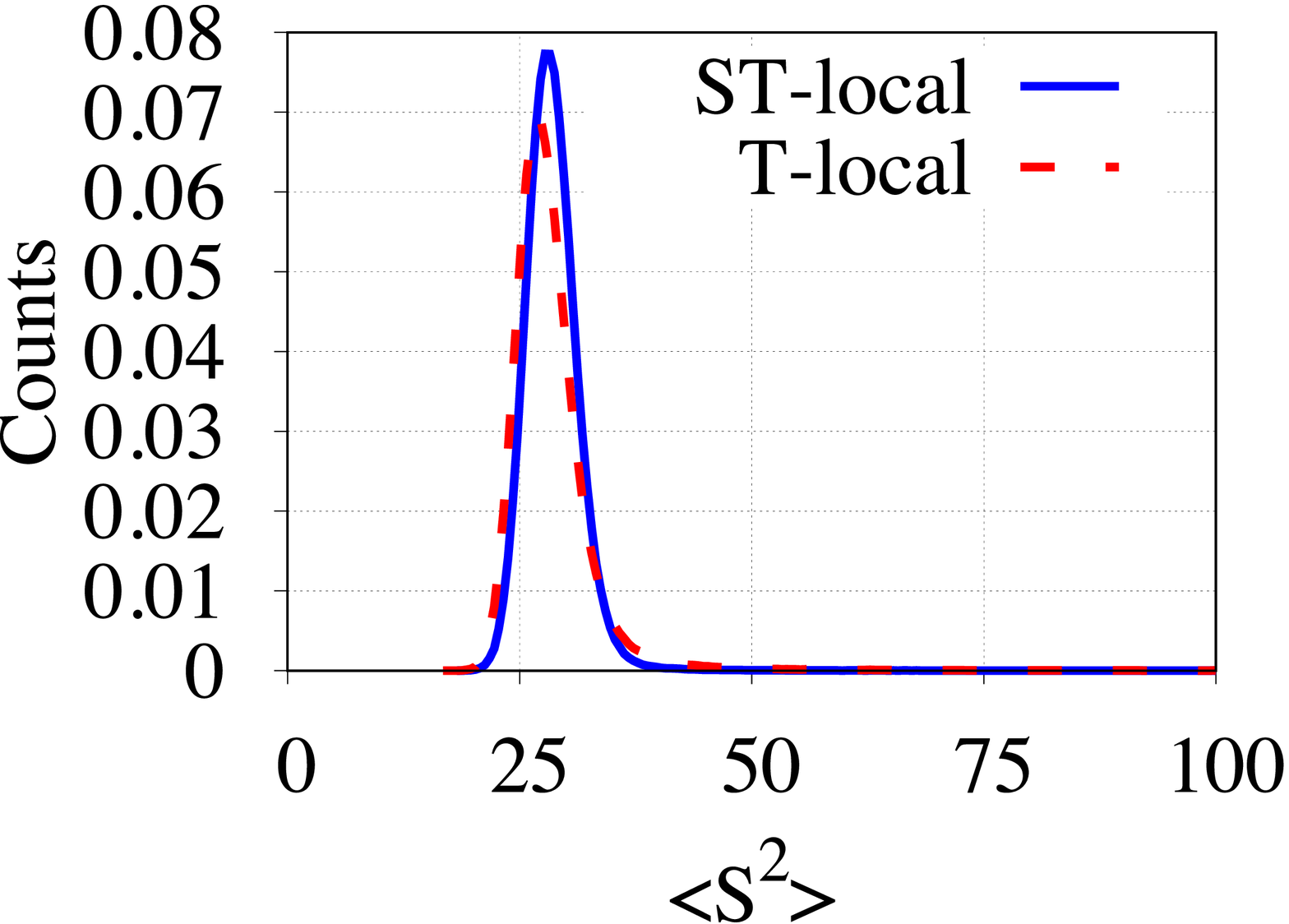}}
 \subfigure[]{\label{fig:histogram_charge_8x8x256sq}\includegraphics[width=0.23\textwidth , angle=0]{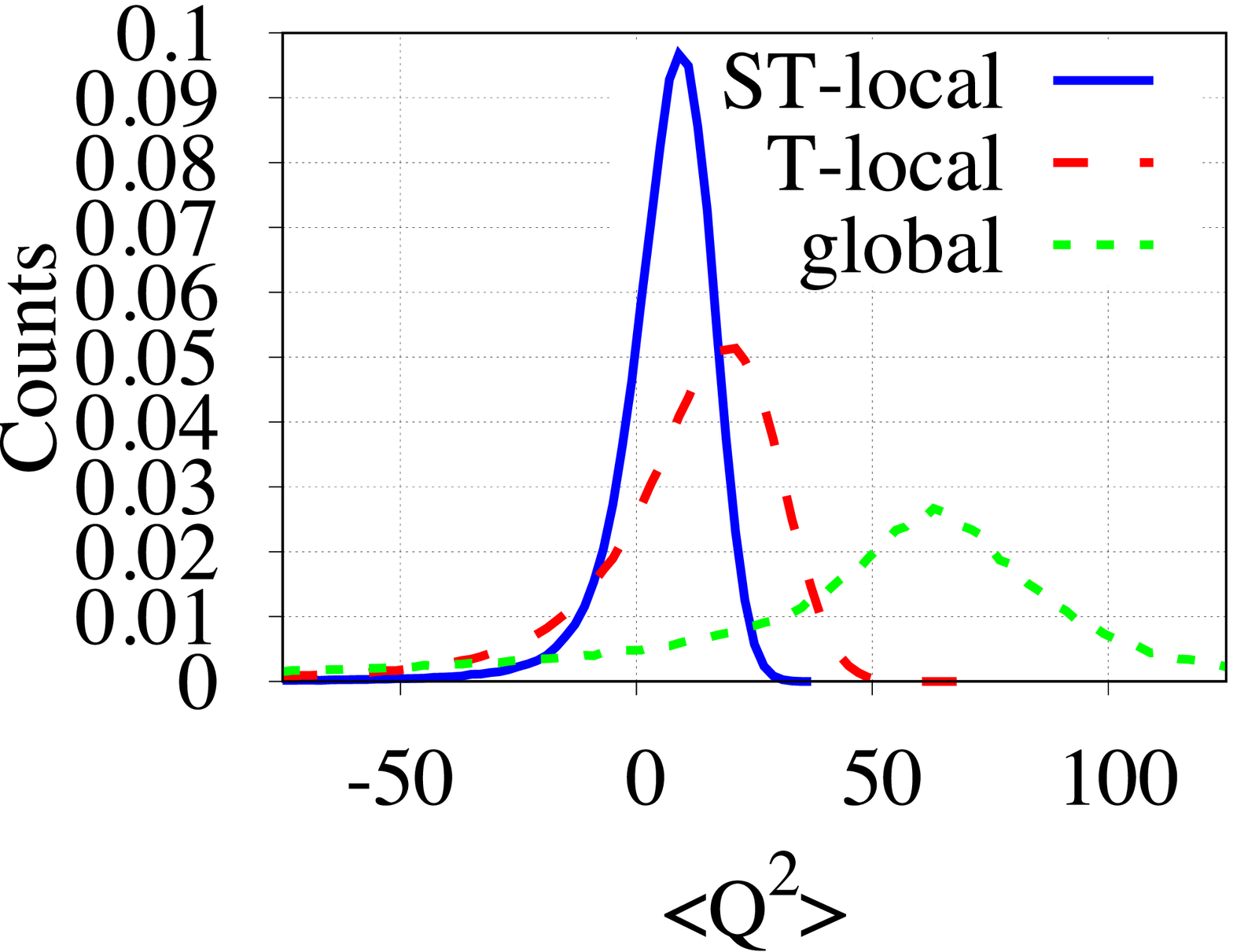}}
  \subfigure[]{\label{fig:histogram_spin_8x8x256sq}\includegraphics[width=0.23\textwidth , angle=0]{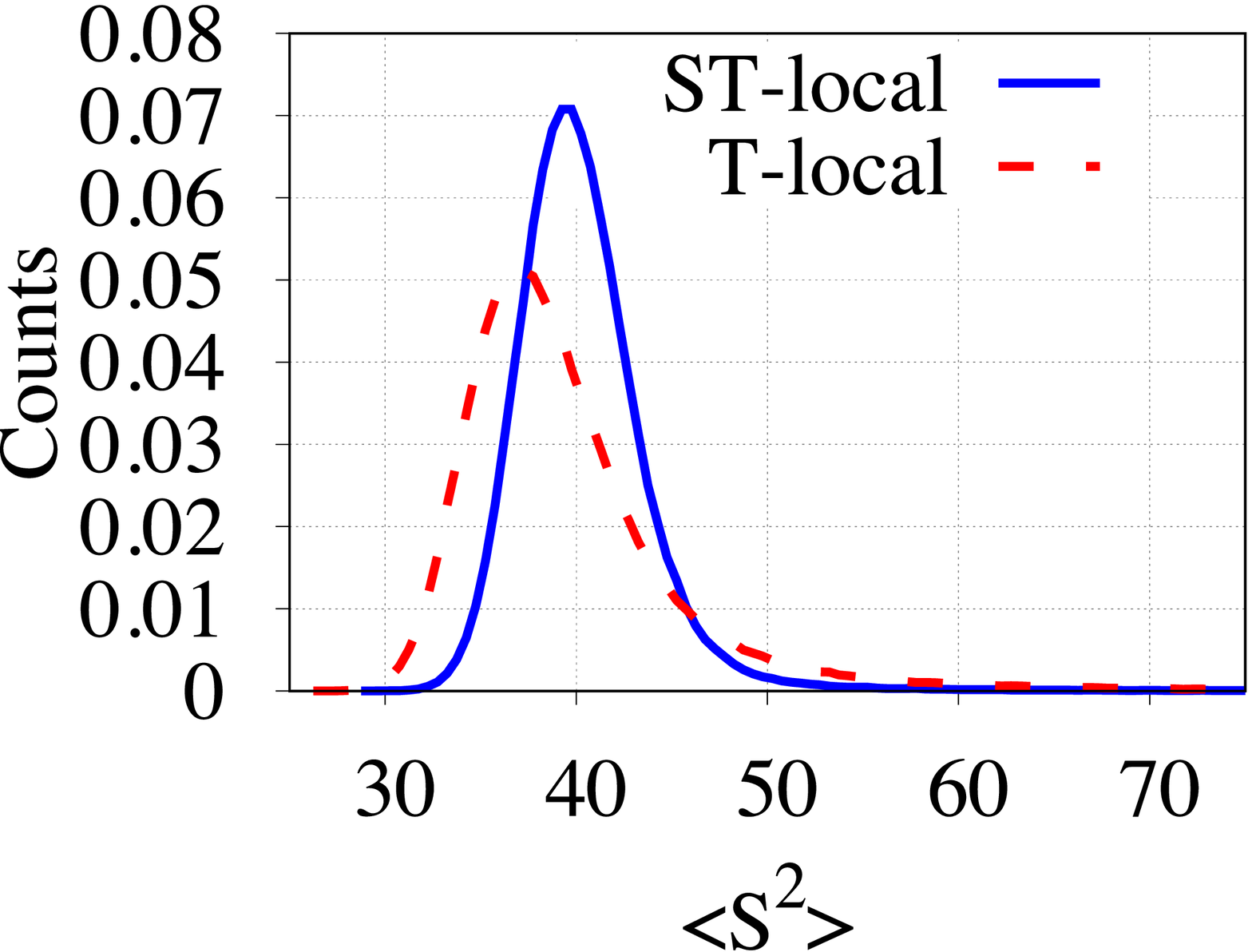}}
            
      \caption{Comparison of the distribution of observables for different lattices and models: left (right) column shows sublattice  charge (spin)   fluctuations. First (upper) row corresponds to a  $12\times12\times256$ hexagonal lattice, the second row shows the results for the $6\times6\times512$ hexagonal lattice with a  twice smaller step in Euclidean time and the third row corresponds to the square lattice Hubbard model on  the $8\times8\times256$ lattice. The Hubbard interaction and temperature are the same in all cases: $U=5.0$ and $\beta=20.0$. For comparison with  the $6\times6\times256$ lattice, one can refer to the upper row in Fig. \ref{fig:full_histograms_6x6x256}. 
     }
   \label{fig:full_histograms_comparison}
\end{figure}

 \section{\label{sec:main_results}QMC results}

We carried  test calculations on several lattices: 
\begin{itemize}
    \item $6\times6$ hexagonal lattices with $L_\tau=256$ and $L_\tau=512$ at the same temperature to test the influence of the step in Euclidean time;
     \item $12\times12$ hexagonal lattice with $L_\tau=256$;
      \item $8\times8$ square lattice with $L_\tau=256$;
\end{itemize}
For all our  calculations,  we  have  fixed the  strength of Hubbard interaction to $U=5.0$ in  units of the hopping matrix  element  and  set the inverse temperature  to  $\beta=20.0$. 

We consider the three schemes of synchronization of updates and measurements, described in the Introduction.
 The double occupancy \ref{eq:squared_d_occ} is a local observable, thus the expressions \ref{eq:Global}, \ref{eq:Tlocal} and \ref{eq:STlocal} are fully applicable. Other observables are only local in Euclidean time, but contain spatial correlations $\hat O_{\ve{i},\tau} \hat O_{\ve{j},\tau}$.
 In this  case,  global and T-local schemes are essentially the same as  for  local observables: in the former case, we compute all correlators at 
 all time slices after updating all fields; in the latter case, all correlators on time slice $\tau$ are computed after  updating all fields in the same time slice.  
The   ST-local scheme is slightly modified: we compute the average for the correlator $\hat O_{\ve{i},\tau} \hat O_{\ve{j},\tau}$ after local updates of 
both  the fields  $s_{\ve{i},\tau}$ and  $s_{\ve{j},\tau}$.
Since we need all possible correlators of spin and charge to measure the  observables \ref{eq:squared_charge}, \ref{eq:squared_spin} and \ref{eq:squared_delta_spin},  we in practice  simply update  the average for  $\hat O_{\ve{i},\tau} \hat O_{\ve{j},\tau}$ $\forall j$ (with $\hat O$ representing operator of local spin or charge) after each local update of the  $s_{\ve{i},\tau}$ field.

We start from the histograms for the distributions of observables. In all cases we take the averages over the 
 full lattice volume after one full sweep of local updates through the field configuration, with corresponding synchronization between updates and measurements. All four observables presented in Fig.~\ref{fig:full_histograms_6x6x256}  are computed on $6\times 6 \times 256$ lattices.
 Distributions for $\langle S^2\rangle$ and $\langle \Delta S^2\rangle$ have heavy tails towards $+\infty$ and the distributions for double occupancy and $\langle Q^2\rangle$ have heavy tails towards $-\infty$. The differences between global, T-local and ST-local schemes are quite noticeable with the width of the distributions decreasing with  synchronization  of measurements and  updates. We are mostly interested in the comparison of ST-local and T-local schemes, since the former one offers some improvement for existing QMC codes \cite{ALF_v2}.  Distributions for double occupancy, $\langle Q^2\rangle$ (Fig. \ref{fig:histogram_double_occ_6x6x256} and \ref{fig:histogram_charge_6x6x256})  shows a notable  improvement   when  considering an
 ST-local  scheme as opposed to a T-local one.  On the other hand,  the  improvement is smaller for  non-local observables   such as  the spin-spin correlations,  Fig. \ref{fig:histogram_spin_6x6x256} and \ref{fig:histogram_delta_spin_6x6x256}.

\begin{figure}[]
    \centering
\subfigure[]{\label{fig:tail_global_charge}\includegraphics[width=0.23\textwidth , angle=0]{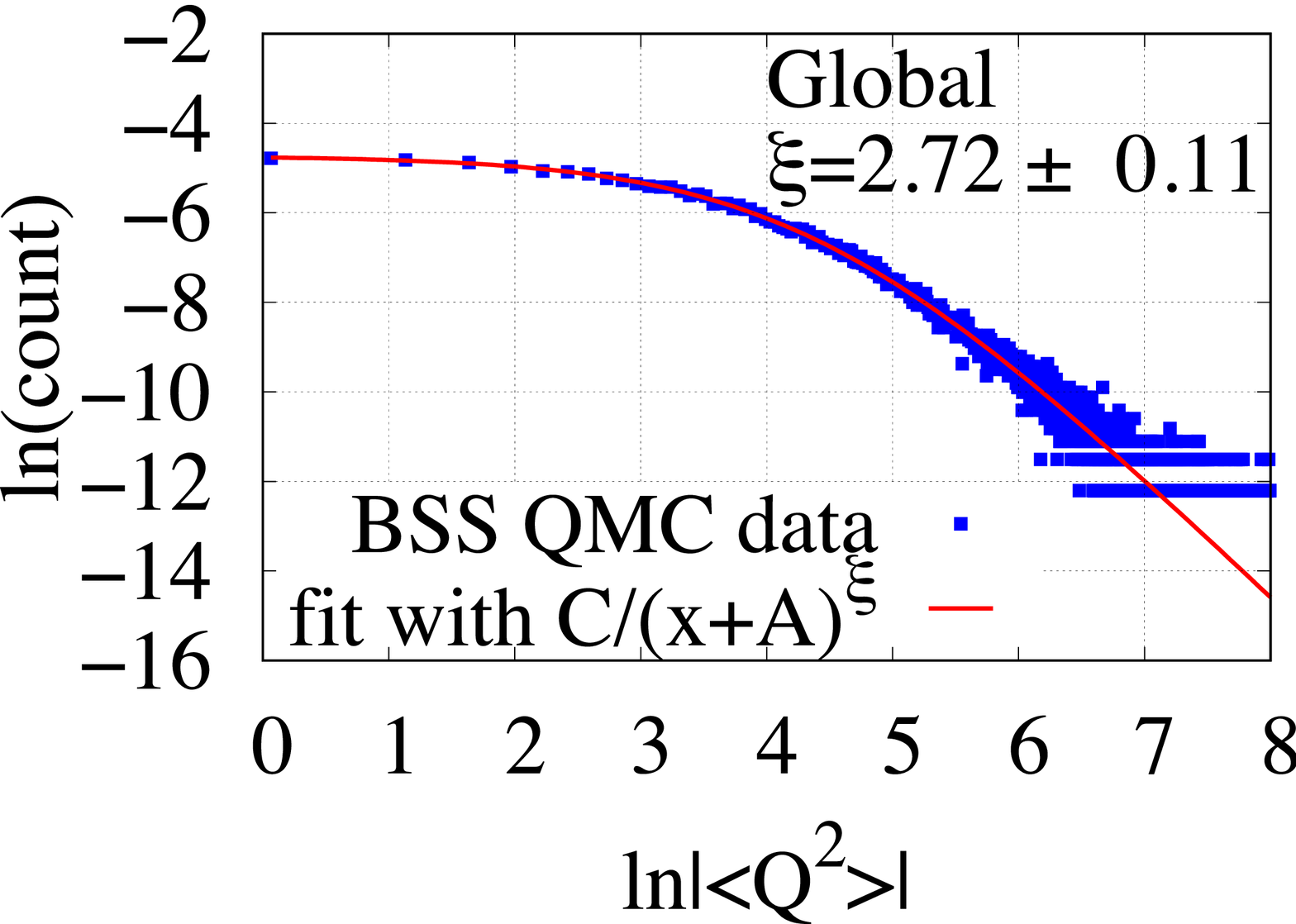}}
\subfigure[]{\label{fig:tail_global_spin}\includegraphics[width=0.23\textwidth , angle=0]{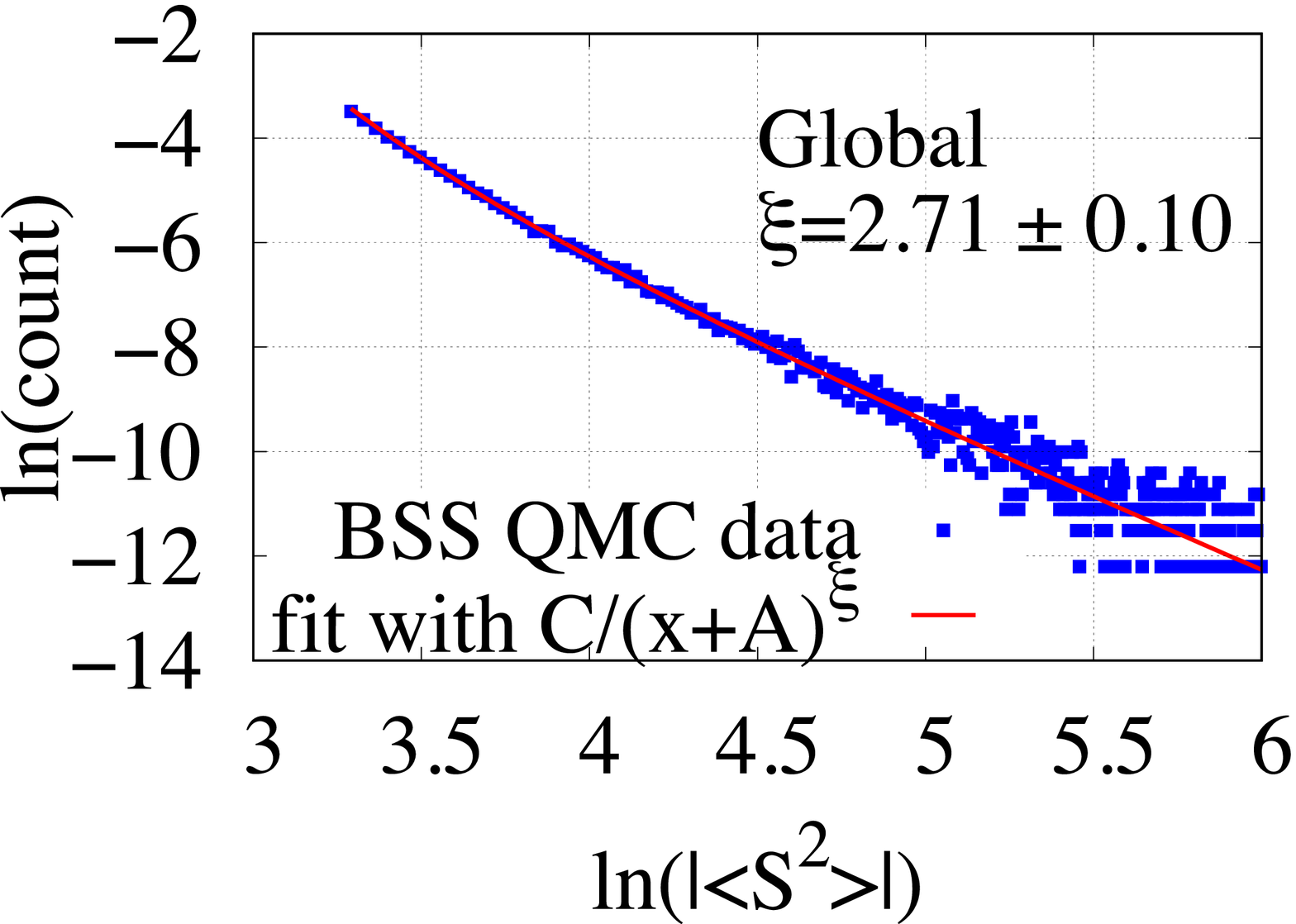}}
 \subfigure[]{\label{fig:tail_tlocal_charge}\includegraphics[width=0.23\textwidth , angle=0]{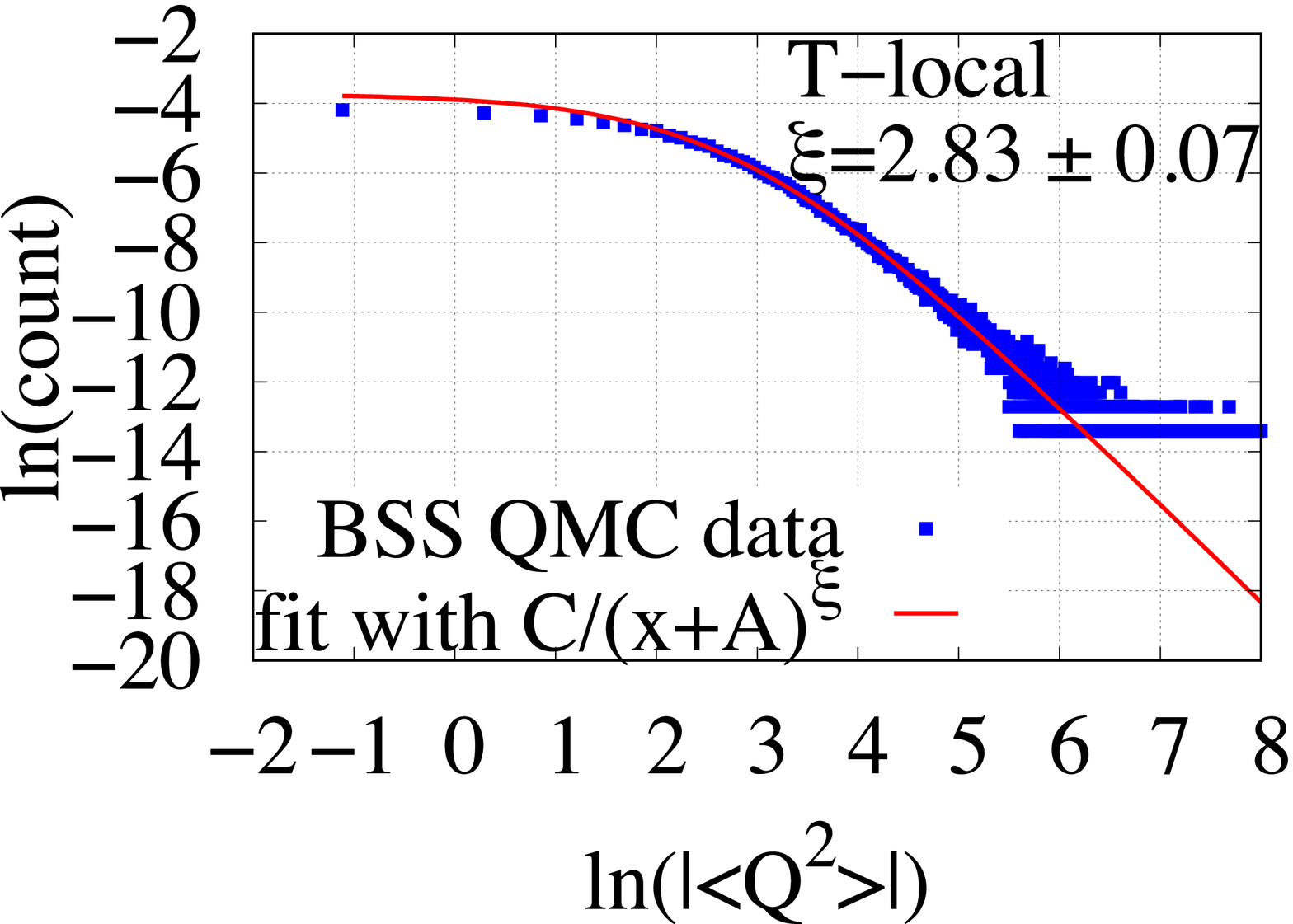}}
 \subfigure[]{\label{fig:tail_tlocal_spin}\includegraphics[width=0.23\textwidth , angle=0]{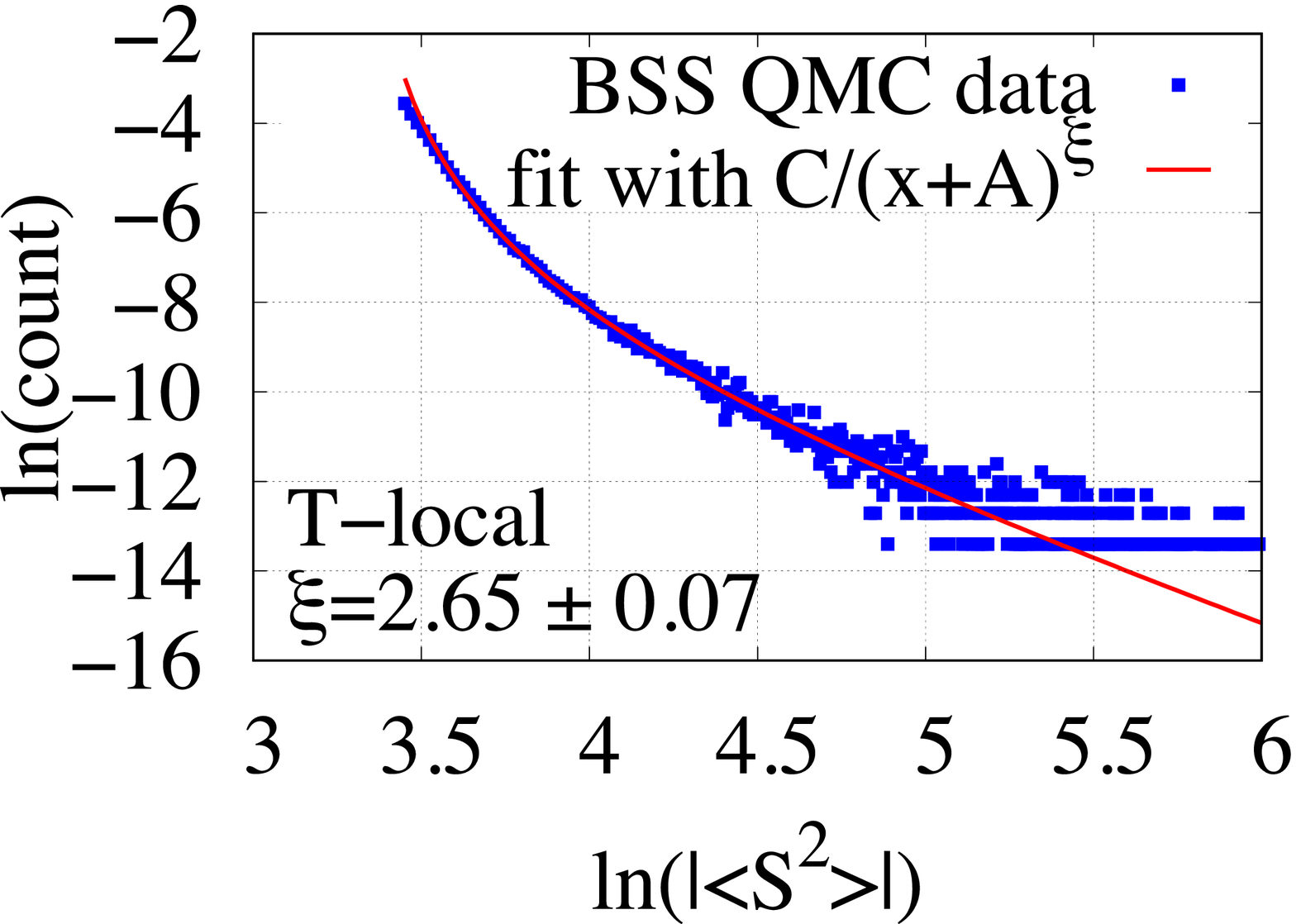}}
\subfigure[]{\label{fig:tail_stlocal_charge}\includegraphics[width=0.23\textwidth , angle=0]{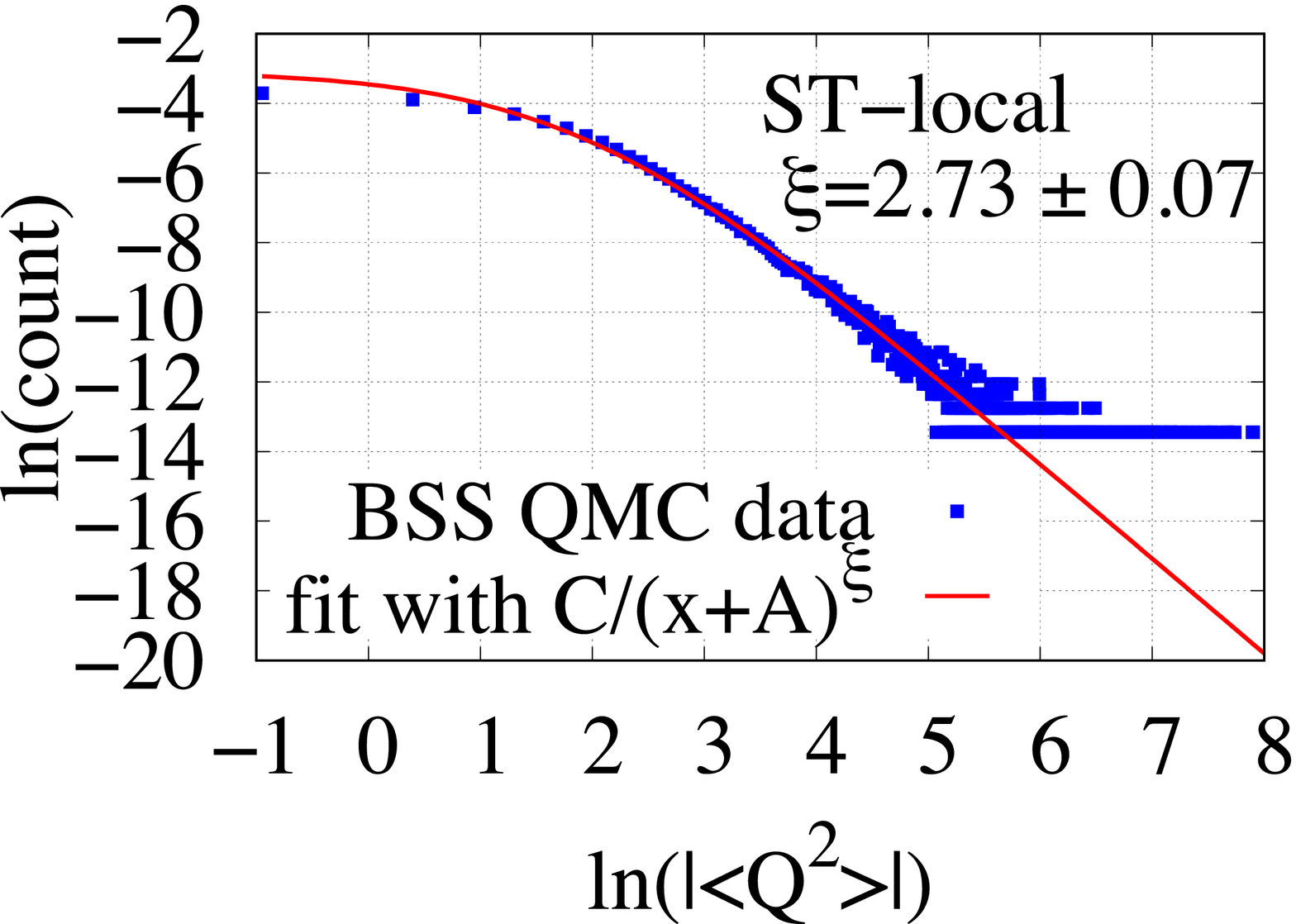}}
 \subfigure[]{\label{fig:tail_stlocal_spin}\includegraphics[width=0.23\textwidth , angle=0]{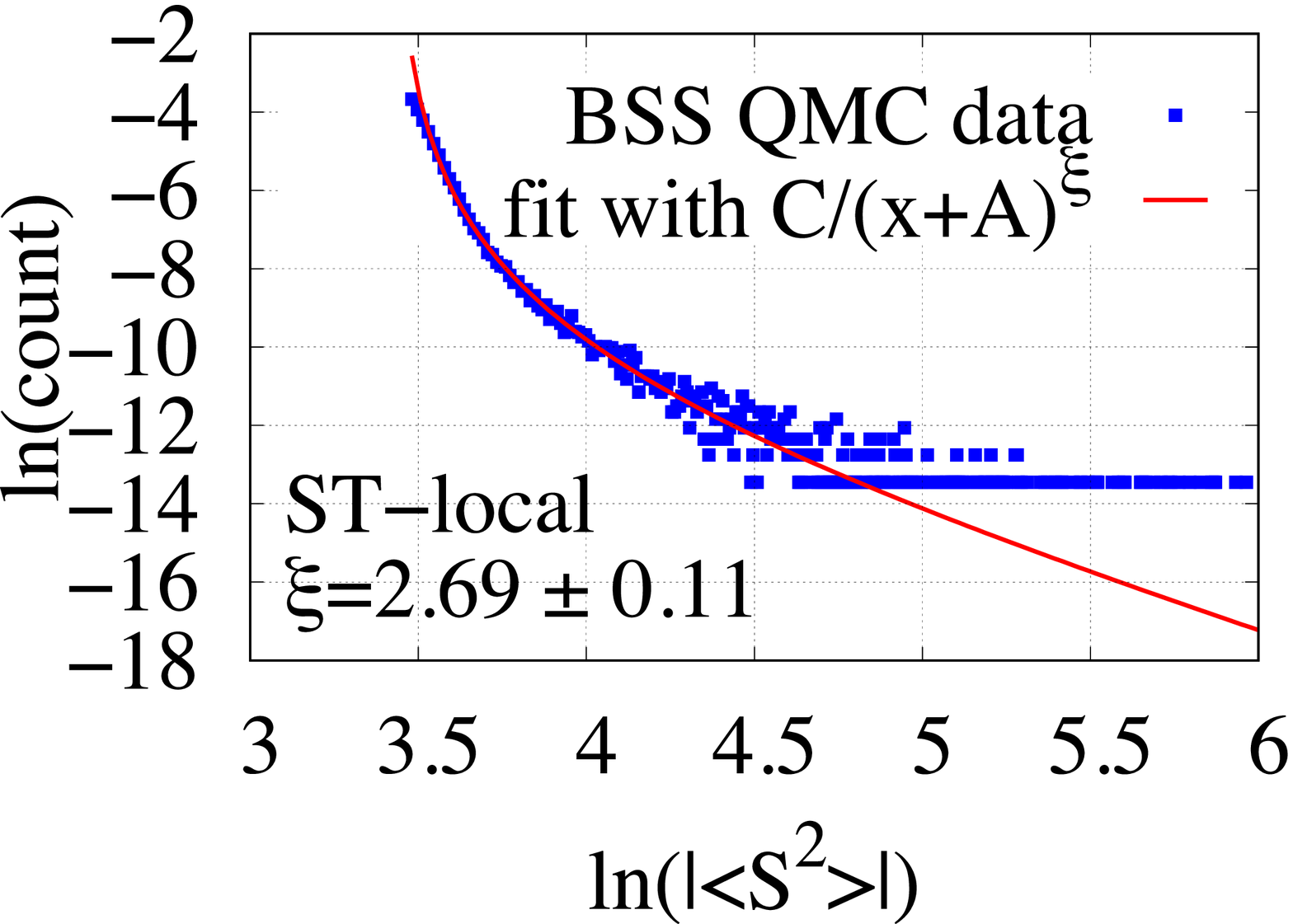}}

      \caption{Fit of the tail of the histograms   to the form $\frac{C}{\left(  x + A \right)^\xi}$. Left  (right) column shows  the sublattice  charge (spin) fluctuations.  First (upper) row corresponds to the global measurement of an observable, the second row shows the T-local computations and the last (lower) row corresponds to the ST-local scheme of measurement.  All data were produced on the  $6\times6\times256$ lattice 
   at  $U=5.0$  and $\beta=20.0$.}
   \label{fig:full_histograms_tail}
\end{figure}

We also check that the observed difference in distribution is not the effect of the changing autocorrelation length. One might propose that the decreased tails in ST-local and T-local schemes are due to larger distance between spikes: if we make more sweeps (collect larger statistics), we will see the appearance of worse spike. From the point of view of the distributions, it means that their form will be dependent on the size of statistics: we did not see the spike (formally: the deviation of observable from the mean value larger than some threshold) on smaller statistics but we will start to see them at larger one. Thus the distributions will start to change with increasing weight of the heavy tail once the statistics is enlarged. However, the figure \ref{fig:autocorrelation_check} shows that it is not the case: the distributions are stable with respect to the statistics size, thus the observed effects can not be attributed to the changing autocorrelation length.

  \begin{figure}[]
   \centering
   \subfigure[]{\label{fig:histogram_mass}\includegraphics[width=0.23\textwidth , angle=0]{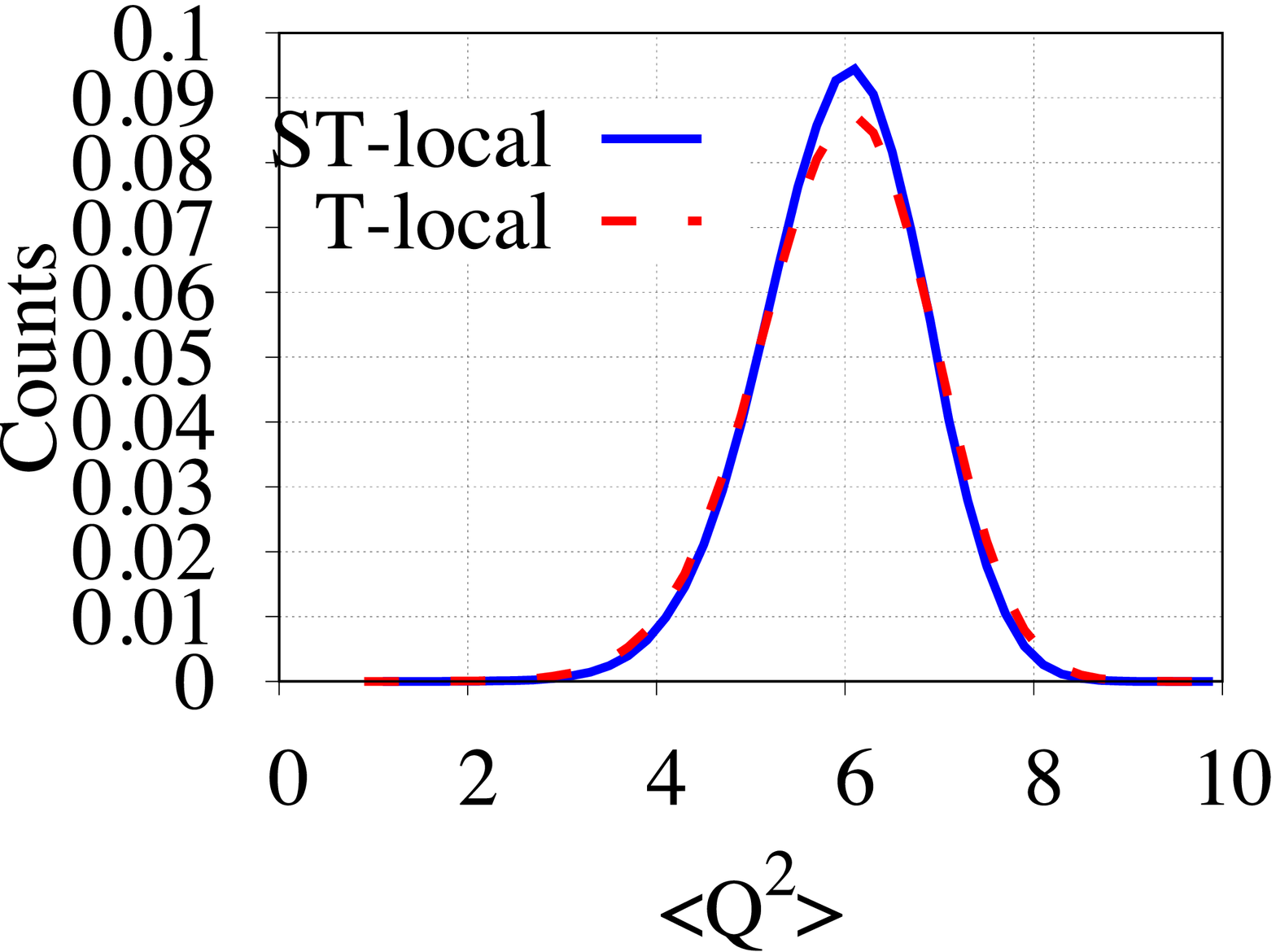}}
         \subfigure[]{\label{fig:N_A_dep}\includegraphics[width=0.23\textwidth , angle=0]{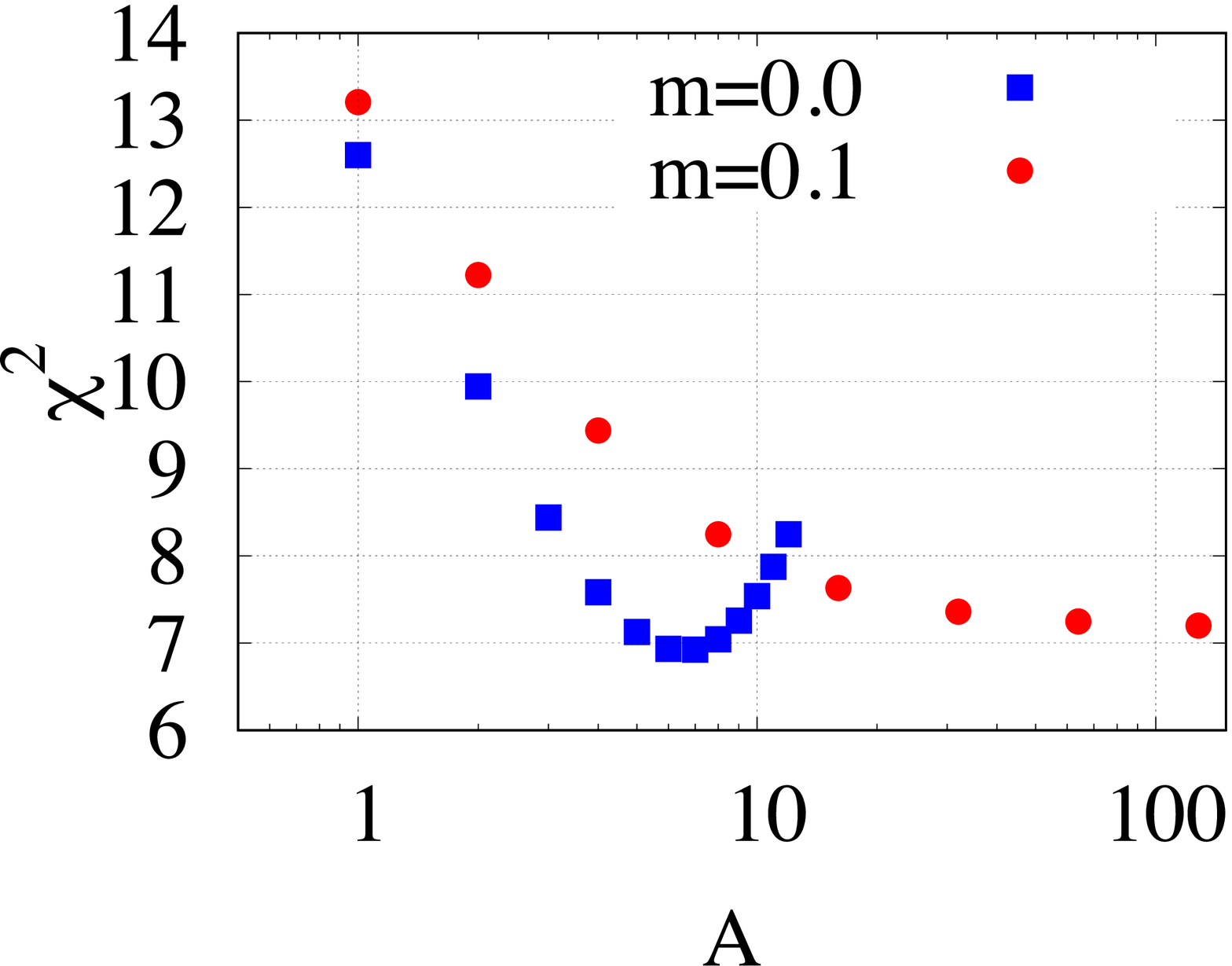}}
         \subfigure[]{\label{fig:tail_mass}\includegraphics[width=0.47\textwidth , angle=00]{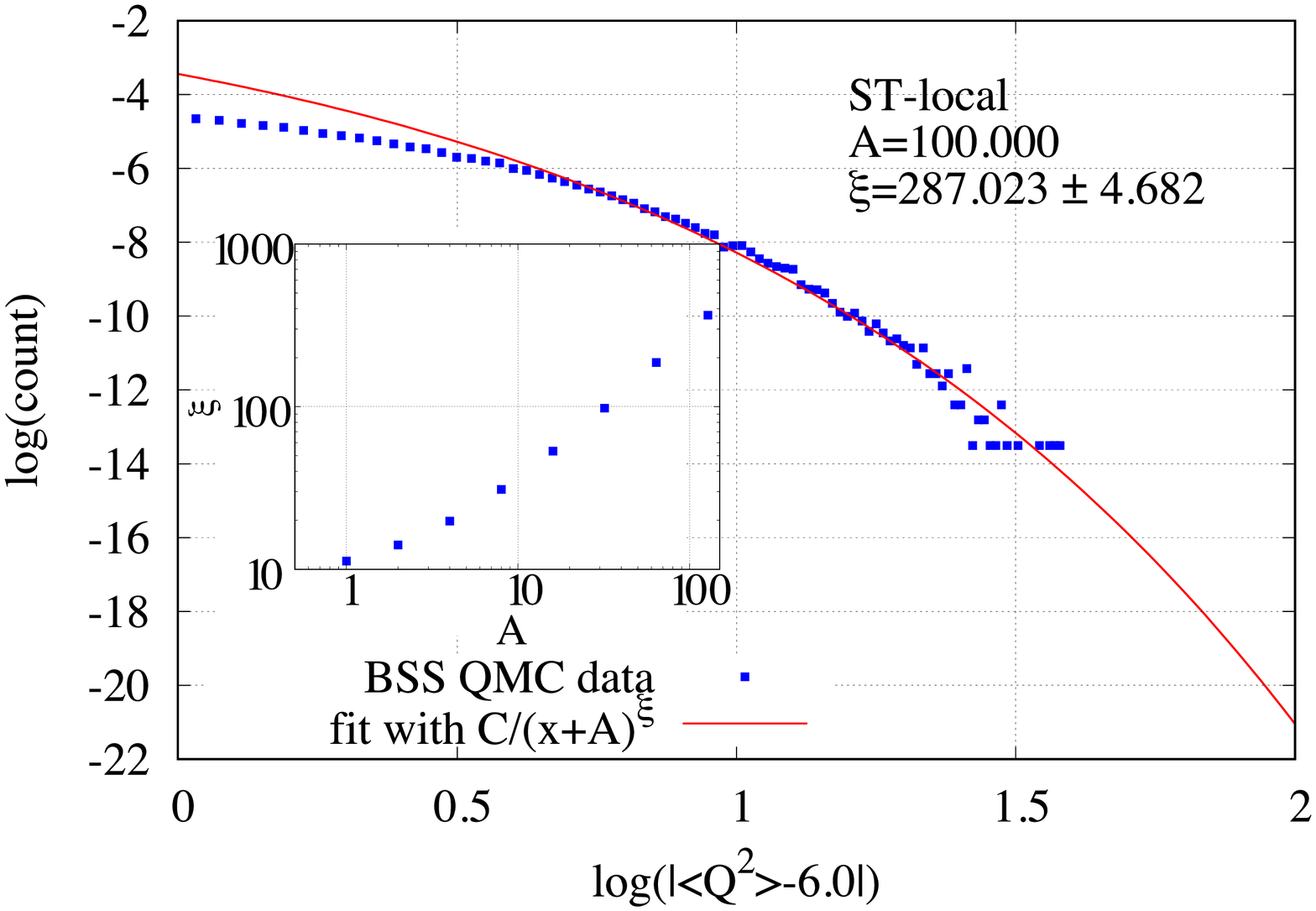}}
      \caption{(a) histogram for the distribution of the sublattice  charge fluctuations in the calculation with the staggered mass, 
      that  eliminates the zeros of the determinant.  (c) an attempt to fit the tail of the histogram by a  power law (at fixed shift parameter $A$). The inset shows the steady increase of the power in the fit upon increasing  $A$. (b) the dependence of $\chi^2$ on the shift $A$ for two different cases: with and without the staggered mass term.  The computations were   carried out on a $6\times6\times256$ lattice at  $\beta=20$, $U=5$,  and mass $m=0.1$. }
   \label{fig:finite_mass_analysis}
\end{figure}

Comparison of the distributions on different lattices (Fig.~\ref{fig:full_histograms_comparison}) shows that the effect of reshuffling increases for larger lattices. This  becomes apparent  when one compares the differences between ST-local and T-local histograms of Figs.~\ref{fig:histogram_charge_6x6x256} and  \ref{fig:histogram_charge_12x12x256}.  In contrast, the effect decreases with decreasing step in Euclidean time.  In particular, the ST-local and T-local histograms are closer to each other in Fig.~\ref{fig:histogram_charge_6x6x512} than in Fig.~\ref{fig:histogram_charge_6x6x256}.  This can be attributed to the decreasing influence of  the single local update in the  continuum limit. Results for the square lattice also demonstrate a somewhat increased difference between ST-local and T-local measurements on lattices of similar size; compare Fig.~\ref{fig:histogram_spin_8x8x256sq} for  the $8\times8\times256$ square lattice and Fig.~\ref{fig:histogram_spin_6x6x256} for  the $6\times6\times256$ hexagonal lattice.

 In Fig.~\ref{fig:full_histograms_tail}   we  fit the  tails of the histograms    for  our  considered  observables and measurement  schemes to the form 
 $\frac{C}{\left(  x + A \right)^\xi}$.    Within  our  error-bars,  $\xi$  is independent  on the measurement scheme, and is reasonably close to the theoretical value $5/2$. The difference  may be explained by the presence of lower-dimensional manifolds in the configurations space, formed by  the intersection of "domain walls" of zeros of the determinant. They can  result in higher-order power law corrections to the asymptotic, that  are quite hard to detect in the noisy numerical data.

One can get   rid  of   the  zeros  of the determinant  by   breaking  symmetries  in the  form of mass terms.    A possible choice  is  
to add  to the  quantum Hamiltonian  Eq.~\ref{eq:Hamiltonian_spin}   the term 
 \begin{equation}
	\hat{H}_m    =  m  \sum_{ \ve{i}\in sublat. 1} (n_{\ve{i},\uparrow}- n_{\ve{i},\downarrow}) + m \sum_{ \ve{i}\in sublat. 2} (n_{\ve{i},\downarrow}- n_{\ve{i},\uparrow}).
	\label{eq:Hamiltonian_mass}
\end{equation}
that  breaks SU(2)  spin  symmetry.     It is known \cite{Polikarpov1206}    that  this  mass term  in conjunction with our  choice of the  HS  transformation
  removes  the   zeros from the  fermion determinant.
 Fig.~\ref{fig:histogram_mass} plots  the  histograms in the presence of the mass  term.  As apparent,  the differences between  the measurement schemes disappear.    
The absence of the heavy tails in this case can be proven if we attempt  to fit the tails of the finite mass histograms with a power law. This fit never converges: there is no local minimum of $\chi^2$ with respect to the shift parameter $A$ in the power law
\begin{equation}
    P^{(L/G)}_O(z)|_{z\rightarrow\infty}\approx C/(z+A)^{\xi} \label{eq:power_law_shift}. 
\end{equation}
Figure \ref{fig:N_A_dep} clearly outlines this difference between finite $m$ and zero $m$ data. Consequently, both the shift $A$ and the power $\xi$ grow indefinitely  in the fitting procedure (see the inset on Fig. \ref{fig:tail_mass}), thus demonstrating that the power law is incompatible with the finite mass data.

This analysis  allows  for  the  claim that the heavy tails  and concomitant reduction of the  observed spikes in various  measurement schemes is  entirely 
caused  by the  zeros of the fermion determinant. It also supports the above mentioned statement that the effect of reshuffling in the reduction of the statistical fluctuations is absent if the second moment of the distribution is well-defined.

  \begin{figure}[]
   \centering
   \subfigure[]{\label{fig:err_comparison_charge_global}\includegraphics[width=0.48\textwidth , angle=0]{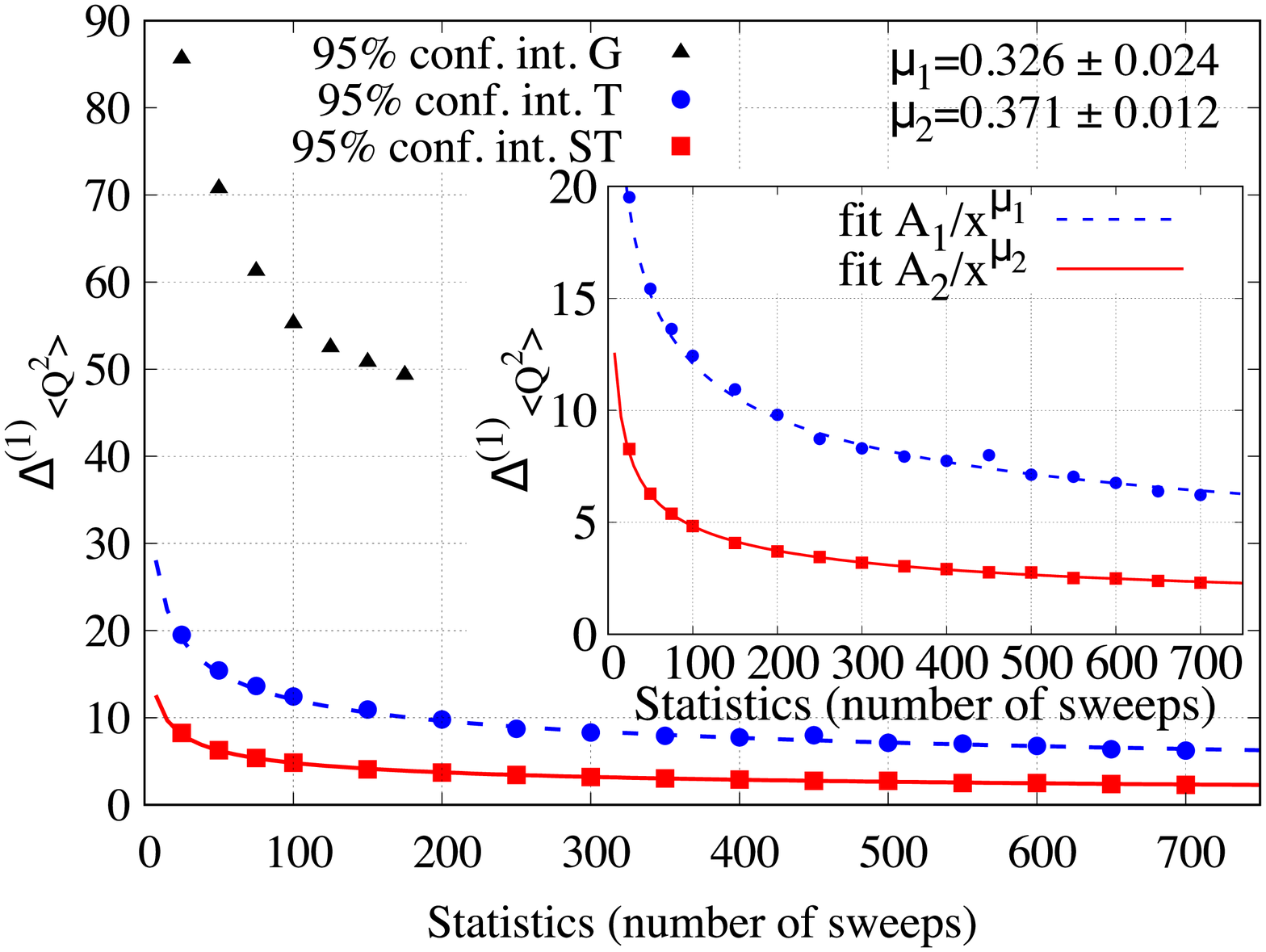}}
   \subfigure[]{\label{fig:err_comparison_spin_global}\includegraphics[width=0.48\textwidth , angle=0]{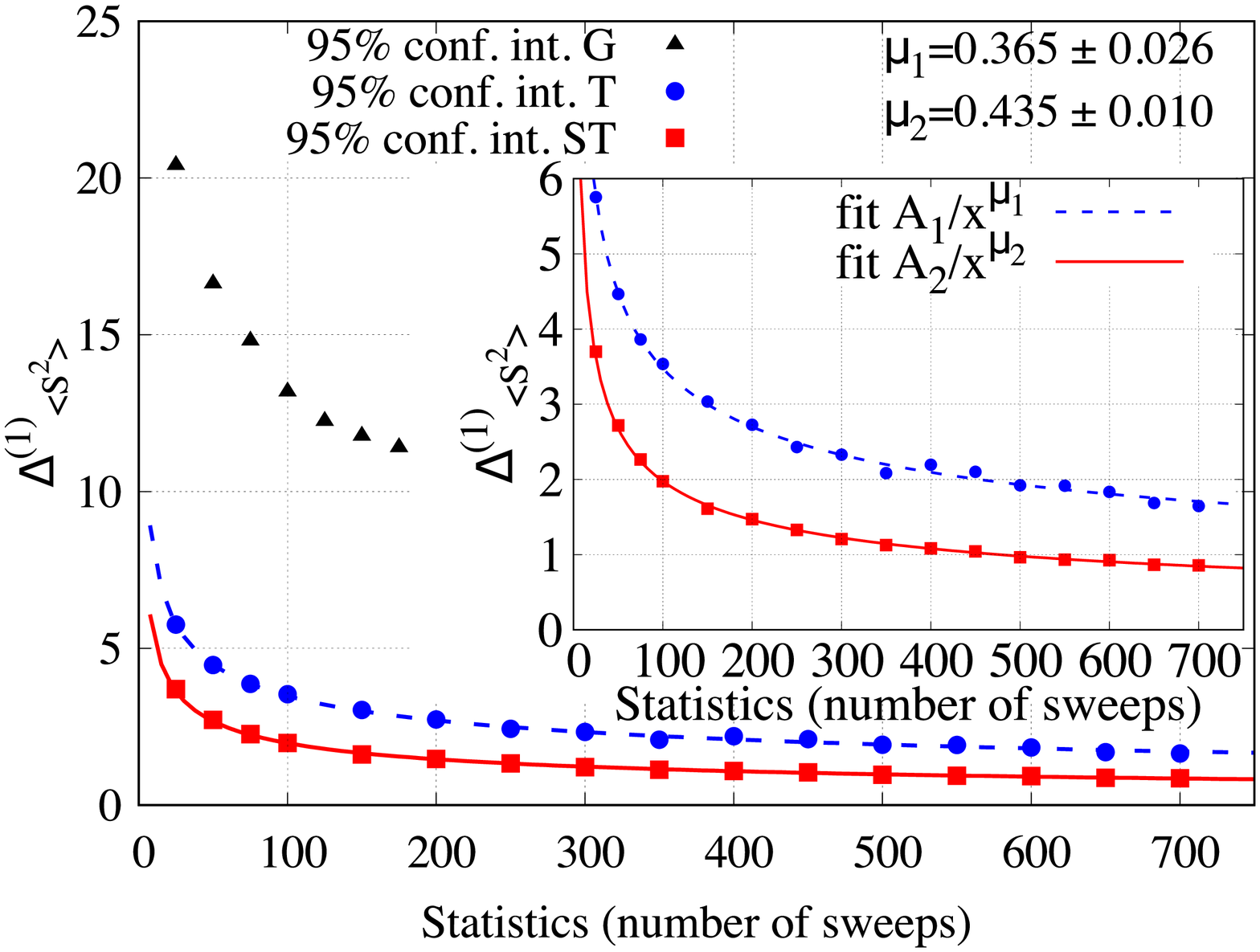}}
   
      \caption{Comparison of error bars (width of the 95\% confidence interval) for different algorithms depending on the number of sweeps per  simulation. We have adjusted the  vertical axis to accommodate all three algorithms: global, T-local and ST-local measurements. The top (bottom) plot shows the results for the sublattice  charge (spin) fluctuations. The data set was produced on a $6\times6\times256$ hexagonal lattice at $U=5.0$ and  $\beta=20.0$. Here, the  staggered mass is equal to zero. Insets in both plots show the same data, but the vertical scale adjusted to included only the T-local and ST-local algorithms. }
    \label{fig:err_comparison_global_6x6x256}
\end{figure}

  \begin{figure}[]
   \centering
 \includegraphics[width=0.3\textwidth , angle=-90]{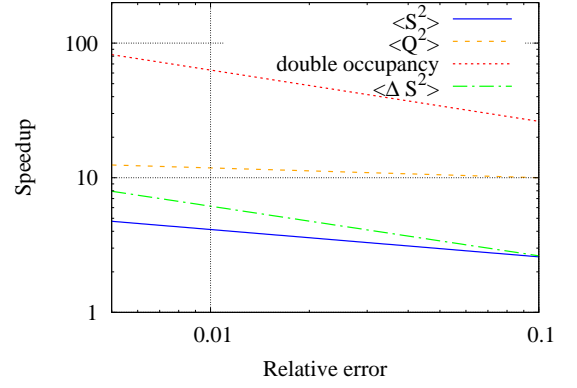}
   
      \caption{Conservative estimate of the speedup: we take the maximal value within the error bars for the power law coefficient $\mu$ in case of the T-local measurements and the minimal value of $\mu$ for the ST-local measurements. The speedup is estimated at fixed relative error of  the corresponding observable. Data for the power law fits are taken from Fig.~\ref{fig:err_comparison_global_6x6x256} and \ref{fig:err_comparison_local_6x6x256},  such that  the speedup pertains to  simulations on the $6\times6\times256$ hexagonal lattice at  $U=5.0$ and  $\beta=20.0$. Here, the staggered mass is equal to zero.}
   \label{fig:average_speedup}
\end{figure}

We now   compare   \textit{error bars}  for  different measurement schemes. Since the second moment  of the  distribution is  not  defined  for  tails  with $\xi\approx5/2$, the central limit theorem is not applicable and we need  to  resort to more elaborate error estimation algorithms. For   a  given  simulation consisting  of  N sweeps  with  N  exceeding  the  auto-correlation  time,   we repeat  it for up to several thousand times and study the properties of distributions for the averages obtained during the individual simulations. The details of this   analysis  are described in  Appendix ~\ref{AppendixA}. Here we present the results for the 95\% confidence interval for the averages depending upon the number of sweeps per  simulation.  Since  our  different  schemes only  reshuffle measurements,    they  come  at the same  computational cost.   Hence the  comparison between measurements schemes  is  carried out at   equal  computational  time.

  \begin{figure}[]
     \centering
\subfigure[]{\label{fig:hist_global_discrete}\includegraphics[width=0.32\textwidth , angle=-90]{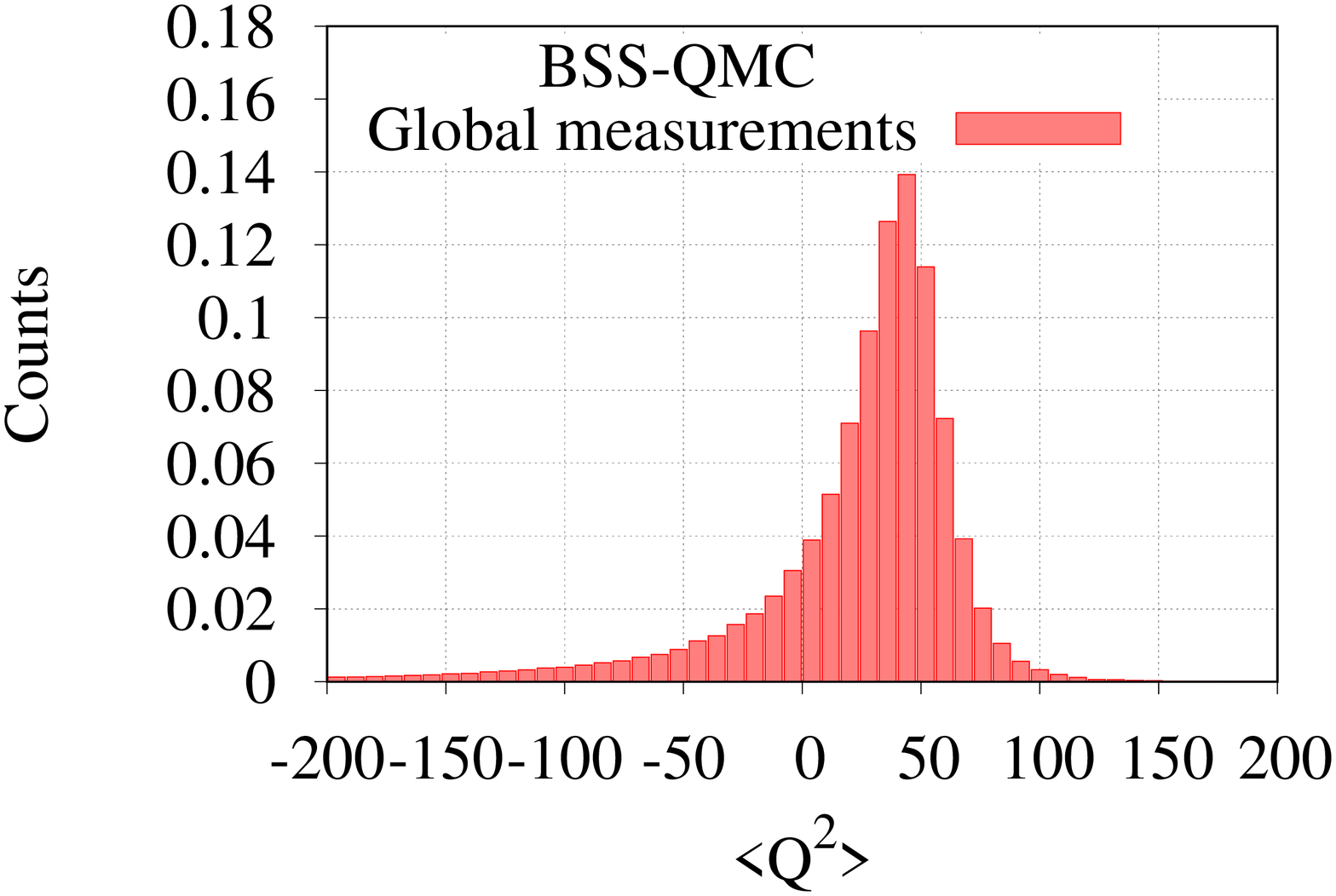}}
  \subfigure[]{\label{fig:hist_global_continuous}\includegraphics[width=0.32\textwidth , angle=-90]{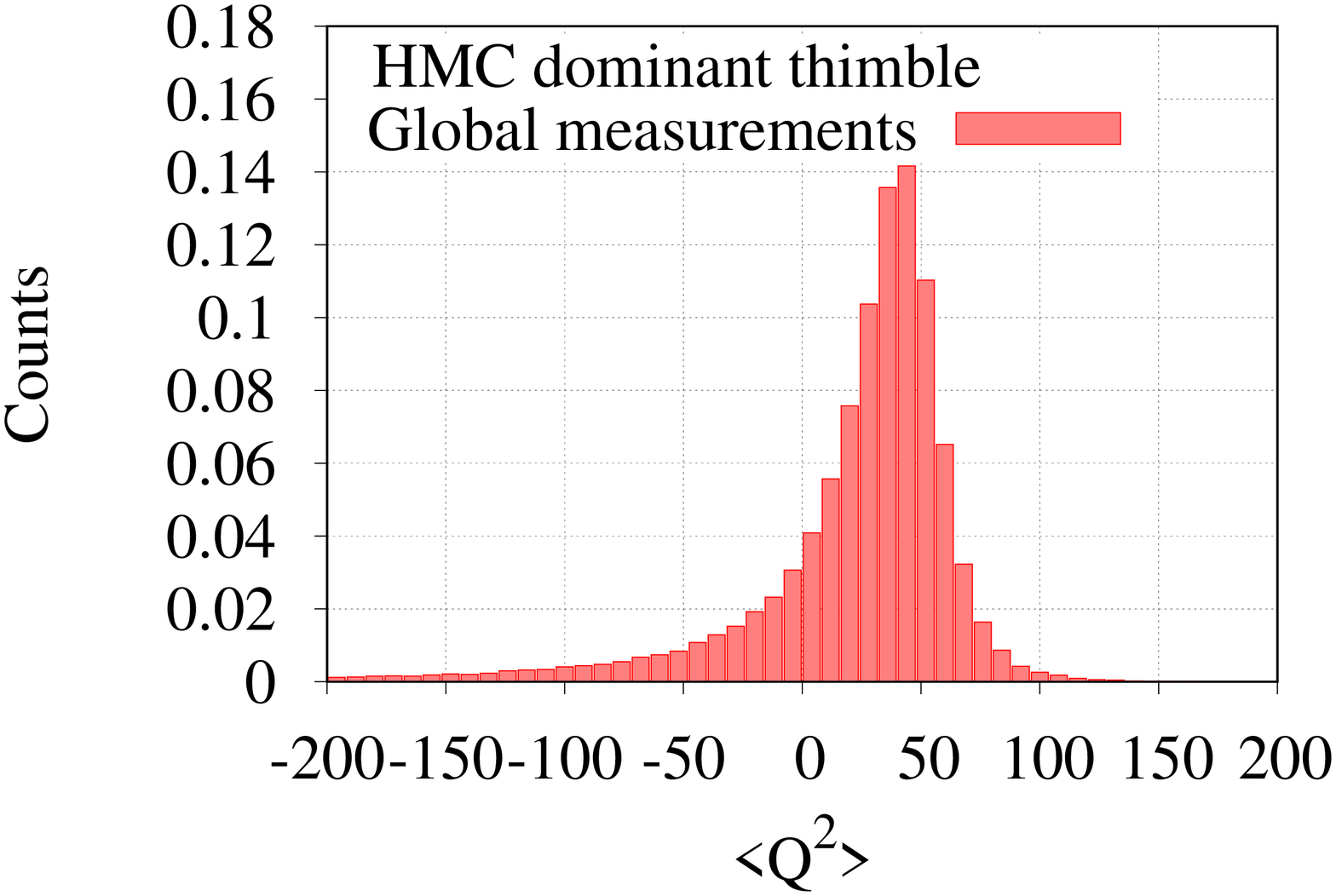}}

      \caption{Comparison of the histograms for the  sublattice  charge fluctuations obtained in QMC calculations with discrete and continuous auxiliary fields. (a) SU(2) symmetrical discrete fields updated using standard local BSS QMC updates. (b) SU(2) symmetrical continuous fields, updated with the Hybrid Monte Carlo algorithm constrained in the dominant thimble \cite{beyond_instanton_gas}. In both cases we employ global measurements: observables are computed on the whole lattice after the update of all fields.  Computations were done on  a $6\times6\times256$   at  $\beta=20$ and $U=5$.}
   \label{fig:discr_cont_comparison}
\end{figure}

Curves for all three measurement schemes are shown in Fig. \ref{fig:err_comparison_global_6x6x256} for  the   sublattice  spin and  charge  fluctuations computed on the $6\times6\times256$ hexagonal lattice. 
In general, the error bars reflect  the  relation between the widths of the histograms for different measurement  schemes. 
The global scheme  stands out  and  is always at  a big disadvantage:  the error bar is at least an order of magnitude larger in comparison to the ST-local scheme. The ST-local scheme also offers an  improvement  in the error bars  by a factor  $2 \cdots 3 $  in comparison with T-local scheme. We  foresee that  this  improvement will   be more pronounced on  larger lattices.  Fig.~\ref{fig:err_comparison_global_6x6x256} shows the results for the  relatively small $6\times6\times256$ lattice,  and plots for larger lattices can be found in Appendix~\ref{AppendixA}. 

The scaling of the errors with the number of sweeps per simulation  is also different for different schemes. Insets in the figures 
\ref{fig:err_comparison_charge_global} and \ref{fig:err_comparison_spin_global} show the power law fit for this scaling, where the error bar is proportional to $1/N_{\text{sweeps}}^{\mu}$.  At the onset, we  notice 
that  the errors  decrease slower than   $N_{\text{sweeps}}^{-1/2}$,  thereby reflecting the  invalidity of the central limit theorem.  The power is consistently higher for the ST-local scheme. Thus the error bar decreases faster for ST-local measurements and the  speedup    increases   with  decreasing   tolerance
on the size of the error  bar.  

Using these fits we can estimate the speedup for a given relative error for various observables. Conservative estimates, based on the fastest (within the error bars) decrease of confidence interval for T-local scheme and the slowest decrease of the  confidence interval for the ST-local scheme are shown in Fig. \ref{fig:average_speedup}. As expected, the largest effect is observed for  the local quantity, the double occupancy. In this case the speedup can 
reach  up to two orders of magnitude if we demand  the relative error to be less than 1\%. However, even for other observables which contain some spatial correlators, the speedup is quite substantial and varies from 5 to 10.

The effectiveness of the ST-local scheme for the double occupancy can be  qualitatively understood by looking into the details of local updates in  BSS-QMC. For the specific  choice  of the  HS  transformation 
\ref{eq:discrete_HS},  and at  the considered  particle-hole  symmetric    case,  the   ratio of  new  to  
old  weight upon  changing  the auxiliary field at  space-time site  $\ve{i},\tau$ from $s_{\ve{i},\tau}$ to $\tilde s_{\ve{i},\tau}$  reads:
\begin{equation}
	R_{\ve{i},\tau}    =  \frac{\gamma(\tilde s_{\ve{i},\tau})}{\gamma(s_{\ve{i},\tau})}   \left|   (1 - E_{\ve{i},\ve{i}} \left( \tau \right)) G_{\ve{i},\ve{i}} \left( \tau \right) +  E_{\ve{i},\ve{i}} \left( \tau \right)\right|^2,
	\label{eq:discrete_HS_update}
\end{equation}
where 
\begin{equation}
	E_{\ve{i},\ve{i}} \left( \tau \right) = e^{i \lambda \eta(\tilde s_{\ve{i},\tau}) \tilde s_{\ve{i},\tau} - i \lambda \eta(s_{\ve{i},\tau}) s_{\ve{i},\tau} }.
	\label{eq:discrete_HS_update1}
\end{equation}
As  a   consequence,  if the Green   function takes a large  value,   due to proximity of a  zero of  the determinant,  the flip will be accepted,  and  the  configuration 
will  be   \textit{ repelled} from   the  zero  mode.     Observables  such as  the double occupancy that  rely only on the  knowledge of 
$ G_{\ve{i},\ve{i}} \left( \tau \right)  $,   after  the  update has  been carried out,  are hence expected to  show  less fluctuations. 
Generally speaking, the performance of the ST-local synchronization scheme lies in the effectiveness of the local updates in driving the field configurations away of zeros of determinant, and in the fact that the local observable $O_{\ve{i},\tau}(C)$ is mostly dependent on the local fields $s_{\ve{i},\tau}$.

Finally,    we  address  the  question  of the  dependence of  our  findings on the choice of the HS  transformation:  continuous, 
Eq.~\ref{eq:continuous_HS_imag},   or discrete, Eq.~\ref{eq:discrete_HS}.   
Since  the  step from  continuous   to  discrete  is    provided  by the Gauss-Hermite quadrature \cite{goth2020higher},   we  can conjecture  that  in 
the  small  imaginary   time step limit,  $\Delta \tau   \rightarrow  0 $,   both  discrete  and   continuous  HS   transformation  will yield   identical 
fat tailed  distributions.   We  however  carry  out  our  simulations at  finite  imaginary  time step such that  further  comparison is  required  between   simulations  based on continuous fields, such  as  Hybrid Monte Carlo (HMC),   and  those based on discrete ones. 
 It is well-known  \cite{Assaad_complex,PhysRevD.101.014508} that  HMC  is not ergodic for the SU(2)-invariant continuous HS transformation for 
 the Hubbard model due to above mentioned domain walls formed by zeros of  the determinant. However,  it  turns out  that an  extremely good  approximation is  obtained when the  simulation  is  started  at  the  dominant  saddle point.  We  refer the  interested  reader to  Refs.~\cite{beyond_instanton_gas,PhysRevD.101.014508}  for a full  account of this 
so  called  \textit{dominant thimble} approximation.    
 In Fig.~\ref{fig:discr_cont_comparison}  we  compare  the  distributions  obtained  from the    HMC  within the \textit{dominant thimble}    approximation and 
 for the   discrete  fields.    
 The distributions are almost identical, thus providing evidence  that the impact of   the  zeros  of the  determinant  on the distribution is  independent  of  the choice of the HS  transformation.

\section*{\label{sec:Conclusion}Conclusion}
In  determinantal    quantum Monte  Carlo methods, the  zeros of the  fermion  determinant    generically lead  to fat tailed    
probability  distributions of  observables.    As a  consequence,  second  or higher  order  moments of  the observable  may  diverge.  This has  the  
consequence    that  the  probability  distribution   acquires    subtle  dependencies  on   how  measurement are   synchronized  with  the   
updating scheme.    Since  the first moment  of the observable  is  defined,    various    synchronization schemes yield  identical  results for the average value of observable. 
Our  approach  becomes  particularly  appealing    when   the  second  moment  is ill  defined.   In this  case, the  central limit  theorem   theorem  does  not apply and the alternative notions    such  as  the    width  of  the  95\%  confidence interval   replaces    generic  error  bars.    
 Our  main result  is to  show  that  the    width  of  the  95\%  confidence interval strongly  depends on  the  synchronization of measurements 
and  updating  scheme in the case when the fat-tail distributions lead to ill-defined second moment of the distribution of observable.
In particular,  our  strategy is to  measure all observables containing the fermionic propagators with source or sink at site $(\ve{i},\tau)$  
\textit{immediately}  after the  local update of the  auxiliary field on this site. 
  This   reshuffling of the measurements    suppresses  spikes in  observables. As  a  consequence,  the statistical uncertainly  is 
    reduced   \textit{ without }  additional  computational effort.  
  Impressive  results are  obtained  for  local observables, and  a  moderate speedup is  seen  for   equal time   two-point  correlations  functions  such   spin
  fluctuations.  It  would  certainly  be  interesting to investigate  this strategy  for time displaced correlation functions. However, this  requires  substantial   reorganization of the  existing  QMC code and is left for  future  work. 

As we can see from our toy model, the dependence of the error bars on the synchronization between updates and measurements in the presence of zeros of  the fermion determinant is a  general phenomena   in  fermion Monte Carlo. It hinges on the  fact  that in these  methods  the determinant of the inverse of the  single particle   Green  function matrix  corresponds  to the   weight of a configuration. The   efficiency of the algorithm,  in the case of local updates synchronised with local measurements  depends on the effectiveness of local updates to drive the field configuration away from zeros of the  determinant. It is hard to predict whether the local updates will  or  will not have this  feature for  other models.   
But  since  the price tag for the implementation of the ST-local scheme is zero  it is  worth  implementing.   We also note that in the absence of  
zeros, as obtained  by including a non-zero mass,   reshuffling  of  measurements   does not alter  the  statistics.  Hence  our  reshuffling scheme  can  
only lead  to  improvements.  Let us also note that we  have checked on  an 8-site $2\times2$ hexagonal lattice that all three measurement schemes give identical  results  and compare  well  to  exact diagonalization.   We are aware of no arguments against the implementation of the proposed synchronization between updates and measurements in existing QMC codes.

Comparison of the distributions of observables for continuous and discrete fields (Fig. \ref{fig:discr_cont_comparison}) also gives  new insight on why simulations with continuous fields suffer worse statistical fluctuations. The observed improvement in spikes after transfer to discrete fields \cite{Assaad_complex}  
 should not  be attributed to the nature of discrete fields themselves, but to the switch from  a global to a T-local measurement scheme.  
This observation will  have consequences for the algorithms  where the global measurement scheme is the  default  such as in  HMC and Langevin  dynamics.  Our results show that  for  such  updating  schemes, it is beneficial to supplement  the  e.g. global HMC update with subsequent series of local updates synchronized with the measurement of observables. This idea   should   however  be  subject of further investigations, since it is unclear, whether it is  possible  or not to  keep the scaling of  the HMC with pseudofermions during these additional local updates.  

\begin{acknowledgments}

Computational resources were provided by the Gauss Centre for Supercomputing e.V. (www.gauss-centre.eu) through the 
John von Neumann Institute for Computing (NIC)
on the GCS Supercomputer JUWELS~\cite{JUWELS} at J\"ulich Supercomputing Centre (JSC).  
MU  thanks  the  DFG   for partial  financial support  under the  project  AS120/14-1.  FFA   acknowledges financial support from the DFG through the W\"urzburg-Dresden Cluster of Excellence on Complexity and Topology in Quantum Matter - ct.qmat (EXC 2147, project-id 39085490).

\end{acknowledgments}

\bibliography{general_bib, new_bib}

\begin{thebibliography}{22}%
\makeatletter
\providecommand \@ifxundefined [1]{%
 \@ifx{#1\undefined}
}%
\providecommand \@ifnum [1]{%
 \ifnum #1\expandafter \@firstoftwo
 \else \expandafter \@secondoftwo
 \fi
}%
\providecommand \@ifx [1]{%
 \ifx #1\expandafter \@firstoftwo
 \else \expandafter \@secondoftwo
 \fi
}%
\providecommand \natexlab [1]{#1}%
\providecommand \enquote  [1]{``#1''}%
\providecommand \bibnamefont  [1]{#1}%
\providecommand \bibfnamefont [1]{#1}%
\providecommand \citenamefont [1]{#1}%
\providecommand \href@noop [0]{\@secondoftwo}%
\providecommand \href [0]{\begingroup \@sanitize@url \@href}%
\providecommand \@href[1]{\@@startlink{#1}\@@href}%
\providecommand \@@href[1]{\endgroup#1\@@endlink}%
\providecommand \@sanitize@url [0]{\catcode `\\12\catcode `\$12\catcode
  `\&12\catcode `\#12\catcode `\^12\catcode `\_12\catcode `\%12\relax}%
\providecommand \@@startlink[1]{}%
\providecommand \@@endlink[0]{}%
\providecommand \url  [0]{\begingroup\@sanitize@url \@url }%
\providecommand \@url [1]{\endgroup\@href {#1}{\urlprefix }}%
\providecommand \urlprefix  [0]{URL }%
\providecommand \Eprint [0]{\href }%
\providecommand \doibase [0]{http://dx.doi.org/}%
\providecommand \selectlanguage [0]{\@gobble}%
\providecommand \bibinfo  [0]{\@secondoftwo}%
\providecommand \bibfield  [0]{\@secondoftwo}%
\providecommand \translation [1]{[#1]}%
\providecommand \BibitemOpen [0]{}%
\providecommand \bibitemStop [0]{}%
\providecommand \bibitemNoStop [0]{.\EOS\space}%
\providecommand \EOS [0]{\spacefactor3000\relax}%
\providecommand \BibitemShut  [1]{\csname bibitem#1\endcsname}%
\let\auto@bib@innerbib\@empty
\bibitem [{\citenamefont {Blankenbecler}\ \emph {et~al.}(1981)\citenamefont
  {Blankenbecler}, \citenamefont {Scalapino},\ and\ \citenamefont
  {Sugar}}]{Blankenbecler:1981jt}%
  \BibitemOpen
  \bibfield  {author} {\bibinfo {author} {\bibfnamefont {R.}~\bibnamefont
  {Blankenbecler}}, \bibinfo {author} {\bibfnamefont {D.~J.}\ \bibnamefont
  {Scalapino}}, \ and\ \bibinfo {author} {\bibfnamefont {R.~L.}\ \bibnamefont
  {Sugar}},\ }\href {\doibase 10.1103/PhysRevD.24.2278} {\bibfield  {journal}
  {\bibinfo  {journal} {Phys. Rev.}\ }\textbf {\bibinfo {volume} {D24}},\
  \bibinfo {pages} {2278} (\bibinfo {year} {1981})}\BibitemShut {NoStop}%
\bibitem [{\citenamefont {Sugiyama}\ and\ \citenamefont
  {Koonin}(1986)}]{Sugiyama86}%
  \BibitemOpen
  \bibfield  {author} {\bibinfo {author} {\bibfnamefont {G.}~\bibnamefont
  {Sugiyama}}\ and\ \bibinfo {author} {\bibfnamefont {S.}~\bibnamefont
  {Koonin}},\ }\href {\doibase http://dx.doi.org/10.1016/0003-4916(86)90107-7}
  {\bibfield  {journal} {\bibinfo  {journal} {Annals of Physics}\ }\textbf
  {\bibinfo {volume} {168}},\ \bibinfo {pages} {1 } (\bibinfo {year}
  {1986})}\BibitemShut {NoStop}%
\bibitem [{\citenamefont {Duane}\ \emph {et~al.}(1987)\citenamefont {Duane},
  \citenamefont {Kennedy}, \citenamefont {Pendleton},\ and\ \citenamefont
  {Roweth}}]{Duane87}%
  \BibitemOpen
  \bibfield  {author} {\bibinfo {author} {\bibfnamefont {S.}~\bibnamefont
  {Duane}}, \bibinfo {author} {\bibfnamefont {A.~D.}\ \bibnamefont {Kennedy}},
  \bibinfo {author} {\bibfnamefont {B.~J.}\ \bibnamefont {Pendleton}}, \ and\
  \bibinfo {author} {\bibfnamefont {D.}~\bibnamefont {Roweth}},\ }\href
  {\doibase 10.1016/0370-2693(87)91197-X} {\bibfield  {journal} {\bibinfo
  {journal} {Phys. Lett.}\ }\textbf {\bibinfo {volume} {B195}},\ \bibinfo
  {pages} {216} (\bibinfo {year} {1987})}\BibitemShut {NoStop}%
\bibitem [{\citenamefont {White}\ \emph {et~al.}(1989)\citenamefont {White},
  \citenamefont {Scalapino}, \citenamefont {Sugar}, \citenamefont {Loh},
  \citenamefont {Gubernatis},\ and\ \citenamefont {Scalettar}}]{White89}%
  \BibitemOpen
  \bibfield  {author} {\bibinfo {author} {\bibfnamefont {S.}~\bibnamefont
  {White}}, \bibinfo {author} {\bibfnamefont {D.}~\bibnamefont {Scalapino}},
  \bibinfo {author} {\bibfnamefont {R.}~\bibnamefont {Sugar}}, \bibinfo
  {author} {\bibfnamefont {E.}~\bibnamefont {Loh}}, \bibinfo {author}
  {\bibfnamefont {J.}~\bibnamefont {Gubernatis}}, \ and\ \bibinfo {author}
  {\bibfnamefont {R.}~\bibnamefont {Scalettar}},\ }\href {\doibase
  10.1103/PhysRevB.40.506} {\bibfield  {journal} {\bibinfo  {journal} {Phys.
  Rev. B}\ }\textbf {\bibinfo {volume} {40}},\ \bibinfo {pages} {506} (\bibinfo
  {year} {1989})}\BibitemShut {NoStop}%
\bibitem [{\citenamefont {Sorella}\ \emph {et~al.}(1989)\citenamefont
  {Sorella}, \citenamefont {Baroni}, \citenamefont {Car},\ and\ \citenamefont
  {Parrinello}}]{Sorella89}%
  \BibitemOpen
  \bibfield  {author} {\bibinfo {author} {\bibfnamefont {S.}~\bibnamefont
  {Sorella}}, \bibinfo {author} {\bibfnamefont {S.}~\bibnamefont {Baroni}},
  \bibinfo {author} {\bibfnamefont {R.}~\bibnamefont {Car}}, \ and\ \bibinfo
  {author} {\bibfnamefont {M.}~\bibnamefont {Parrinello}},\ }\href {\doibase
  10.1209/0295-5075/8/7/014} {\bibfield  {journal} {\bibinfo  {journal}
  {Europhysics Letters ({EPL})}\ }\textbf {\bibinfo {volume} {8}},\ \bibinfo
  {pages} {663} (\bibinfo {year} {1989})}\BibitemShut {NoStop}%
\bibitem [{\citenamefont {Assaad}\ and\ \citenamefont
  {Evertz}(2008)}]{Assaad08_rev}%
  \BibitemOpen
  \bibfield  {author} {\bibinfo {author} {\bibfnamefont {F.}~\bibnamefont
  {Assaad}}\ and\ \bibinfo {author} {\bibfnamefont {H.}~\bibnamefont
  {Evertz}},\ }in\ \href {\doibase 10.1007/978-3-540-74686-7_10} {\emph
  {\bibinfo {booktitle} {Computational Many-Particle Physics}}},\ \bibinfo
  {series} {Lecture Notes in Physics}, Vol.\ \bibinfo {volume} {739},\ \bibinfo
  {editor} {edited by\ \bibinfo {editor} {\bibfnamefont {H.}~\bibnamefont
  {Fehske}}, \bibinfo {editor} {\bibfnamefont {R.}~\bibnamefont {Schneider}}, \
  and\ \bibinfo {editor} {\bibfnamefont {A.}~\bibnamefont {Wei{\ss}e}}}\
  (\bibinfo  {publisher} {Springer},\ \bibinfo {address} {Berlin Heidelberg},\
  \bibinfo {year} {2008})\ pp.\ \bibinfo {pages} {277--356}\BibitemShut
  {NoStop}%
\bibitem [{\citenamefont {Shi}\ and\ \citenamefont {Zhang}(2016)}]{Hao16}%
  \BibitemOpen
  \bibfield  {author} {\bibinfo {author} {\bibfnamefont {H.}~\bibnamefont
  {Shi}}\ and\ \bibinfo {author} {\bibfnamefont {S.}~\bibnamefont {Zhang}},\
  }\href {\doibase 10.1103/PhysRevE.93.033303} {\bibfield  {journal} {\bibinfo
  {journal} {Phys. Rev. E}\ }\textbf {\bibinfo {volume} {93}},\ \bibinfo
  {pages} {033303} (\bibinfo {year} {2016})}\BibitemShut {NoStop}%
\bibitem [{\citenamefont {{ALF Collaboration}}\ \emph
  {et~al.}(2021)\citenamefont {{ALF Collaboration}}, \citenamefont {Assaad},
  \citenamefont {Bercx}, \citenamefont {Goth}, \citenamefont {G{\"o}tz},
  \citenamefont {Hofmann}, \citenamefont {Huffman}, \citenamefont {Liu},
  \citenamefont {{Parisen Toldin}}, \citenamefont {Portela},\ and\
  \citenamefont {Schwab}}]{ALF_v2}%
  \BibitemOpen
  \bibfield  {author} {\bibinfo {author} {\bibnamefont {{ALF Collaboration}}},
  \bibinfo {author} {\bibfnamefont {F.~F.}\ \bibnamefont {Assaad}}, \bibinfo
  {author} {\bibfnamefont {M.}~\bibnamefont {Bercx}}, \bibinfo {author}
  {\bibfnamefont {F.}~\bibnamefont {Goth}}, \bibinfo {author} {\bibfnamefont
  {A.}~\bibnamefont {G{\"o}tz}}, \bibinfo {author} {\bibfnamefont {J.~S.}\
  \bibnamefont {Hofmann}}, \bibinfo {author} {\bibfnamefont {E.}~\bibnamefont
  {Huffman}}, \bibinfo {author} {\bibfnamefont {Z.}~\bibnamefont {Liu}},
  \bibinfo {author} {\bibfnamefont {F.}~\bibnamefont {{Parisen Toldin}}},
  \bibinfo {author} {\bibfnamefont {J.~S.~E.}\ \bibnamefont {Portela}}, \ and\
  \bibinfo {author} {\bibfnamefont {J.}~\bibnamefont {Schwab}},\ }\href@noop {}
  {\bibfield  {journal} {\bibinfo  {journal} {arXiv:2012.11914}\ } (\bibinfo
  {year} {2021})},\ \Eprint
  {http://arxiv.org/abs/https://arxiv.org/abs/2012.11914}
  {https://arxiv.org/abs/2012.11914 [cond-mat.str-el]} \BibitemShut {NoStop}%
\bibitem [{\citenamefont {Batrouni}\ \emph {et~al.}(1985)\citenamefont
  {Batrouni}, \citenamefont {Katz}, \citenamefont {Kronfeld}, \citenamefont
  {Lepage}, \citenamefont {Svetitsky},\ and\ \citenamefont
  {Wilson}}]{Batrouni85}%
  \BibitemOpen
  \bibfield  {author} {\bibinfo {author} {\bibfnamefont {G.~G.}\ \bibnamefont
  {Batrouni}}, \bibinfo {author} {\bibfnamefont {G.~R.}\ \bibnamefont {Katz}},
  \bibinfo {author} {\bibfnamefont {A.~S.}\ \bibnamefont {Kronfeld}}, \bibinfo
  {author} {\bibfnamefont {G.~P.}\ \bibnamefont {Lepage}}, \bibinfo {author}
  {\bibfnamefont {B.}~\bibnamefont {Svetitsky}}, \ and\ \bibinfo {author}
  {\bibfnamefont {K.~G.}\ \bibnamefont {Wilson}},\ }\href {\doibase
  10.1103/PhysRevD.32.2736} {\bibfield  {journal} {\bibinfo  {journal} {Phys.
  Rev. D}\ }\textbf {\bibinfo {volume} {32}},\ \bibinfo {pages} {2736}
  (\bibinfo {year} {1985})}\BibitemShut {NoStop}%
\bibitem [{\citenamefont {Goetz}\ \emph {et~al.}(2021)\citenamefont {Goetz},
  \citenamefont {Beyl}, \citenamefont {Hohenadler},\ and\ \citenamefont
  {Assaad}}]{Goetz21}%
  \BibitemOpen
  \bibfield  {author} {\bibinfo {author} {\bibfnamefont {A.}~\bibnamefont
  {Goetz}}, \bibinfo {author} {\bibfnamefont {S.}~\bibnamefont {Beyl}},
  \bibinfo {author} {\bibfnamefont {M.}~\bibnamefont {Hohenadler}}, \ and\
  \bibinfo {author} {\bibfnamefont {F.~F.}\ \bibnamefont {Assaad}},\ }\href
  {https://arxiv.org/abs/2102.08899} {\  (\bibinfo {year} {2021})},\ \Eprint
  {http://arxiv.org/abs/2102.08899} {arXiv:2102.08899 [cond-mat.str-el]}
  \BibitemShut {NoStop}%
\bibitem [{\citenamefont {Beyl}\ \emph {et~al.}(2018)\citenamefont {Beyl},
  \citenamefont {Goth},\ and\ \citenamefont {Assaad}}]{Assaad_complex}%
  \BibitemOpen
  \bibfield  {author} {\bibinfo {author} {\bibfnamefont {S.}~\bibnamefont
  {Beyl}}, \bibinfo {author} {\bibfnamefont {F.}~\bibnamefont {Goth}}, \ and\
  \bibinfo {author} {\bibfnamefont {F.}~\bibnamefont {Assaad}},\ }\href@noop {}
  {\bibfield  {journal} {\bibinfo  {journal} {Phys. Rev.}\ }\textbf {\bibinfo
  {volume} {B97}},\ \bibinfo {pages} {085144} (\bibinfo {year} {2018})},\
  \Eprint {http://arxiv.org/abs/1708.03661} {arXiv:1708.03661
  [cond-mat.str-el]} \BibitemShut {NoStop}%
\bibitem [{\citenamefont {Ulybyshev}\ \emph {et~al.}(2020)\citenamefont
  {Ulybyshev}, \citenamefont {Winterowd},\ and\ \citenamefont
  {Zafeiropoulos}}]{PhysRevD.101.014508}%
  \BibitemOpen
  \bibfield  {author} {\bibinfo {author} {\bibfnamefont {M.}~\bibnamefont
  {Ulybyshev}}, \bibinfo {author} {\bibfnamefont {C.}~\bibnamefont
  {Winterowd}}, \ and\ \bibinfo {author} {\bibfnamefont {S.}~\bibnamefont
  {Zafeiropoulos}},\ }\href {\doibase 10.1103/PhysRevD.101.014508} {\bibfield
  {journal} {\bibinfo  {journal} {Phys. Rev. D}\ }\textbf {\bibinfo {volume}
  {101}},\ \bibinfo {pages} {014508} (\bibinfo {year} {2020})}\BibitemShut
  {NoStop}%
\bibitem [{\citenamefont {Bercx}\ \emph {et~al.}(2017)\citenamefont {Bercx},
  \citenamefont {Goth}, \citenamefont {Hofmann},\ and\ \citenamefont
  {Assaad}}]{ALF2017}%
  \BibitemOpen
  \bibfield  {author} {\bibinfo {author} {\bibfnamefont {M.}~\bibnamefont
  {Bercx}}, \bibinfo {author} {\bibfnamefont {F.}~\bibnamefont {Goth}},
  \bibinfo {author} {\bibfnamefont {J.~S.}\ \bibnamefont {Hofmann}}, \ and\
  \bibinfo {author} {\bibfnamefont {F.~F.}\ \bibnamefont {Assaad}},\ }\href
  {\doibase 10.21468/SciPostPhys.3.2.013} {\bibfield  {journal} {\bibinfo
  {journal} {SciPost Phys.}\ }\textbf {\bibinfo {volume} {3}},\ \bibinfo
  {pages} {013} (\bibinfo {year} {2017})},\ \Eprint
  {http://arxiv.org/abs/1704.00131} {arXiv:1704.00131} \BibitemShut {NoStop}%
\bibitem [{\citenamefont {Sokal}(1997)}]{Sokal97}%
  \BibitemOpen
  \bibfield  {author} {\bibinfo {author} {\bibfnamefont {A.}~\bibnamefont
  {Sokal}},\ }\enquote {\bibinfo {title} {Monte carlo methods in statistical
  mechanics: Foundations and new algorithms},}\ in\ \href {\doibase
  10.1007/978-1-4899-0319-8_6} {\emph {\bibinfo {booktitle} {Functional
  Integration: Basics and Applications}}},\ \bibinfo {editor} {edited by\
  \bibinfo {editor} {\bibfnamefont {C.}~\bibnamefont {DeWitt-Morette}},
  \bibinfo {editor} {\bibfnamefont {P.}~\bibnamefont {Cartier}}, \ and\
  \bibinfo {editor} {\bibfnamefont {A.}~\bibnamefont {Folacci}}}\ (\bibinfo
  {publisher} {Springer US},\ \bibinfo {address} {Boston, MA},\ \bibinfo {year}
  {1997})\ pp.\ \bibinfo {pages} {131--192}\BibitemShut {NoStop}%
\bibitem [{\citenamefont {Buividovich}\ \emph {et~al.}(2019)\citenamefont
  {Buividovich}, \citenamefont {Smith}, \citenamefont {Ulybyshev},\ and\
  \citenamefont {von Smekal}}]{Conformal_PhysRevB.99.205434}%
  \BibitemOpen
  \bibfield  {author} {\bibinfo {author} {\bibfnamefont {P.}~\bibnamefont
  {Buividovich}}, \bibinfo {author} {\bibfnamefont {D.}~\bibnamefont {Smith}},
  \bibinfo {author} {\bibfnamefont {M.}~\bibnamefont {Ulybyshev}}, \ and\
  \bibinfo {author} {\bibfnamefont {L.}~\bibnamefont {von Smekal}},\ }\href
  {\doibase 10.1103/PhysRevB.99.205434} {\bibfield  {journal} {\bibinfo
  {journal} {Phys. Rev.}\ }\textbf {\bibinfo {volume} {B99}},\ \bibinfo {pages}
  {205434} (\bibinfo {year} {2019})},\ \Eprint
  {http://arxiv.org/abs/1812.06435} {arXiv:1812.06435 [cond-mat.str-el]}
  \BibitemShut {NoStop}%
\bibitem [{\citenamefont {Buividovich}\ \emph {et~al.}(2018)\citenamefont
  {Buividovich}, \citenamefont {Smith}, \citenamefont {Ulybyshev},\ and\
  \citenamefont {von Smekal}}]{Buividovich:2018yar}%
  \BibitemOpen
  \bibfield  {author} {\bibinfo {author} {\bibfnamefont {P.}~\bibnamefont
  {Buividovich}}, \bibinfo {author} {\bibfnamefont {D.}~\bibnamefont {Smith}},
  \bibinfo {author} {\bibfnamefont {M.}~\bibnamefont {Ulybyshev}}, \ and\
  \bibinfo {author} {\bibfnamefont {L.}~\bibnamefont {von Smekal}},\ }\href
  {\doibase 10.1103/PhysRevB.98.235129} {\bibfield  {journal} {\bibinfo
  {journal} {Phys. Rev.}\ }\textbf {\bibinfo {volume} {B98}},\ \bibinfo {pages}
  {235129} (\bibinfo {year} {2018})},\ \Eprint
  {http://arxiv.org/abs/1807.07025} {arXiv:1807.07025 [cond-mat.str-el]}
  \BibitemShut {NoStop}%
\bibitem [{\citenamefont {Hirsch}(1983)}]{Hirsch83}%
  \BibitemOpen
  \bibfield  {author} {\bibinfo {author} {\bibfnamefont {J.}~\bibnamefont
  {Hirsch}},\ }\href {\doibase 10.1103/PhysRevB.28.4059} {\bibfield  {journal}
  {\bibinfo  {journal} {Phys. Rev. B}\ }\textbf {\bibinfo {volume} {28}},\
  \bibinfo {pages} {4059} (\bibinfo {year} {1983})}\BibitemShut {NoStop}%
\bibitem [{\citenamefont {Goth}(2020)}]{goth2020higher}%
  \BibitemOpen
  \bibfield  {author} {\bibinfo {author} {\bibfnamefont {F.}~\bibnamefont
  {Goth}},\ }\href@noop {} {\enquote {\bibinfo {title} {Higher order auxiliary
  field quantum monte carlo methods},}\ } (\bibinfo {year} {2020}),\ \Eprint
  {http://arxiv.org/abs/2009.04491} {arXiv:2009.04491 [cond-mat.str-el]}
  \BibitemShut {NoStop}%
\bibitem [{\citenamefont {Ulybyshev}\ and\ \citenamefont
  {Valgushev}(2017)}]{Ulybyshev:2017}%
  \BibitemOpen
  \bibfield  {author} {\bibinfo {author} {\bibfnamefont {M.~V.}\ \bibnamefont
  {Ulybyshev}}\ and\ \bibinfo {author} {\bibfnamefont {S.~N.}\ \bibnamefont
  {Valgushev}},\ }\href@noop {} {\  (\bibinfo {year} {2017})},\ \Eprint
  {http://arxiv.org/abs/1712.02188} {arXiv:1712.02188 [cond-mat.str-el]}
  \BibitemShut {NoStop}%
\bibitem [{\citenamefont {Buividovich}\ and\ \citenamefont
  {Polikarpov}(2012)}]{Polikarpov1206}%
  \BibitemOpen
  \bibfield  {author} {\bibinfo {author} {\bibfnamefont {P.~V.}\ \bibnamefont
  {Buividovich}}\ and\ \bibinfo {author} {\bibfnamefont {M.~I.}\ \bibnamefont
  {Polikarpov}},\ }\href {\doibase 10.1103/PhysRevB.86.245117} {\bibfield
  {journal} {\bibinfo  {journal} {Phys. Rev. B}\ }\textbf {\bibinfo {volume}
  {86}},\ \bibinfo {pages} {245117} (\bibinfo {year} {2012})}\BibitemShut
  {NoStop}%
\bibitem [{\citenamefont {Ulybyshev}\ and\ \citenamefont
  {Assaad}(2021)}]{beyond_instanton_gas}%
  \BibitemOpen
  \bibfield  {author} {\bibinfo {author} {\bibfnamefont {M.}~\bibnamefont
  {Ulybyshev}}\ and\ \bibinfo {author} {\bibfnamefont {F.}~\bibnamefont
  {Assaad}},\ }\href@noop {} {\bibfield  {journal} {\bibinfo  {journal} {To be
  published}\ } (\bibinfo {year} {2021})}\BibitemShut {NoStop}%
\bibitem [{\citenamefont {{J\"{u}lich Supercomputing Centre}}(2019)}]{JUWELS}%
  \BibitemOpen
  \bibfield  {author} {\bibinfo {author} {\bibnamefont {{J\"{u}lich
  Supercomputing Centre}}},\ }\href {\doibase 10.17815/jlsrf-5-171} {\bibfield
  {journal} {\bibinfo  {journal} {Journal of large-scale research facilities}\
  }\textbf {\bibinfo {volume} {5}} (\bibinfo {year} {2019}),\
  10.17815/jlsrf-5-171}\BibitemShut {NoStop}%
\end{thebibliography}%

\pagebreak
\clearpage
\newpage

\appendix
\counterwithin{figure}{section}

\section{\label{sec:AppendixA}Statistical processing of QMC data}
\label{AppendixA}

\begin{figure}
    \centering
    \subfigure[]{\includegraphics[width=0.32\textwidth,angle=-90]{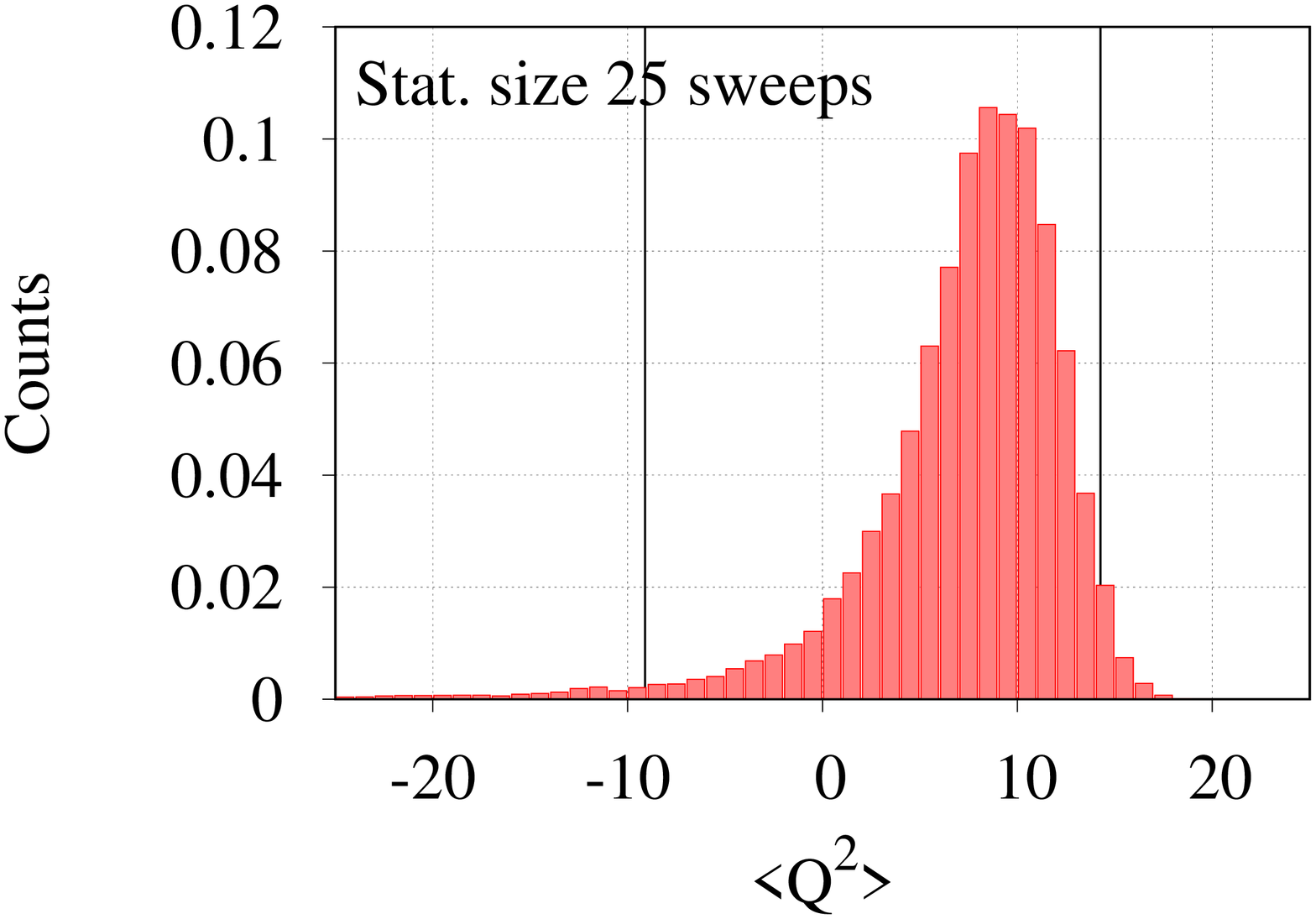}\label{fig:hist_bin25}} 
    \subfigure[]{\includegraphics[width=0.32\textwidth,angle=-90]{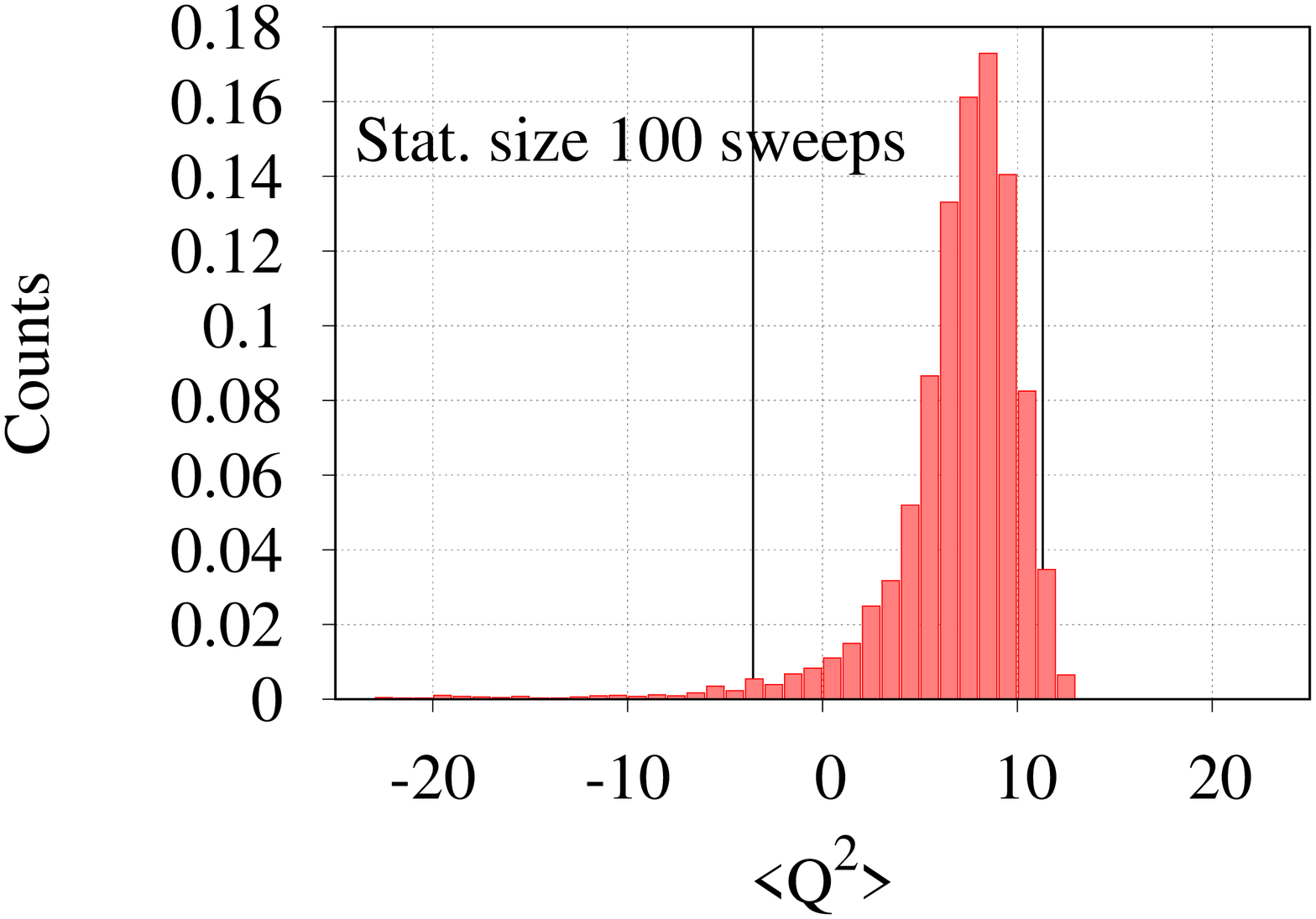}\label{fig:hist_bin100}} 
    \caption{Histograms for the distribution of averages of the sublattice  charge fluctuations  over 25 (figure (a)) and 100 (figure (b)) sweeps.  Vertical lines show the borders of 95\% credible interval obtained from Eq. \ref{eq:z1_def} and \ref{eq:z2_def}. Computations were done on  the $6\times6\times256$ lattice at $\beta=20$ and  $U=5$. T-local measurements were used to compute the observable.}
   \label{fig:histogram_bins}
\end{figure}

In this appendix we describe the statistical analysis of the heavy-tailed Monte Carlo data.  To  obtain  an   error estimation  on observables measured in  simulations with  statistics size of $N$ sweeps,  we  repeat  the  simulation   several times   with  different  random seeds so as  determine  the  distribution
of the  average  observable for  a given simulation.  In the absence of  fat tails,  and  provided that the number of sweeps  exceeds the  autocorrelation  time,    the   distribution will  follow  a  Gauss  law  as  expected  from the central limit  theorem.    
As   we  cannot rely  on  this theorem  some  care  has to be taken  in determining the error. In particular  we  carry  out the following analysis. 
\begin{itemize}
    \item Individual simulations with $N$ sweeps are repeated $M$ times, where $M$ is large enough to study the tails of the averages obtained in individual simulations. 
    \item For any observable $O$, the average over the whole set of  $M$ simulations with $N$ sweeps each is used as an estimate for the "true average" $\langle \langle O \rangle \rangle$.
    \item For each $i-$th simulation, we compute average $\langle O \rangle_i$ and variance for all observables.
    \item Distribution of these individual averages and the variances are used in further error estimation.
\end{itemize}
The following quantities are used to estimate the error bar:
\begin{itemize}
    \item 95\% confidence interval. Its width $\Delta^{(1)}_O$ is defined by the fact that the true average $\langle \langle O \rangle \rangle$ fits inside 95\% of the intervals $\left( \langle O \rangle_i - (1-r)\Delta^{(1)}_O; \, \langle O \rangle_i + r\Delta^{(1)}_O \right)$.  Since the distributions of observables are highly non-symmetrical, we also use non-symmetrical confidence intervals, with the parameter $r\in (0;1)$ defining the deviation of the confidence interval from symmetrical form. The width of the confidence interval $\Delta^{(1)}_O$ is optimized using this non-symmetry: we take the value of $r$, where the width $\Delta^{(1)}_O$ is minimal.
    \item 95\% credible interval. Its width $\Delta^{(2)}_O$ is defined by cutting 2.5\% tails of the probability distributions of averages over  individual simulations  $\langle O \rangle_i$:
    \begin{equation}
    \Delta^{(2)}_O=z_2-z_1,
    \end{equation}
    where
    \begin{eqnarray}
    \int_{z_2}^\infty dz P_N(z) = 0.025, \label{eq:z2_def}  \\ 
    \int_{-\infty}^{z_1} dz P_N(z) = 0.025 \label{eq:z1_def}
    \end{eqnarray}
    and $P_N(z)$ is the probability distribution for individual averages over simulations (each simulation contains  $N$ sweeps).  In practice, these probability distributions are defined numerically using the histograms plotted for the corresponding data sets. It means that $M$ should be large enough to have enough data points in the tails of these histograms. 
    \item Average variance multiplied by the  corresponding Student's t critical value $t_{\infty, 0.025}\approx1.960$: 
    \begin{equation}
    \Delta^{(3)}_O=2 t_{\infty, 0.025} \frac{1}{M} \sum_{i=1}^M \left( \sqrt{ \frac{  \sum_{j=1}^N \left( O_i^j - \langle O \rangle_i \right)  }{N(N-1)}  } \right),    \label{eq:av_variation}
    \end{equation}
    where $O_i^j$ is the value of observable $O$ for $j$-th sweep in $i$-th simulation. In the normal situation, when the central limit theorem holds, this quantity should scale as $1/\sqrt{N}$. In our case, the tails are heavy enough to make the central limit theorem not applicable, and we will see that $\Delta^{(3)}_O$ deviates from this scaling.
\end{itemize}

\begin{figure*}[]
    \centering
\subfigure[]{\label{fig:err_stlocal_charge}\includegraphics[width=0.32\textwidth , angle=-90]{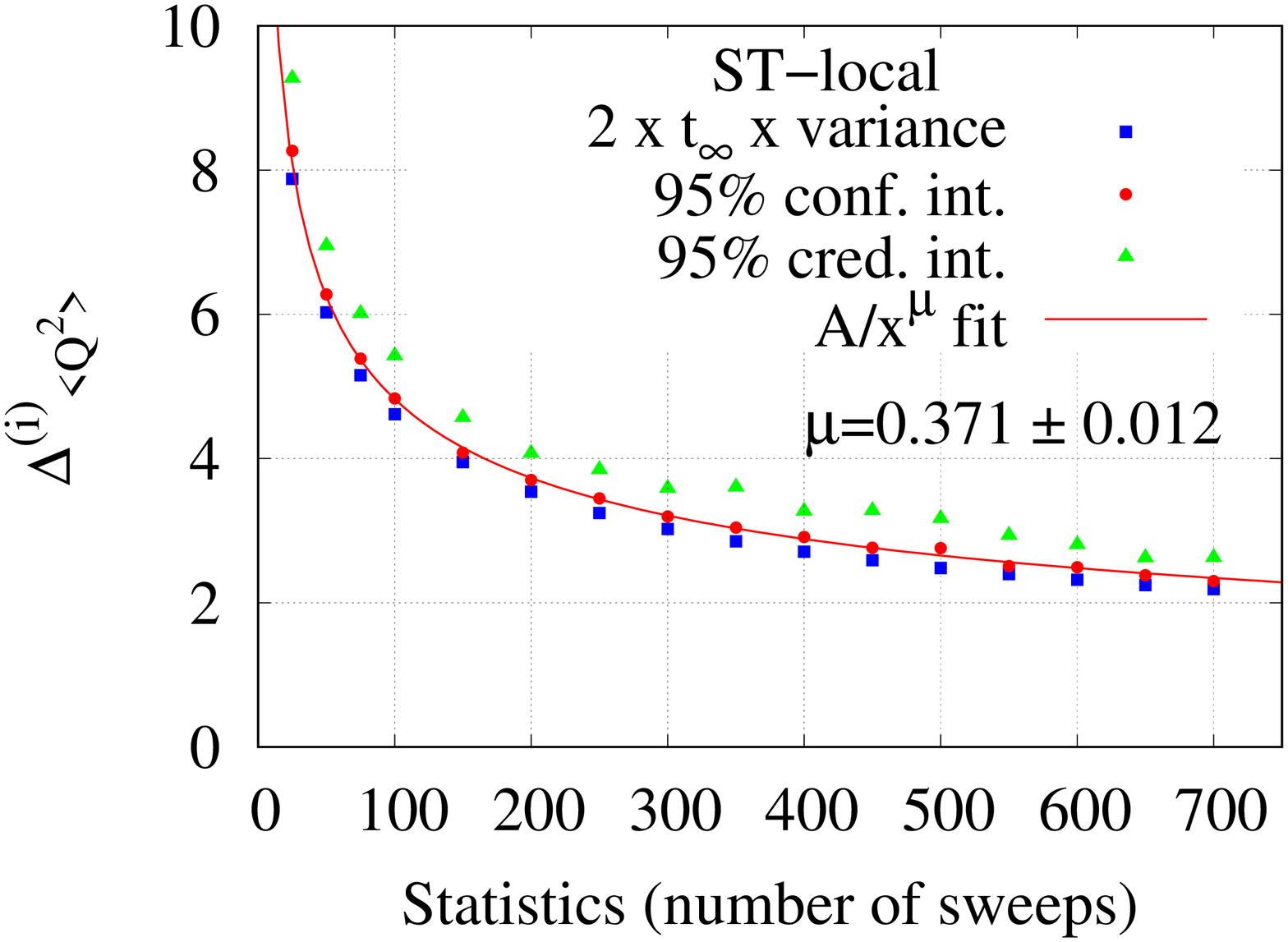}}
 \subfigure[]{\label{fig:err_tlocal_charge}\includegraphics[width=0.32\textwidth , angle=-90]{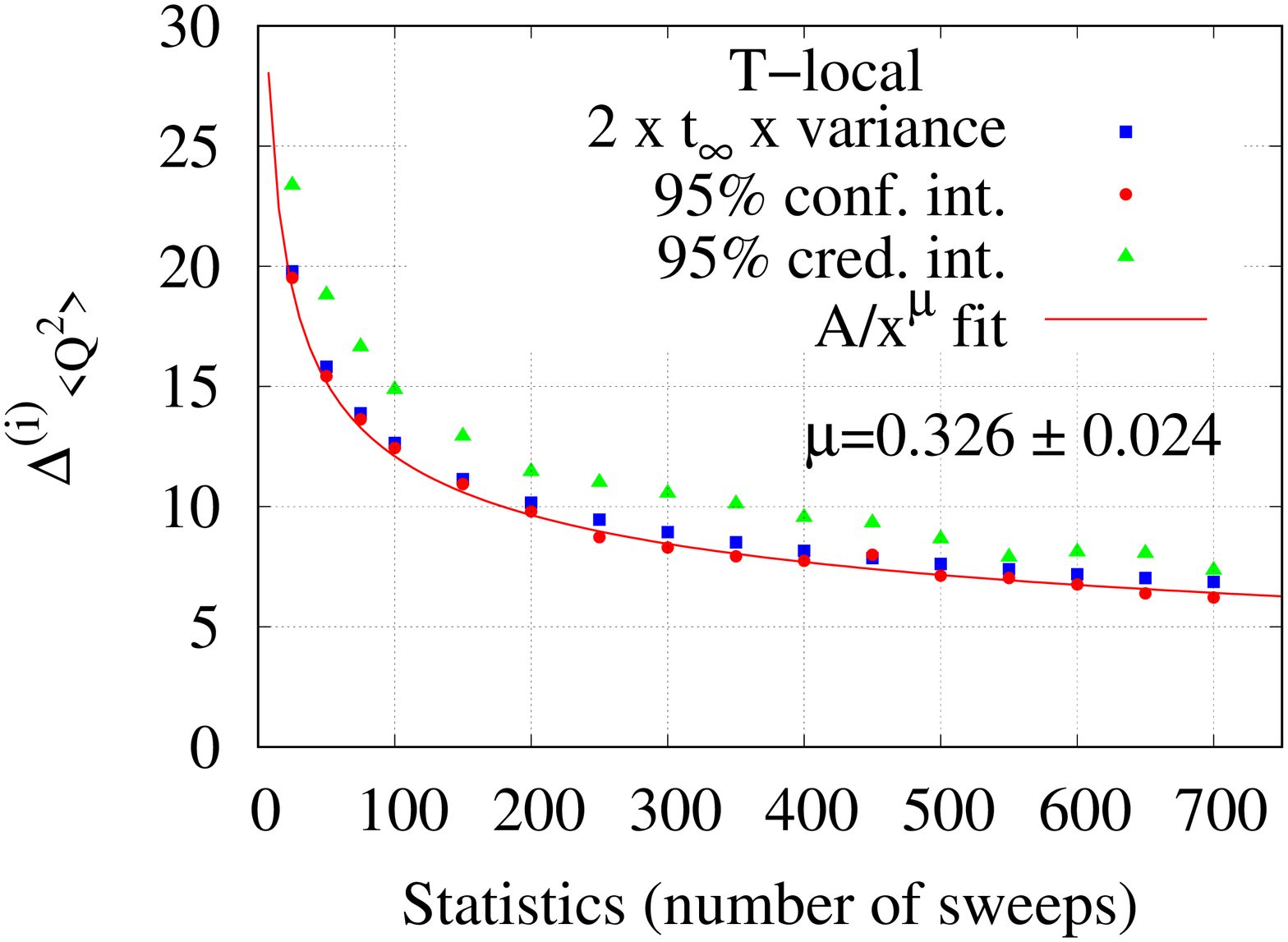}}
 \subfigure[]{\label{fig:err_stlocal_spin}\includegraphics[width=0.32\textwidth , angle=-90]{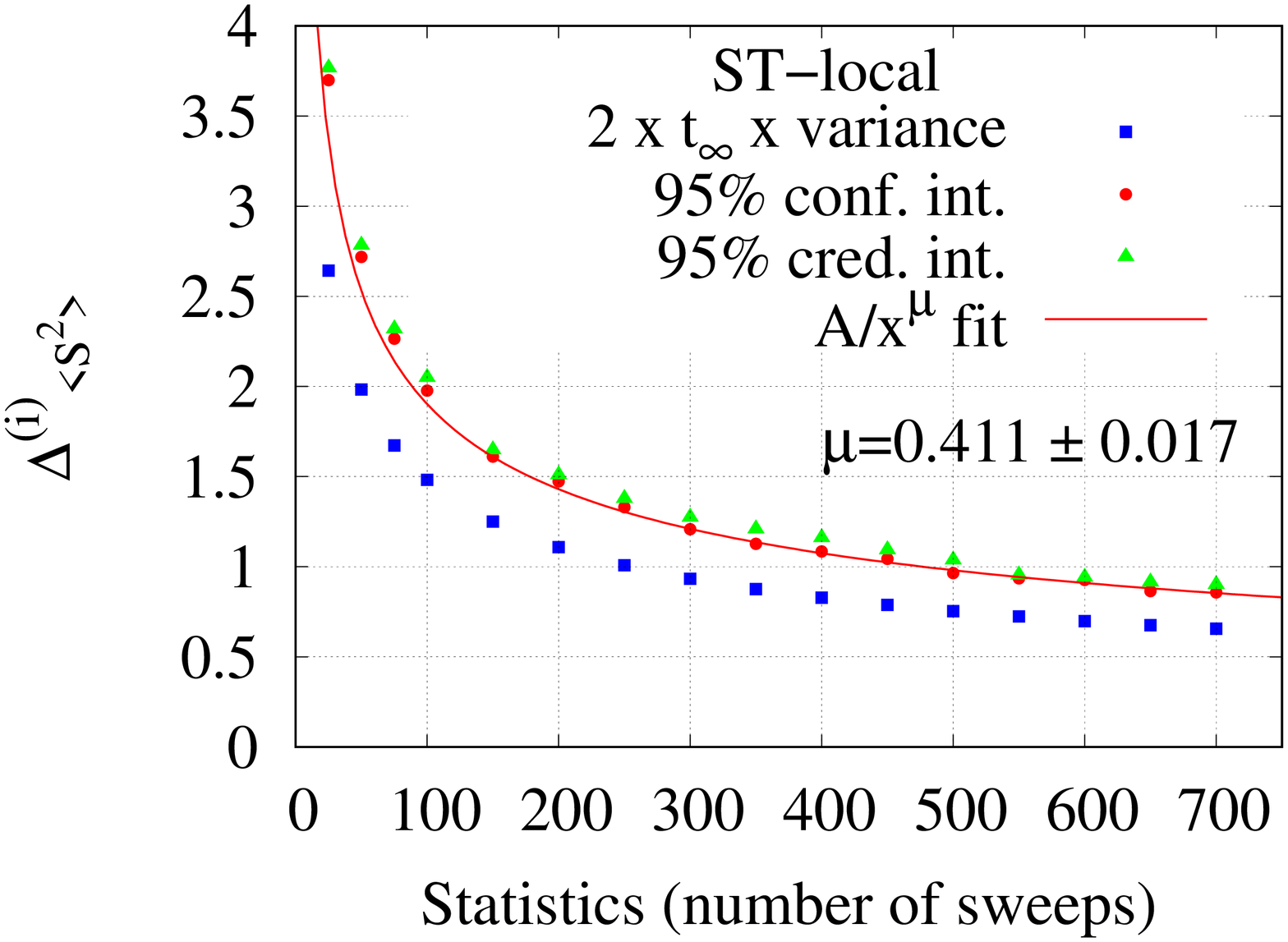}}
 \subfigure[]{\label{fig:err_tlocal_spin}\includegraphics[width=0.32\textwidth , angle=-90]{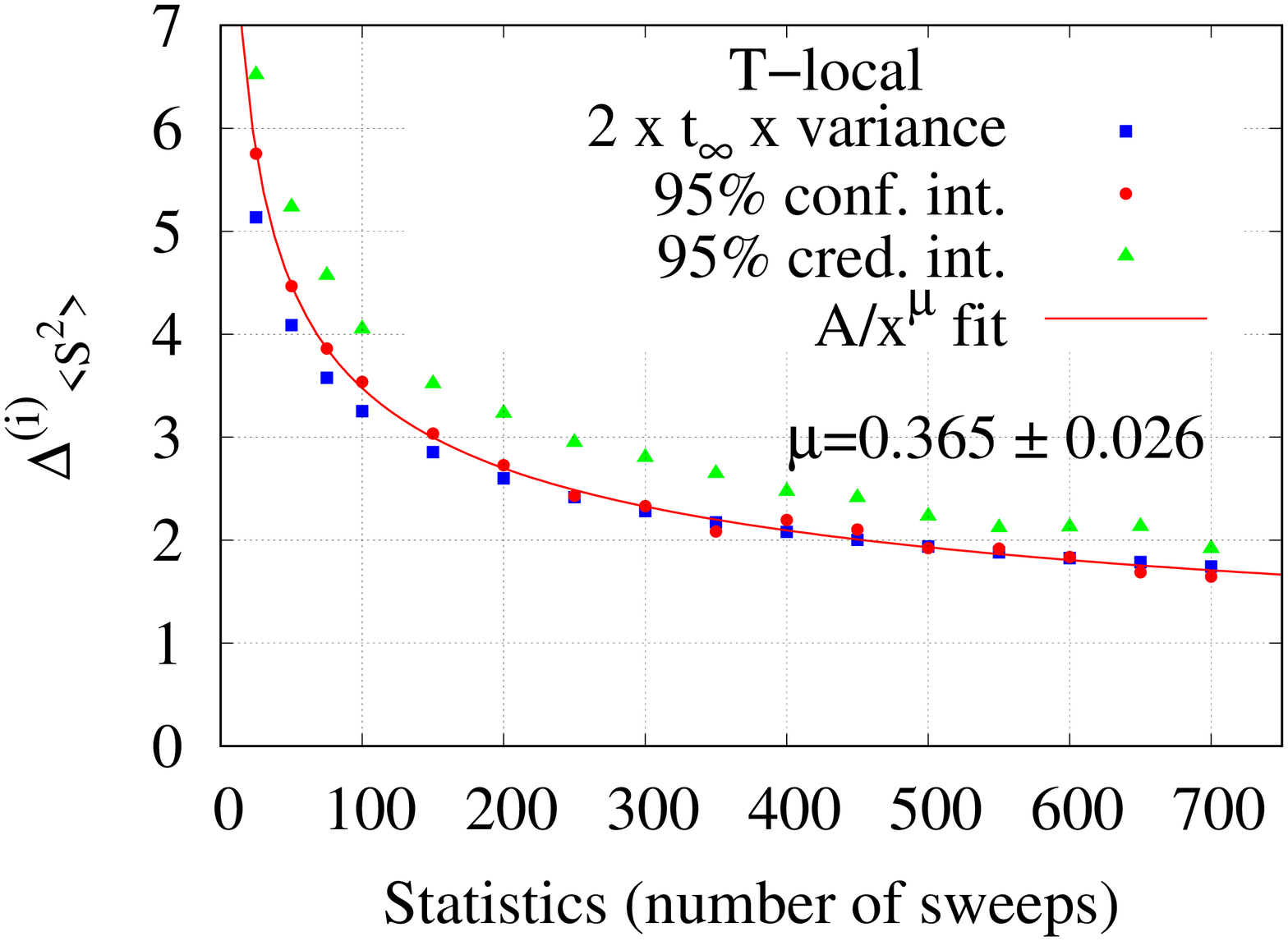}}
\subfigure[]{\label{fig:err_stlocal_d_occ}\includegraphics[width=0.32\textwidth , angle=-90]{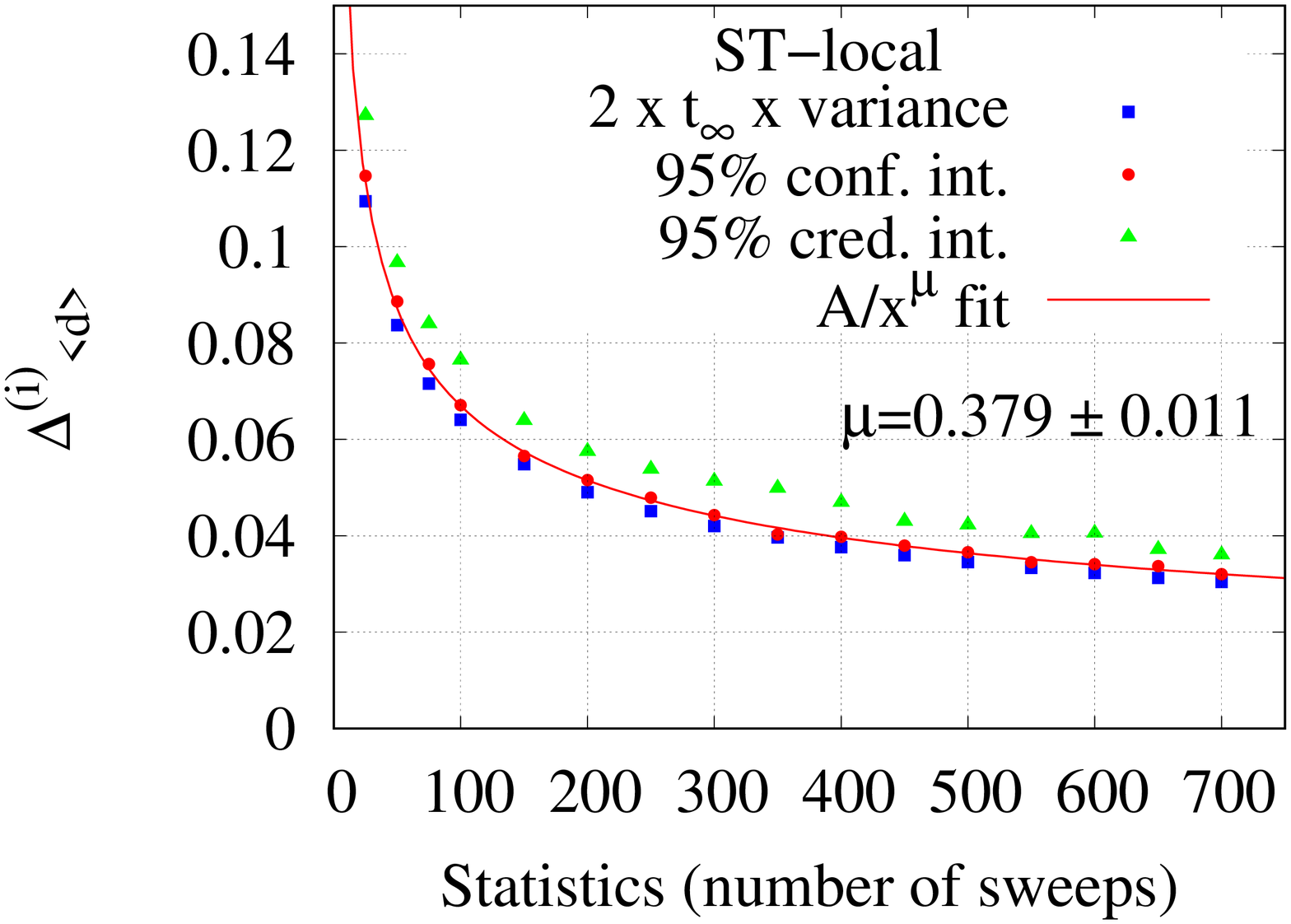}}
 \subfigure[]{\label{fig:err_tlocal_d_occ}\includegraphics[width=0.32\textwidth , angle=-90]{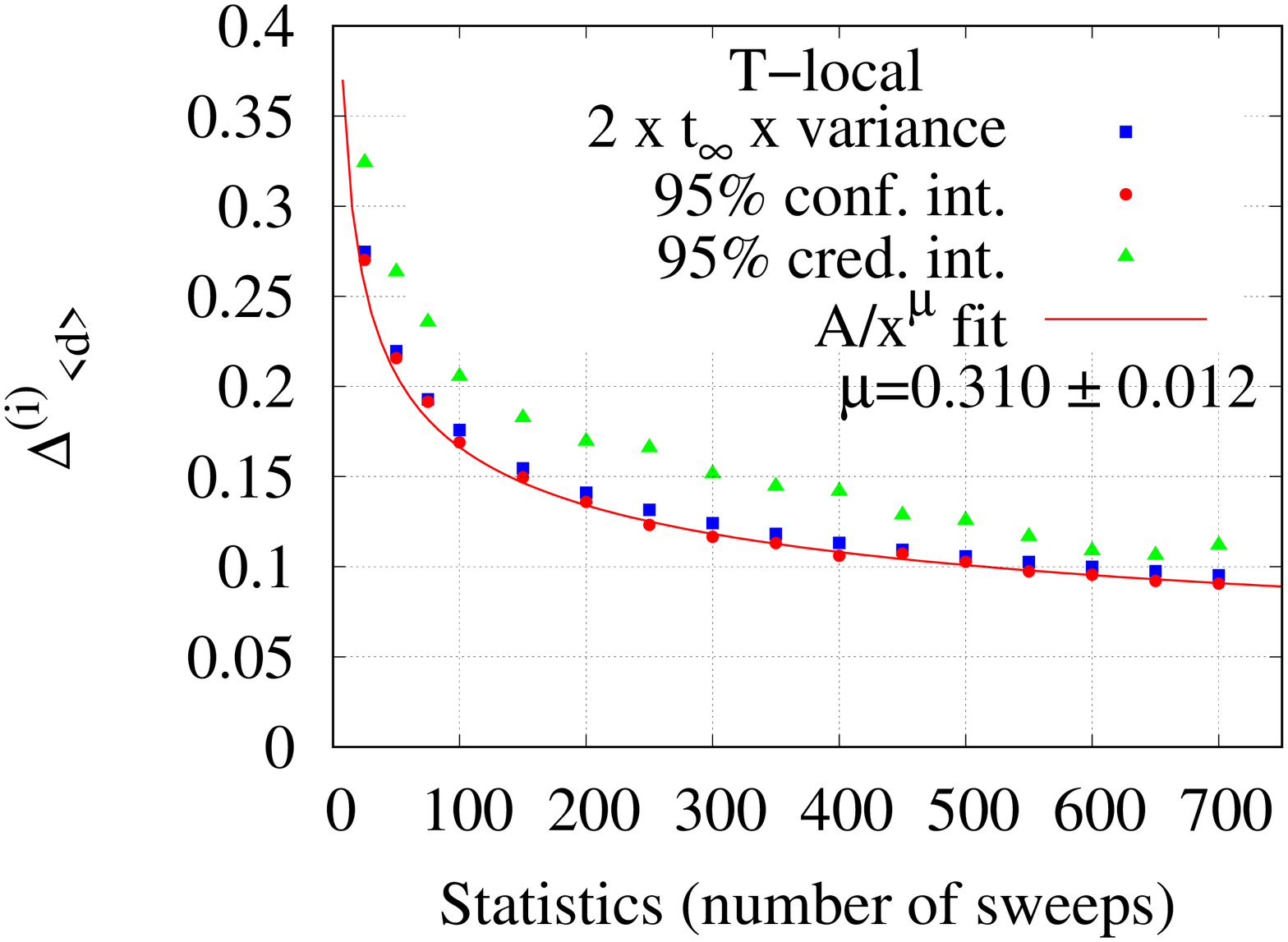}}
      \caption{Dependence of various error  estimates on the  number of sweeps per  simulation.  We consider the 95\% confidence interval, the 95\% credible interval, and the average variance  \ref{eq:av_variation}. In all cases we added a power law fit for the confidence interval data set.  Left column corresponds to the ST-local measurements and right column corresponds to the T-local measurements. Different rows show the results for different observables: the first (upper) row demonstrates the error bar for  the sublattice charge fluctuations, the second row for  the  sublattice  spin fluctuations  and the last (bottom) row for  the double occupancy. All data was produced on  the $6\times6\times256$ hexagonal lattice at $U=5.0$ and $\beta=20.0$.  The staggered mass was set to zero.}
   \label{fig:err_analysis}
\end{figure*}

\begin{figure*}[]
    \centering
\subfigure[]{\label{fig:err_comp_charge}\includegraphics[width=0.32\textwidth , angle=-90]{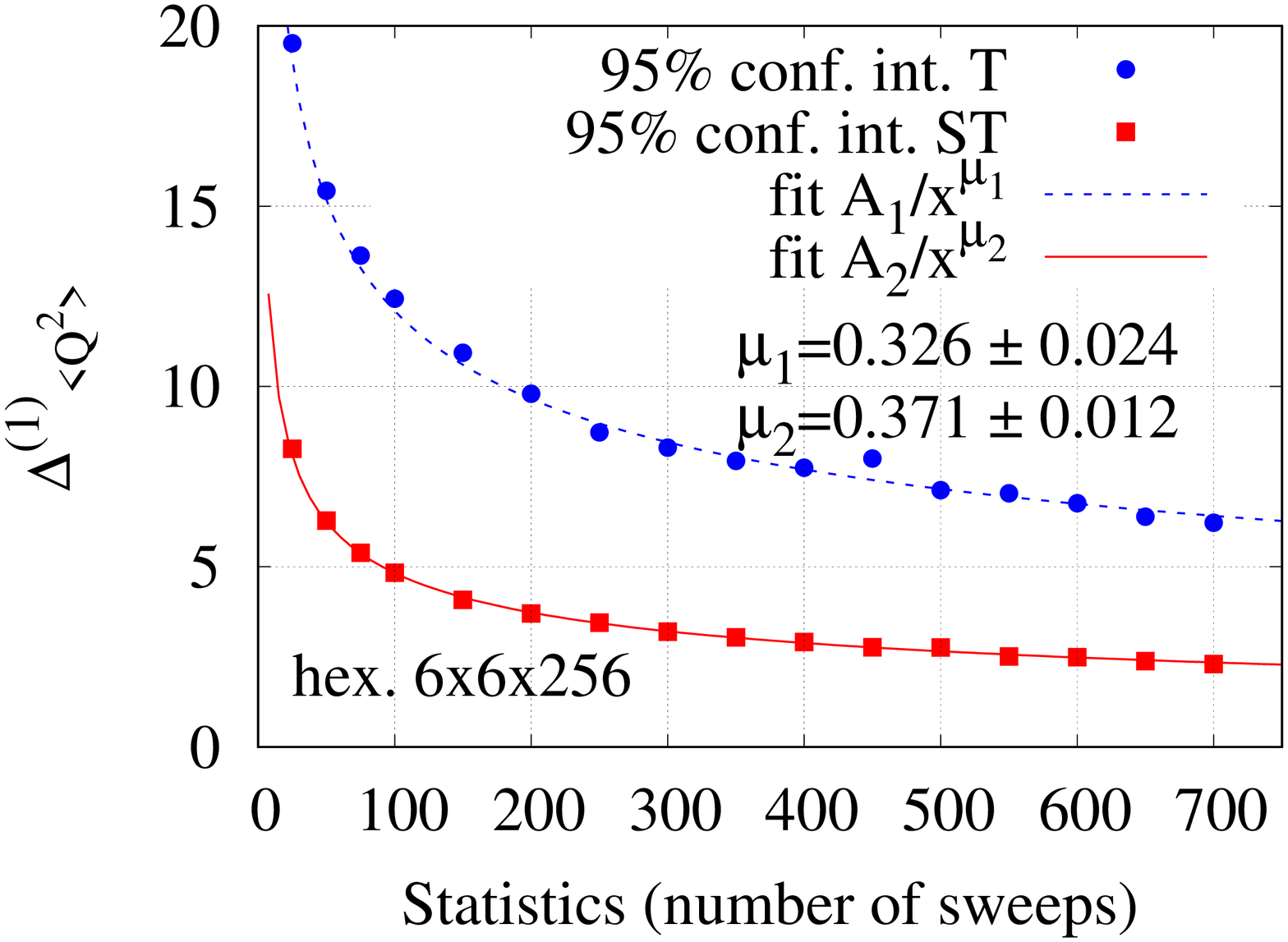}}
 \subfigure[]{\label{fig:err_comp_spin}\includegraphics[width=0.32\textwidth , angle=-90]{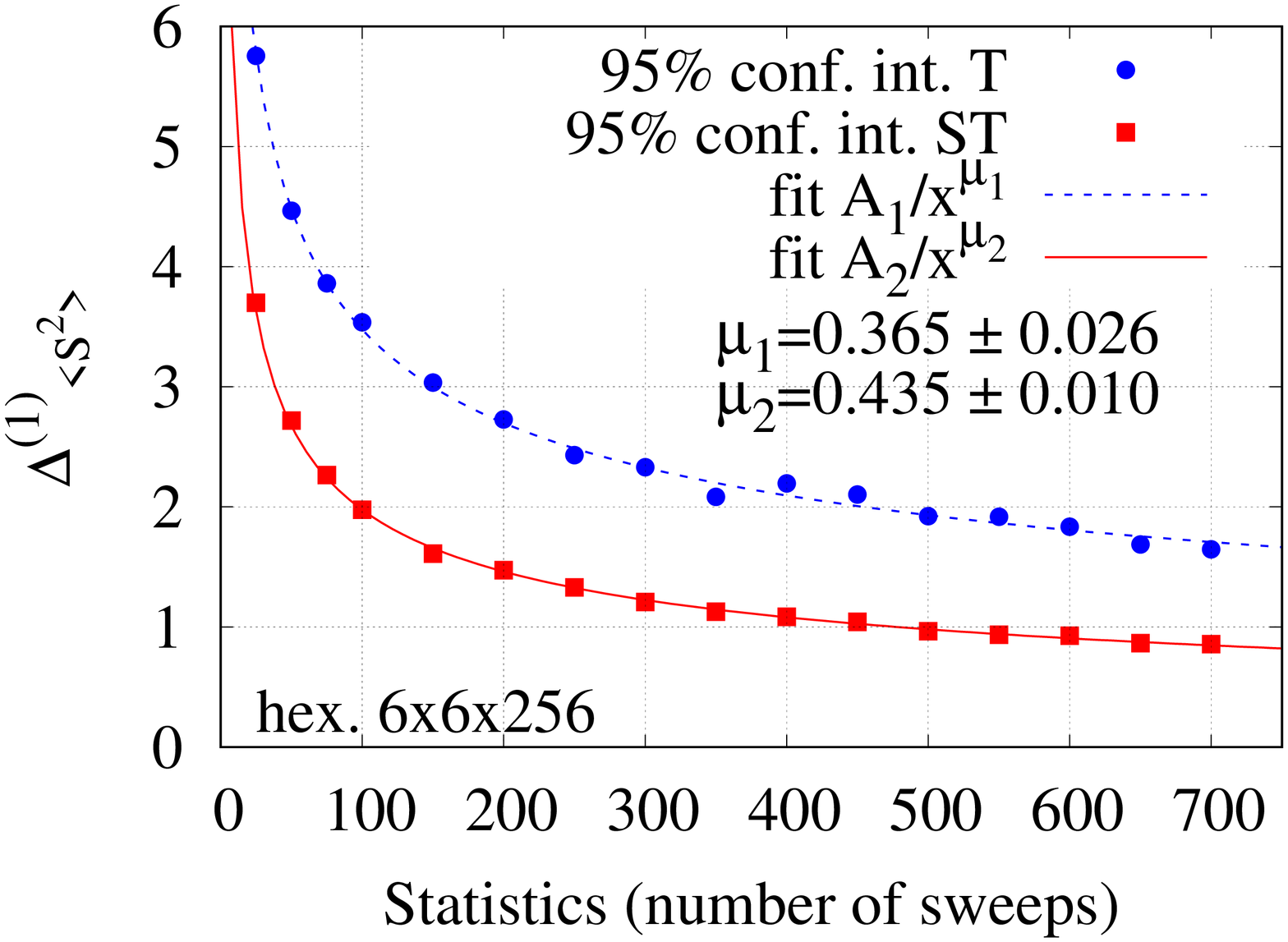}}
 \subfigure[]{\label{fig:err_comp_delta_d_occ}\includegraphics[width=0.32\textwidth , angle=-90]{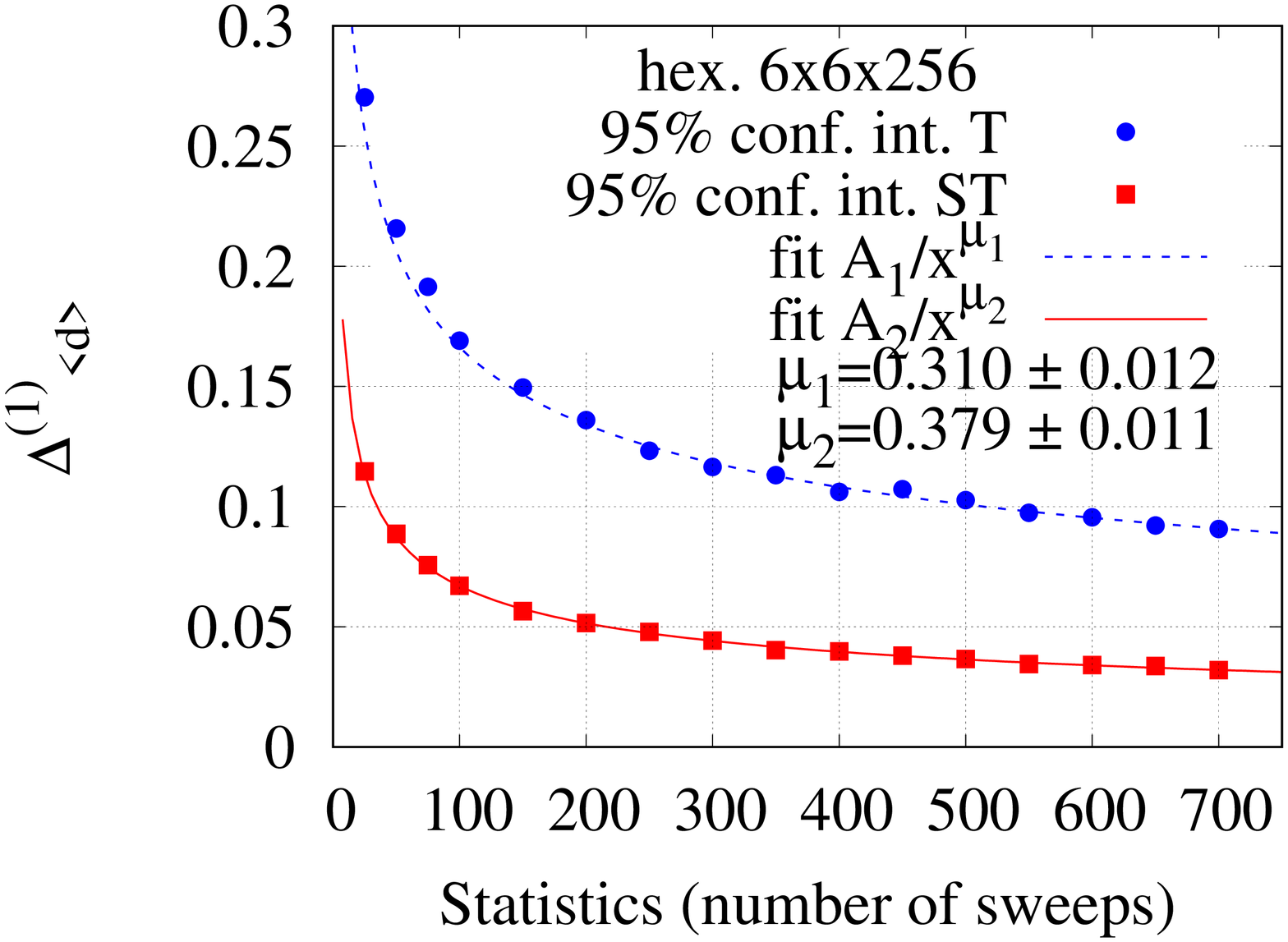}}
 \subfigure[]{\label{fig:err_comp_delta_spin}\includegraphics[width=0.32\textwidth , angle=-90]{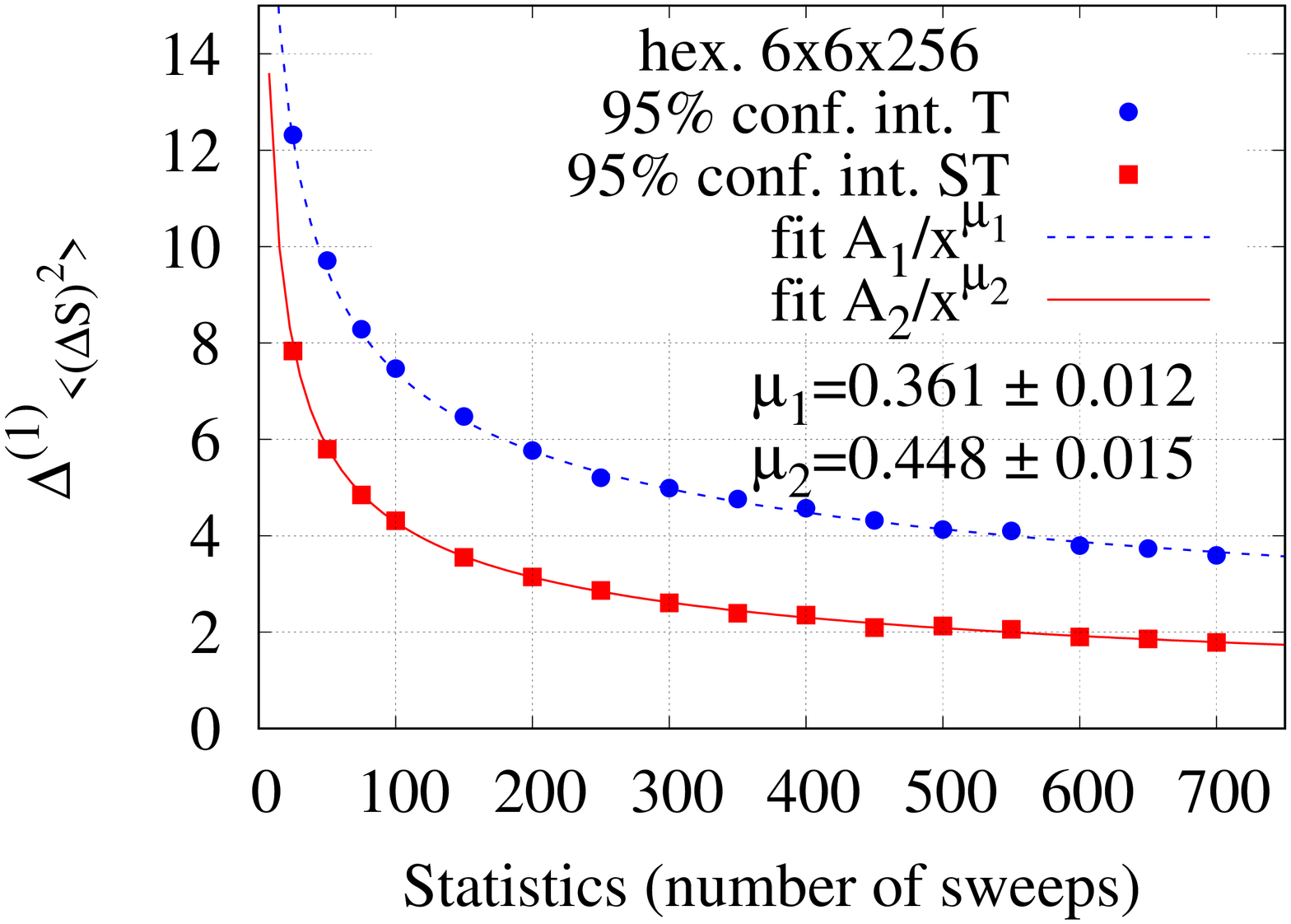}}
      \caption{Comparison of error bars (width of the 95\% confidence interval) for different observables depending on the number of sweeps $N$  per simulation and on the employed algorithm:  T-local or ST-local measurements. The  top left plot shows the sublattice  charge fluctuations, the top right corresponds  to the sublattice  spin fluctuations, bottom left is the double occupancy and bottom right  the squared staggered moment. In all cases we perform power law fits. The  values of the powers $\mu$ for all cases are inserted as labels in the plots. The data set was produced on the  $6\times6\times256$ hexagonal lattice at $U=5.0$ and $\beta=20.0$.  The staggered mass is  set  to zero.}
   \label{fig:err_comparison_local_6x6x256}
\end{figure*}

\begin{figure*}[]
    \centering
\subfigure[]{\label{fig:err_comp_charge_Ns12Nt256}\includegraphics[width=0.32\textwidth , angle=-90]{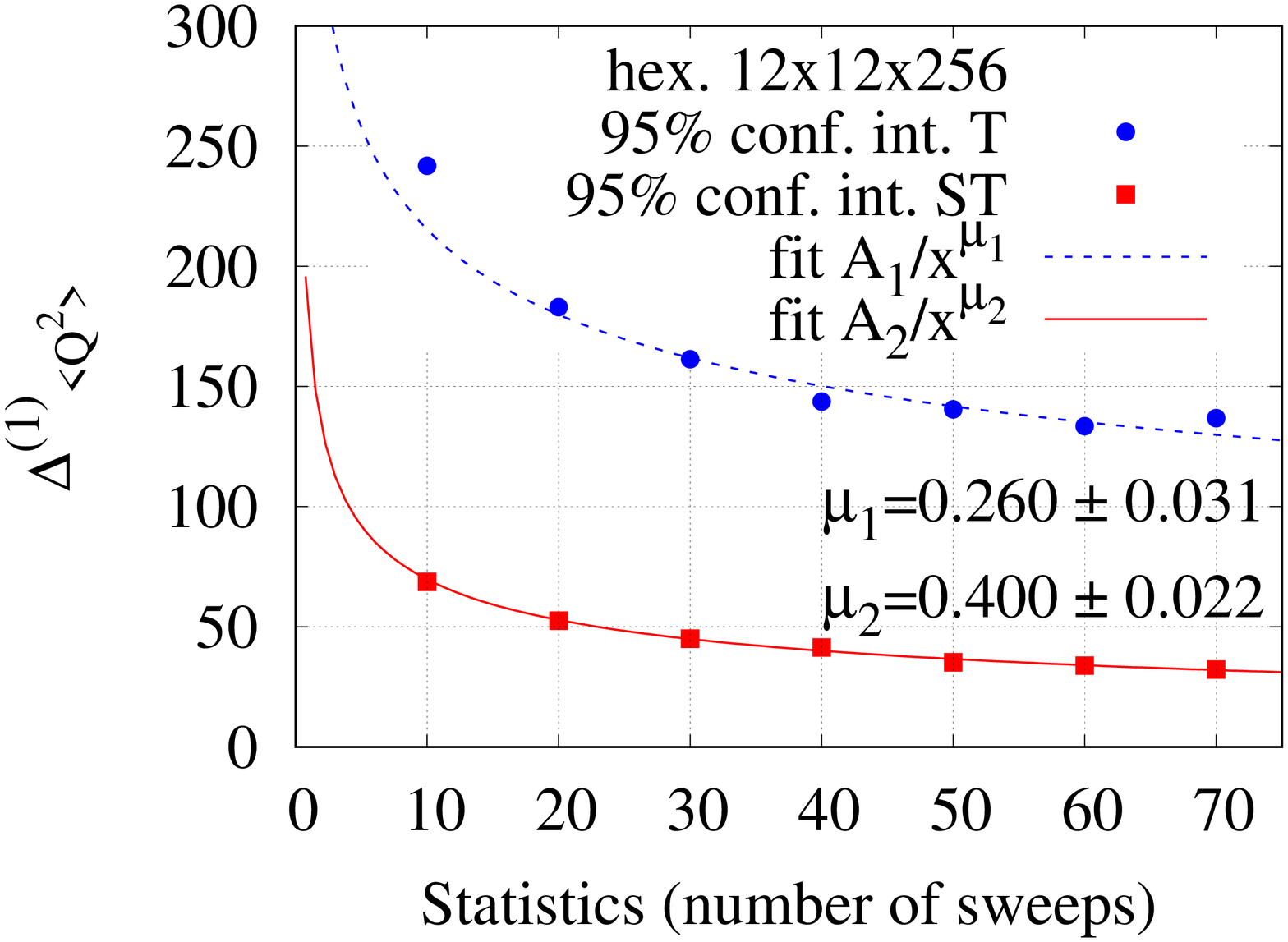}}
 \subfigure[]{\label{fig:err_comp_spin_Ns12Nt256}\includegraphics[width=0.32\textwidth , angle=-90]{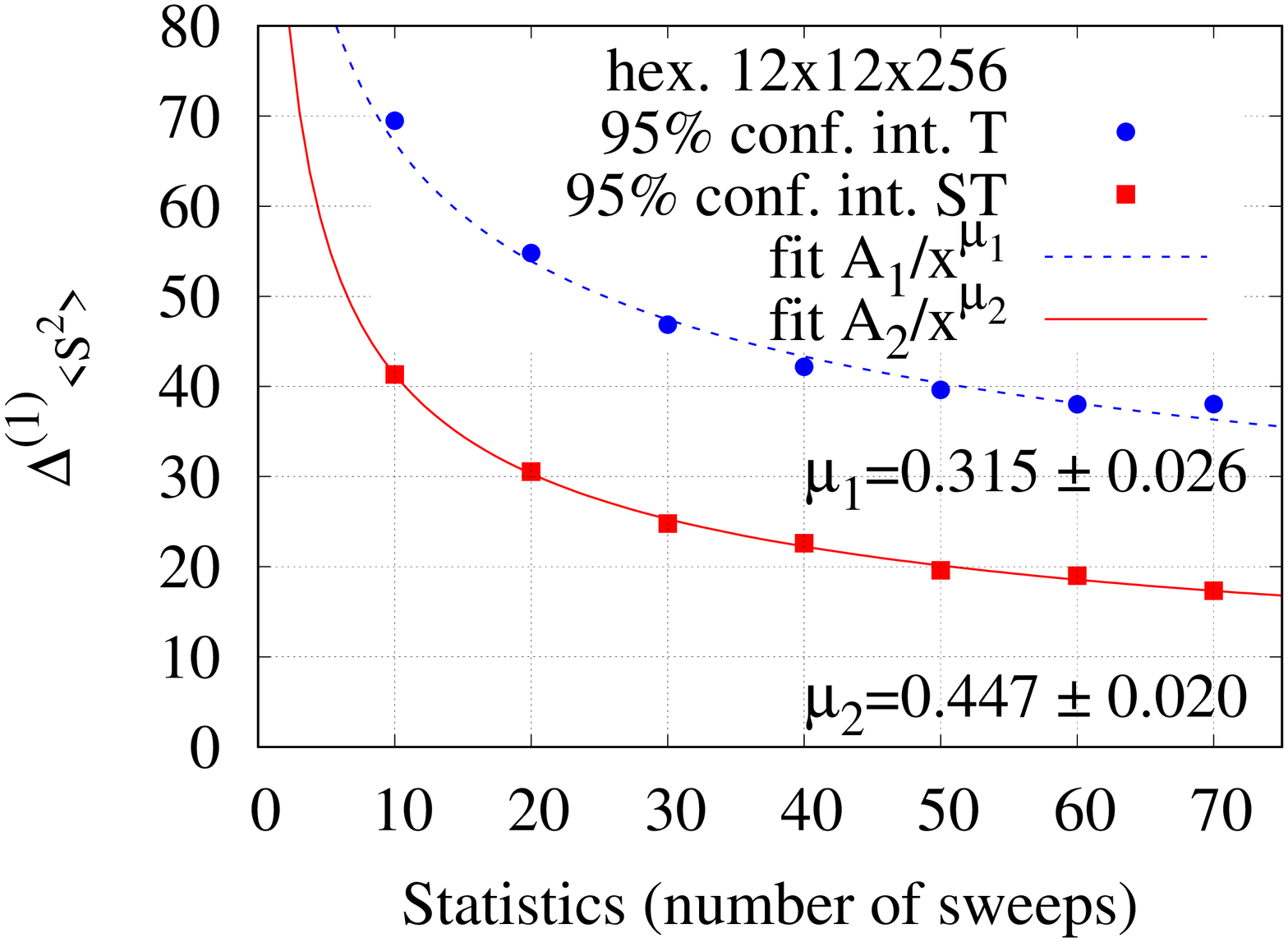}}
\subfigure[]{\label{fig:err_comp_charge_Ns6Nt512}\includegraphics[width=0.32\textwidth , angle=-90]{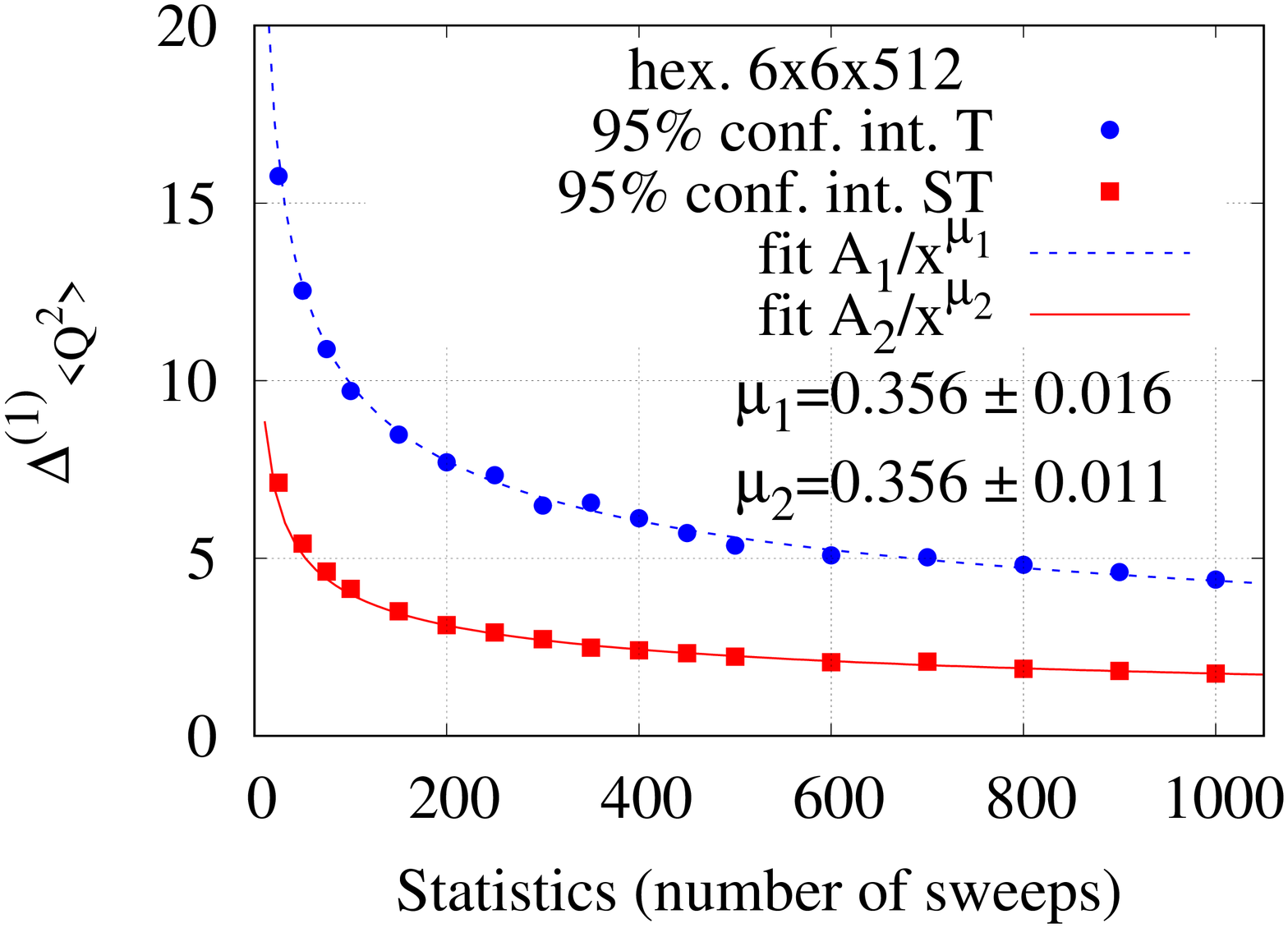}}
 \subfigure[]{\label{fig:err_comp_spin_Ns6Nt512}\includegraphics[width=0.32\textwidth , angle=-90]{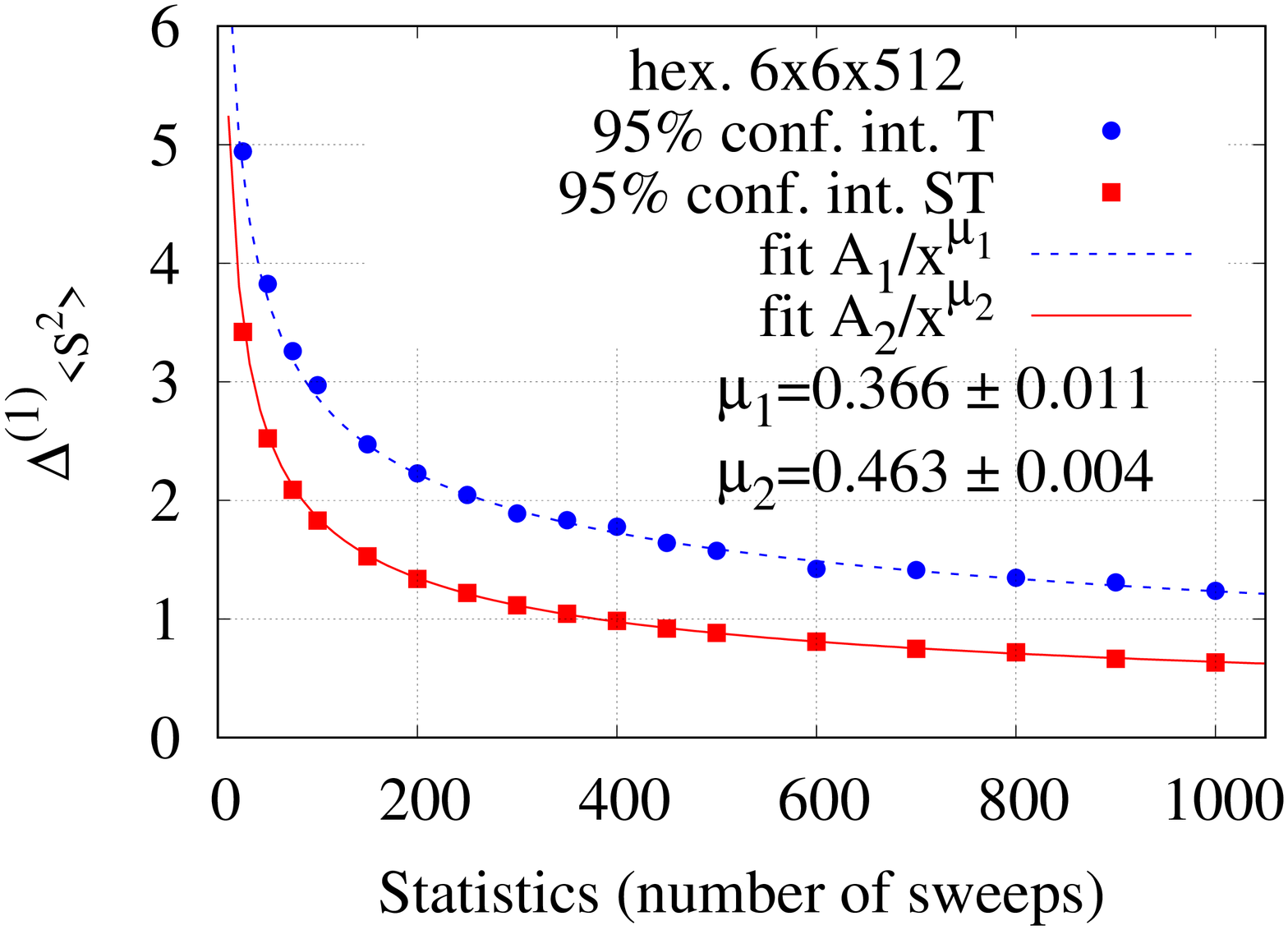}}
 \subfigure[]{\label{fig:err_comp_charge_Ns8Nt256sq}\includegraphics[width=0.32\textwidth , angle=-90]{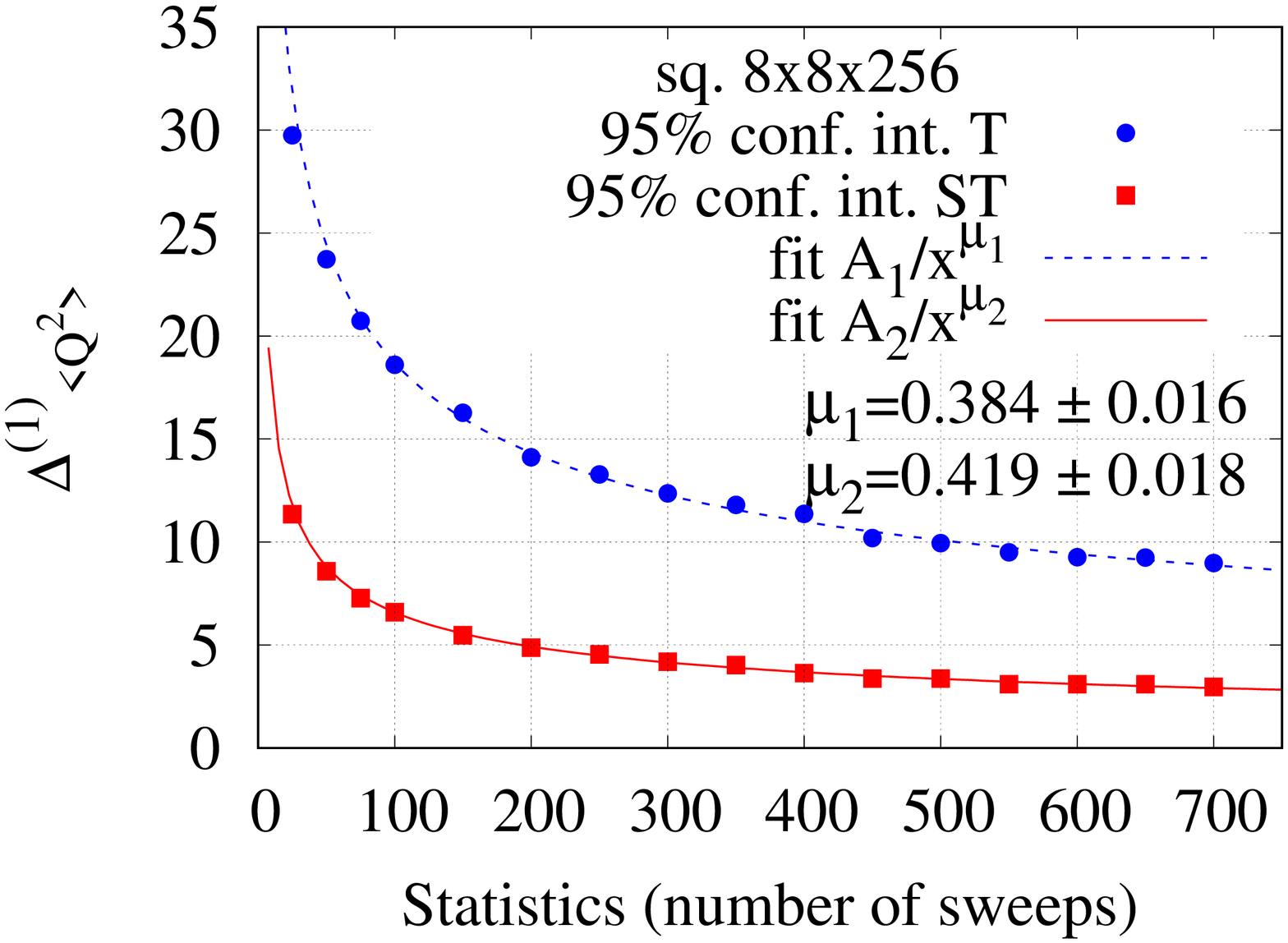}}
 \subfigure[]{\label{fig:err_comp_spin_Ns8Nt256sq}\includegraphics[width=0.32\textwidth , angle=-90]{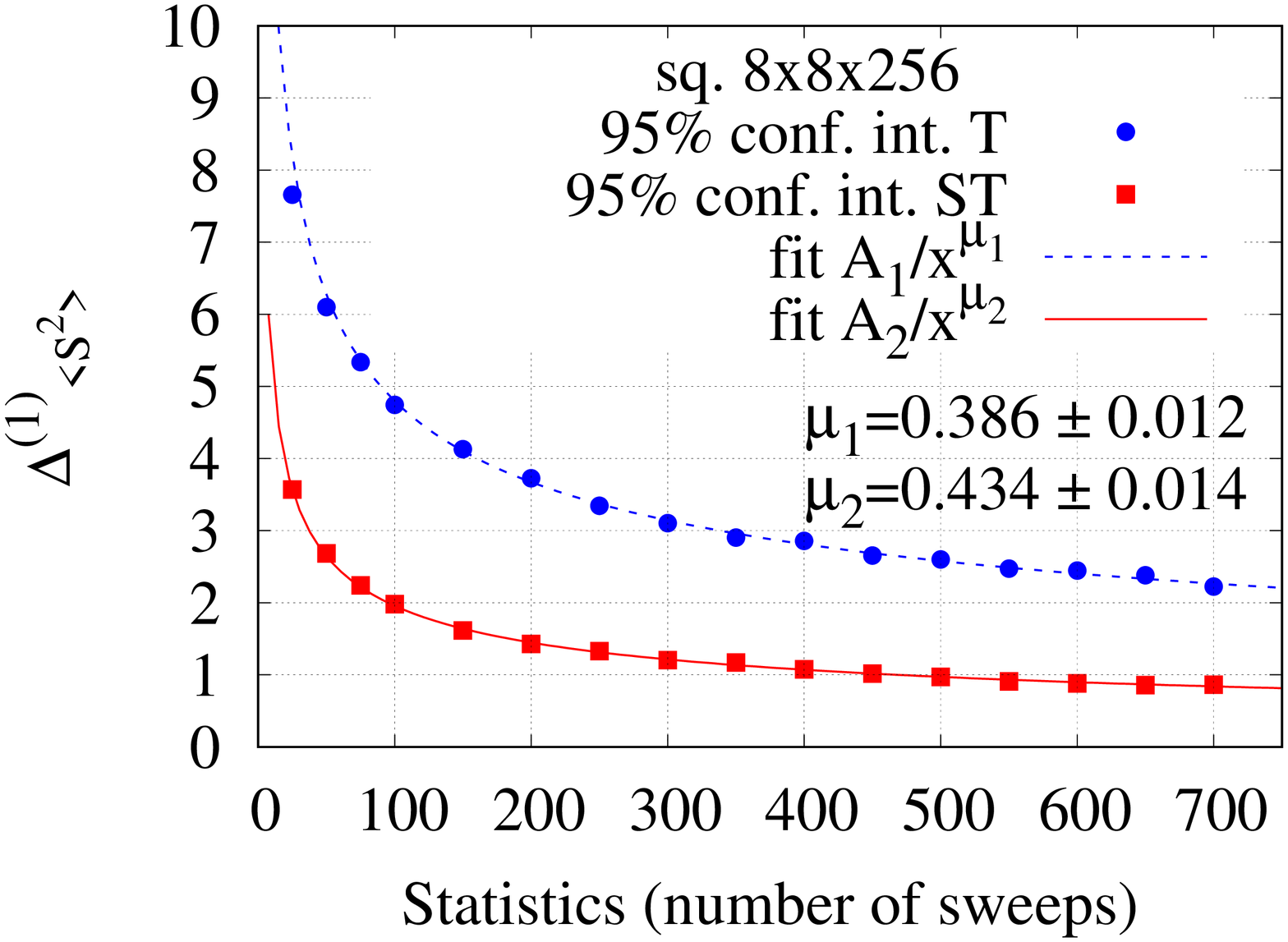}}
      \caption{Comparison of error bars (width of the 95\% confidence interval) for the sublattice charge  fluctuations  (left column) and sublattice spin fluctuations  (right column) depending on the number  of  sweeps per  simulation and  for different models and lattices. First (upper) row corresponds to a $12\times12\times256$ hexagonal lattice, the second row shows the results for  a $6\times6\times512$ hexagonal lattice and the third row corresponds to the square lattice Hubbard model on an $8\times8\times256$ lattice. Here we consider $U=5.0$ and $\beta=20.0$. For comparison with the  $6\times6\times256$ lattice, one can refer to the upper row in Fig.~\ref{fig:err_comparison_local_6x6x256}.}
   \label{fig:err_comparison_all}
\end{figure*}

Fig.~\ref{fig:histogram_bins} shows two examples of histograms for the averages over individual simulations. Fig.~\ref{fig:histogram_bins}\textcolor{red}{(a)} corresponds to the individual simulations with $N=25$ sweeps and Fig.~\ref{fig:histogram_bins}\textcolor{red}{(b)} corresponds to the simulations with $N=100$ sweeps. One can see how the probability distribution becomes more narrow, while remaining heavy tailed towards negative values of the observable.  
Vertical lines in the figures correspond to the borders of the 95\% credible interval $z_1$ and $z_2$ defined in Eq.\ref{eq:z1_def} and \ref{eq:z2_def}. Since the distribution becomes more narrow, the width of the credible interval also decreases from approximately 23.2 for the simulation size 25 to 14.9 for the simulation size 100. However, it does not decrease by two times as expected for the case when the central limit theorem holds.

The dependence of various error bars estimators on the number of sweeps, $N$, in the simulation  is shown in Fig.\ref{fig:err_analysis}. In all cases we compare the three measures: confidence interval $\Delta^{(1)}_O$, credible interval $\Delta^{(2)}_O$ and average variation $\Delta^{(3)}_O$. It appears that all three quantities give quite similar estimate for the error bar. The largest discrepancy is around 30\% and it is observed only in the single case of sublattice  spin  fluctuations  for ST-local measurements. In all other cases the discrepancy is smaller than 20\% and is typically observed as a deviation of credible interval, while the  average variance and confidence interval are in next to  perfect agreement. 
For further consideration we will exclusively use the 95\% confidence interval as a measure of the error bar.  The coincidence with the  average variance allows us to use the variance as an error estimate in realistic situation, when we do not repeat simulations $M$ times in order to directly observe the distribution of averages over individual simulations. Fitting the confidence interval data with the power law function $A/N^\mu$ shows that the error bars decrease noticeably slower than the $1/\sqrt{N}$ law. Typical values of $\mu$ are between 0.3 and 0.4.

Finally, we can compare different schemes of measurements using the 95\% confidence interval as an estimate of  the error bar. This comparison is done in Fig.~\ref{fig:err_comparison_local_6x6x256} for the $6\times6\times256$ lattice and for four different observables. In Fig.~\ref{fig:err_comparison_all} the  comparison is  carried out  for  the sublattice  spin and  charge  fluctuations and  on different lattices.
We generally compare only ST-local and T-local measurements, since the global  measurement scheme  is not competitive.  As one can see, the error bar is typically two to four times smaller for the ST-local scheme.  This  speedup is larger for the double occupancy and for the sublattice charge fluctuations, but smaller for observables involving spin:  compare right and left columns in Fig.~\ref{fig:err_comparison_local_6x6x256}.  As apparent  from the comparison of Figs.~\ref{fig:err_comparison_local_6x6x256}\textcolor{red}{(a)} and \ref{fig:err_comparison_all}\textcolor{red}{(a)}, the speedup  increases with the system size.

  \begin{figure}[]
   \centering
 \includegraphics[width=0.3\textwidth , angle=-90]{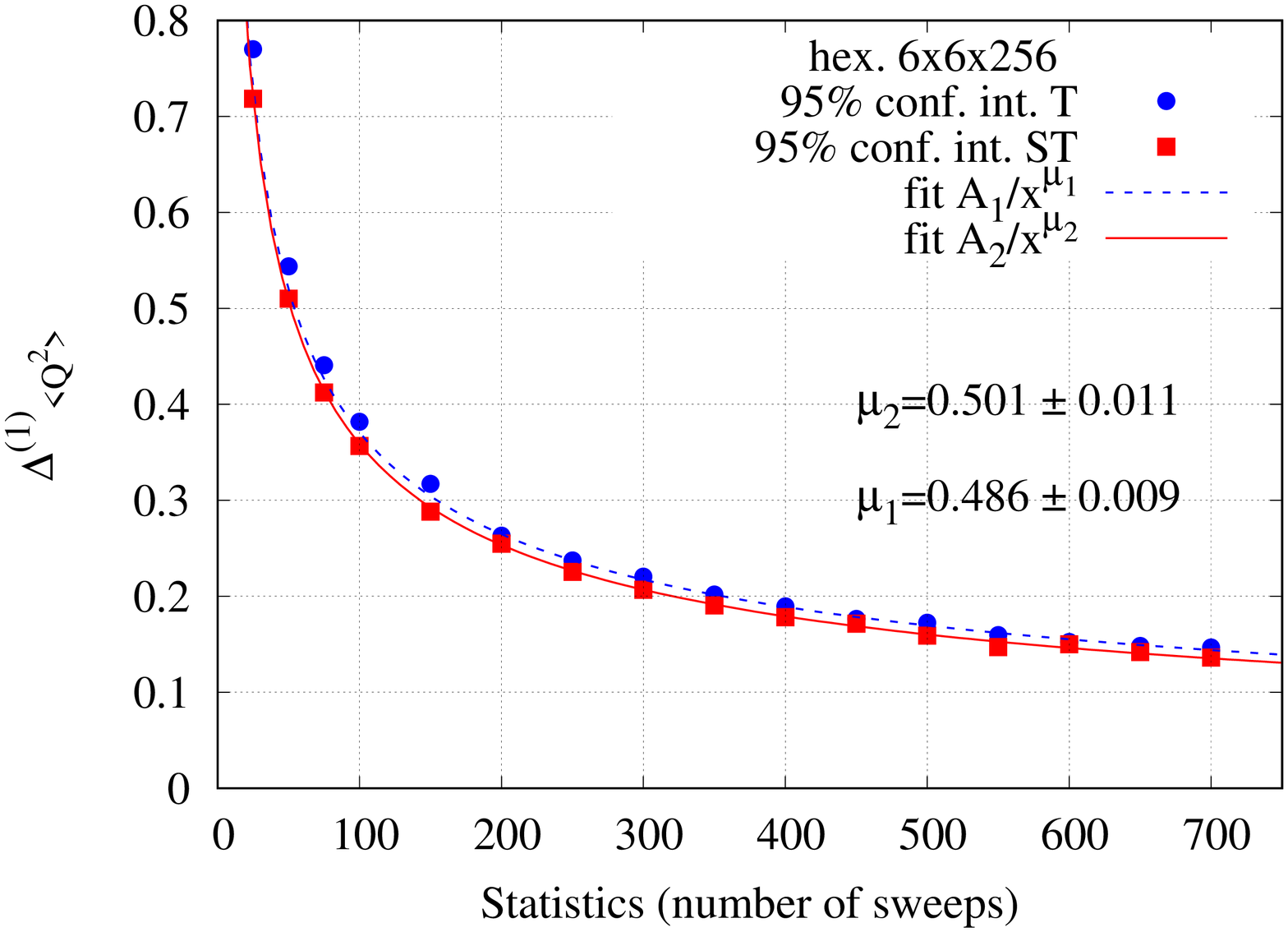} 
      \caption{Error bars (width of the 95\% confidence interval) for the  sublattice charge fluctuations as a function of the number of  sweeps 
      per  simulation.   Calculations were done on  the $6\times6\times256$ hexagonal lattice with the mass term $m=0.1$. The mass term   removes the zeros from the fermion  determinant.  Power law fits with their respective coefficients $\mu$ are shown in the plot. Here we  consider $U=5.0$ and $\beta=20.0$.}
   \label{fig:err_comparison_finite_mass}
\end{figure}

Generically, the dependence of the error bar on  the  number of  sweeps, N,  in a given  simulation can be described by the power law 
\begin{equation}
\Delta^{(1)}_O=\frac{A}{N^\mu}, \label{eq:err_power_law}
\end{equation}
with  power $\mu$ and prefactor $A$ dependent on the  measurement scheme, the lattice size, etc. An important observation is that $\mu$ is consistently larger for ST-local measurements, thus the confidence interval  decreases faster in that case and the speedup will increase with decreasing error bar. For the speedup presented in Fig.~\ref{fig:average_speedup} in the main text, we use  a conservative estimate corresponding to the largest $\mu$ within the error bars for T-local measurements and the smallest $\mu$ within the error bars for the ST-local measurements.

  \begin{figure}[]
     \centering
\subfigure[]{\label{fig:Tlocal_corr}\includegraphics[width=0.32\textwidth , angle=-90]{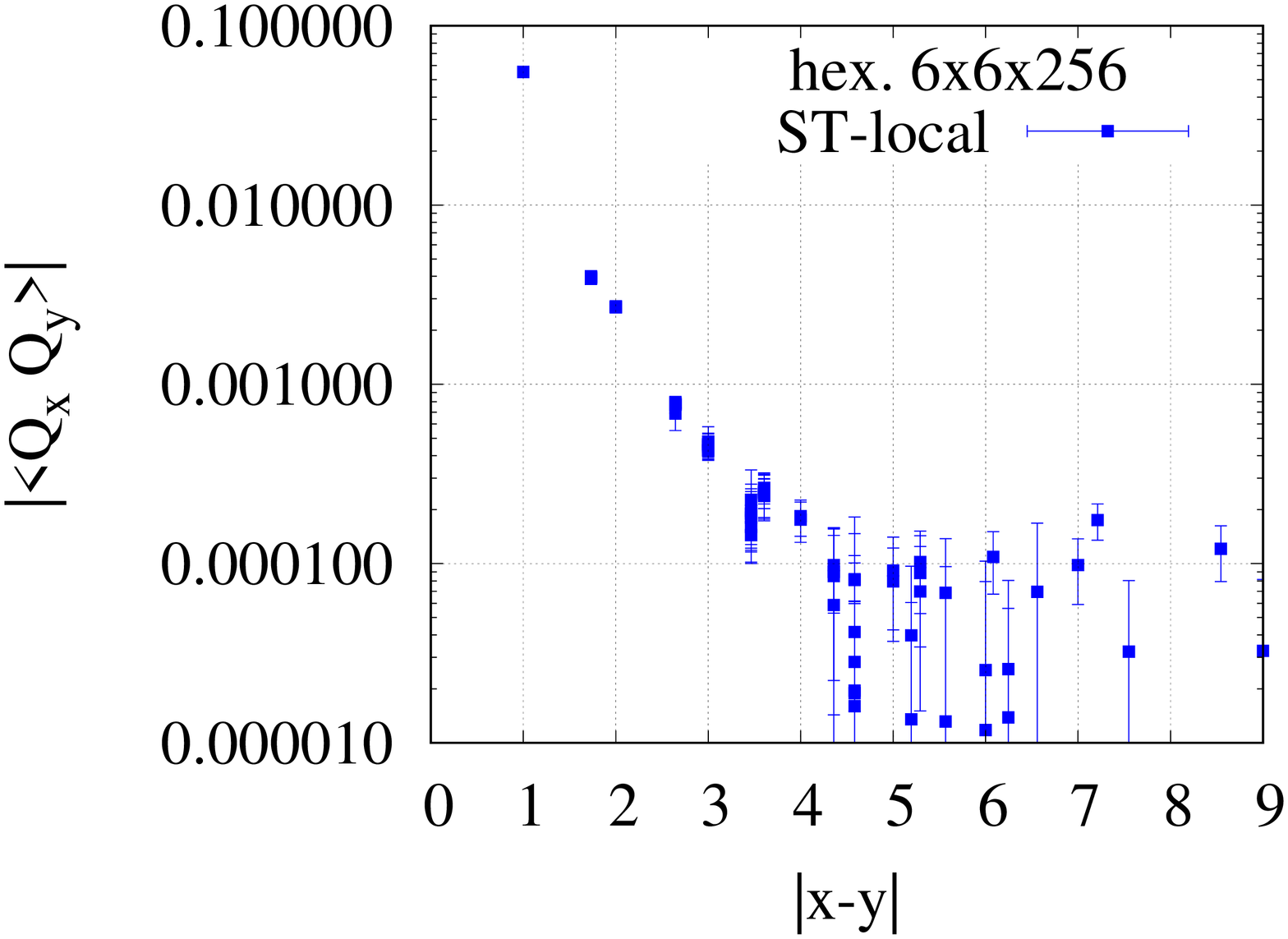}}
  \subfigure[]{\label{fig:STlocal_corr}\includegraphics[width=0.32\textwidth , angle=-90]{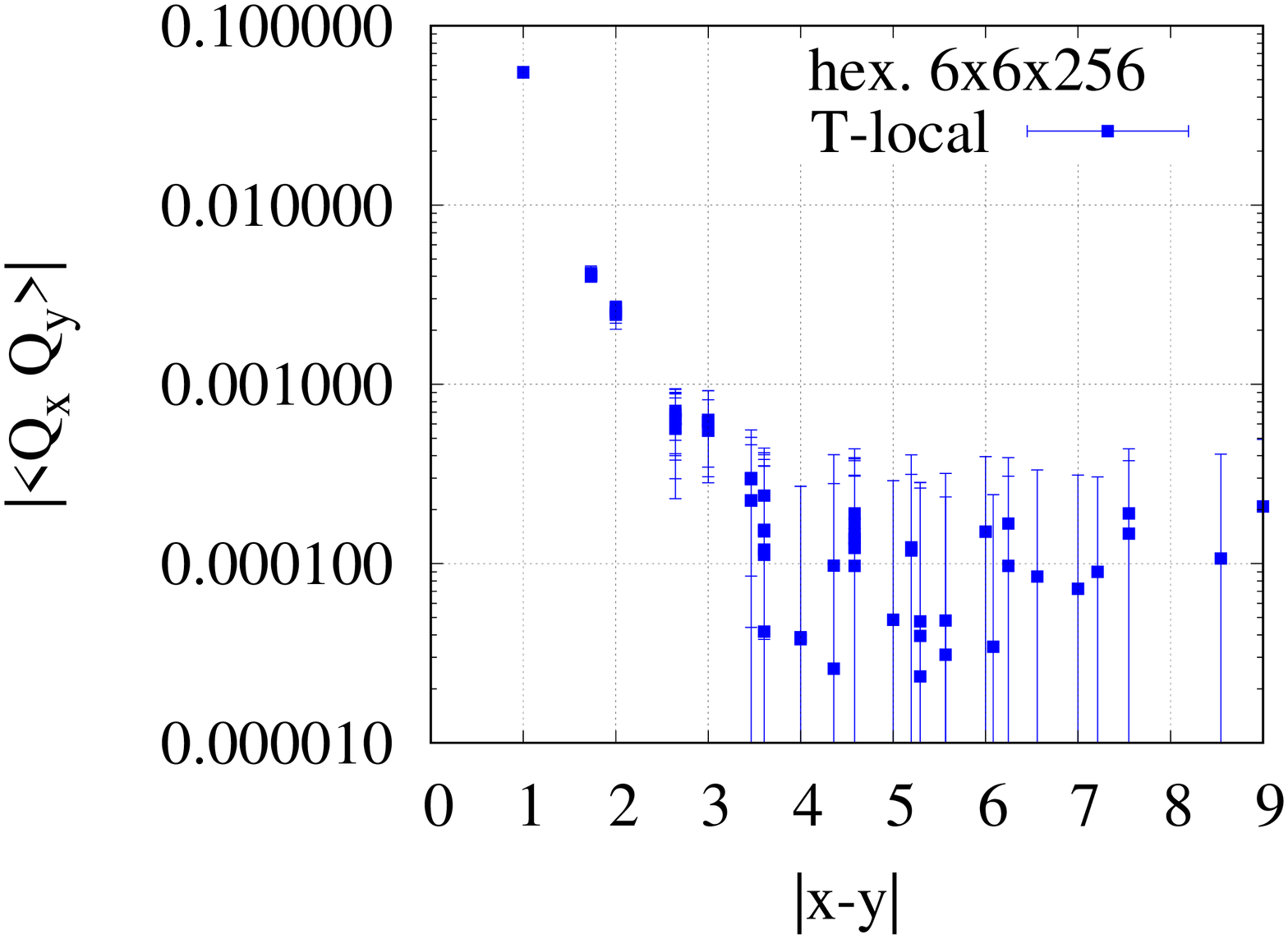}}
      \caption{Comparison of charge-charge correlators computed with the same statistics (1000 sweeps through the lattice) using ST-local measurements (a) and T-local measurements (b). The distance between sites is in  units of the separation between nearest neighbours. Here we  consider a  $6\times6\times256$ lattice at  $\beta=20$ and $U=5$.}
   \label{fig:correlators}
\end{figure}

Up to  now, we  have  considered  cases where the  fermion  determinant posses  zeros.  Fig.~\ref{fig:err_comparison_finite_mass} shows the case of non-zero explicit mass term \ref{eq:Hamiltonian_mass} inserted in the Hamiltonian \ref{eq:Hamiltonian}.  The differences between T-local and ST-local measurements disappear: both constant prefactor $A$ and the power $\mu$ are  identical within  the quoted  uncertainty. 

Another characteristic example of the reduction of error bars is shown in Fig. \ref{fig:correlators}. We compare the charge-charge correlator computed 
within the  T-local (Fig. \ref{fig:correlators}\textcolor{red}{(a)}) and ST-local (Fig. \ref{fig:correlators}\textcolor{red}{(b)}) schemes. The improvement of  the error bars is noticeable.  In the latter figure we can reliably trace the correlator up to the distance of 4 lattice units,  while  for the T-local scheme we control   the correlator data up to $|x-y|=3$.

\section{\label{sec:AppendixB} Power law in distributions of observables}
\label{AppendixB}
In this appendix we derive the power law \ref{eq:toy_asymptotic} for the distribution of observables in the case of real QMC simulations for continuous auxiliary fields. 

Let us consider the complementary cumulative distribution function for some four-fermion observable $\Theta$. Among other terms, $\Theta$ always includes the second power of Euclidean fermionic propagator $g^2$. This is the most important term in the vicinity of the configurations where the fermion determinant is zero, since the propagator diverges there as $1/\Delta$,  where $\Delta$ is the distance to the zero point of the determinant in the space of auxiliary fields $\phi_{{\ve{i}},\tau}$. Due to the divergence of observables near the zero points of the determinant, exactly these field configurations define the complementary cumulative distribution function $\bar F_\Theta (\theta)$ at large values of the observable. The asymptotic behaviour of the function $\bar F$ can be described by the integral: 
\begin{equation}
    \bar F_{\Theta}(\theta) =\mathcal{P} (\Theta>\theta) =\int_{V:\Theta>\theta} d^N \phi P (\phi),
    \label{eq:cumul_distr1}
\end{equation}
where $P(\phi)$ is the probability distribution for the $\phi$ fields and $N$ corresponds  to the   total  number of HS  fields.

First we consider the case where the zeros of determinant form ``domain walls'' in configuration space. If $\theta$ is sufficiently large, the volume $V$ is just some thin layer in the vicinity of the ``domain wall'', where  $P(\phi)=0$.  Now we change the variables so that $x_2...x_N$ correspond to the shift parallel to the ``domain wall'' while the coordinate $x_1$ is perpendicular to it. The ``domain wall'' itself corresponds to $x_1=0$. Thus
\begin{equation}
    \bar F_{\Theta}(\theta)|_{\theta \rightarrow \infty} =\int_{V:\Theta>\theta} d^N x \frac{\mathcal{D}(\phi)}{\mathcal{D}(x)} x_1^2 f(x_2...x_N),
     \label{eq:cumul_distr2}
\end{equation}
and we used Eq.~\ref{eq:Z_continuous} in order to estimate the probability distribution $P(x)$ in the vicinity of ``domain walls'' as $P(x) \approx x_1^2 f(x_2...x_N)$. This estimation follows from the equivalence of fermion determinants for electrons and holes at half filling. Since the observable diverges as we approach the ``domain wall'':
\begin{equation}
    \Theta\approx \frac{C(x_2, ... x_N)}{x_1^2},
\end{equation}
the integral over $x_1$ in \ref{eq:cumul_distr2} has the limits
\begin{equation}
x_1 \in [-C_1(x_2, ... x_N)/\sqrt{\theta},C_2(x_2, ... x_N)/\sqrt{\theta} ],  
\end{equation}
where $C_1,C_2>0$. If the Jacobian doesn't have any divergences in the limit $x_1\rightarrow 0$, the integral over $x_1$ in \ref{eq:cumul_distr2} can be computed separately. Thus the asymptotic behaviour of the function $\bar F$ is described by the expression:
\begin{eqnarray}
    \bar F_{\Theta}(\theta)|_{\theta \rightarrow \infty} \approx \frac{\mathcal{C} }{\theta^{3/2}}, \\
     \mathcal{C} =  \frac{1}{3}  \int dx_2 ... dx_N \left.{\frac{\mathcal{D}(\phi)}{\mathcal{D}(x)}}\right|_{{x_1}=0}\times \\ \times \left({C_1(x_2,...x_N)}^3 + {C_2(x_2,...x_N)}^3\right). \nonumber
     \label{eq:cumul_distr_fin1}
\end{eqnarray}
Conversion to the probability distribution gives
\begin{eqnarray}
    P_{\Theta}(\theta)|_{\theta\rightarrow \infty} \sim \frac{1}{\theta^{5/2}}.
    \label{eq:asympt1}
\end{eqnarray}
The derivation can also be modified to accommodate the case when the dimensionality of the manifolds with zero determinant is reduced to $N-2$, where $N$ is the total number of auxiliary fields $\phi$.  We should now separate two coordinates $x_1$ and $x_2$ which corresponds to the shift perpendicular to the ``line'' with zero fermion determinant, while all other coordinates $x_3,..x_N$ again correspond to the shift parallel to this ``line''. After it $x_1$ and $x_2$ are changed to polar coordinates $(\rho, \varphi)$ and the resulting asymptotic behaviour for the complementary cumulative distribution function is described by the expression
\begin{eqnarray}
    \bar F_{\Theta}(\theta)|_{\theta \rightarrow \infty}  \approx  \int d\varphi dx_3 ... dx_N \\ \int_0^{C(\varphi,x_3,..x_N)/\sqrt{\theta}} \rho^3 d\rho   \left.{\frac{\mathcal{D}(\phi)}{\mathcal{D}(x)}}\right|_{{\rho}=0} f(\varphi, x_3...x_N). \nonumber
     \label{eq:cumul_distr_fin2}
\end{eqnarray}
An additional power of $\rho$ appears from the transfer to polar coordinates. Alternatively, one can say that this power appears from the Jacobian if the transfer to polar coordinate is included in the general change of variables $\phi \rightarrow x$. The probability distribution for the observable $\Theta$ has now the asymptote
\begin{eqnarray}
    P_{\Theta}(\theta)|_{\theta\rightarrow \infty} \sim \frac{1}{\theta^3}.
    \label{eq:asympt2}
\end{eqnarray}
Similar derivations can be repeated for lower dimensions of the manifolds with zero fermion determinant with larger powers of $\rho$ appearing from the Jacobian. Generally, the lower the dimensionality of the manifolds with zero fermion determinant leads to the larger absolute value of the power in the tail of the distribution.

\end{document}